\documentclass[11pt]{combine}
\usepackage{amsmath,amsfonts,amssymb,amsthm,epsfig, graphicx, hyperref}
\usepackage[dvipsnames,table,dvipsnames*, svgnames*, hyperref]{xcolor}
\usepackage{natbib,amsmath,amssymb,amsthm,graphicx,setspace,paralist,booktabs,rotating,subcaption,float,color}
\usepackage{multirow}
\usepackage{fancyhdr}
\usepackage{bibunits}
\usepackage[small]{titlesec}

\newtheorem{theorem}{Theorem}
\newtheorem{corollary}{Corollary}
\newtheorem{proposition}{Proposition}
\newtheorem{assumption}{Assumption}
\newtheorem{lemma}{Lemma}
{
\theoremstyle{definition}
\newtheorem{definition}{Definition}
\newtheorem{example}{Example}
\newtheorem{remark}{Remark}
}
\newcommand{\beq}{\begin{equation}}
\newcommand{\eeq}{\end{equation}}
\newcommand{\beas}{\begin{align*}}
\newcommand{\eeas}{\end{align*}}
\newcommand{\bea}{\begin{align}}
\newcommand{\eea}{\end{align}}
\newcommand{\bei}{\begin{itemize}}
\newcommand{\eei}{\end{itemize}}
\newcommand{\ben}{\begin{enumerate}}
\newcommand{\een}{\end{enumerate}}
\newcommand{\bet}{\begin{theorem}}
\newcommand{\eet}{\end{theorem}}
\newcommand{\bel}{\begin{lemma}}
\newcommand{\eel}{\end{lemma}}
\newcommand{\bep}{\begin{proposition}}
\newcommand{\eep}{\end{proposition}}

\newcommand{\bed}{\begin{definition}}
\newcommand{\eed}{\end{definition}}
\newcommand{\bec}{\begin{corollary}}
\newcommand{\eec}{\end{corollary}}
\newcommand{\bex}{\begin{example}}
\newcommand{\eex}{\end{example}}

\newcommand{\E}{\mathbb{E}}

\newtheorem{lem1}{Lemma S.}
\def\T{{ \mathrm{\scriptscriptstyle T} }}

\def\E{{\mathbb E}}
\def\p{{\text{pr}}}
\numberwithin{equation}{section}

\usepackage{algorithm,algorithmic}

\begin{document}


\title{\scshape Heterogeneity-aware and communication-efficient distributed statistical inference}
\author{Rui Duan$^1$, Yang Ning$^2$ and Yong Chen$^3$ \\
$^1$Department of Biostatistics, Harvard T.H. Chan Shool of Public Health,\\
Boston, MA 02115\\
$^2$Department of Statistics and Data Science,\\ Cornell University, Ithaca, NY, 14853\\
$^3$Department of Biostatistics, Epidemiology and Informatics, \\
University of Pennsylvania,
Philadelphia, PA 19104\\
}
\date{}
\maketitle
\thispagestyle{empty}

	\begin{abstract}
In multicenter research, individual-level data are often protected against sharing across sites. To overcome the barrier of data sharing, many distributed algorithms, which only require sharing aggregated information, have been developed. The existing distributed algorithms usually assume the data are homogeneously distributed across sites. This assumption ignores the important fact that the data collected at different sites may come from various sub-populations and environments,  which can lead to heterogeneity in the distribution of the data. Ignoring the heterogeneity may lead to erroneous statistical inference. In this paper, we propose distributed algorithms which account for the heterogeneous distributions by allowing site-specific nuisance parameters. The proposed methods extend the surrogate likelihood approach \citep{wang2017efficient,jordan2018communication} to the heterogeneous setting by applying a novel density ratio tilting method to the efficient score function. The proposed algorithms maintain the same communication cost as the existing communication-efficient algorithms. We establish a non-asymptotic risk bound for the proposed distributed estimator and its limiting distribution in the two-index asymptotic setting which allows both sample size per site and the number of sites to go to infinity. In addition, we show that the asymptotic variance of the estimator attains the Cram\'er-Rao lower bound when the number of sites is in rate smaller than the sample size at each site. Finally, we use  simulation studies and a real data application to demonstrate the validity and feasibility of the proposed methods.
	\bigskip
	
	\noindent\emph{KEY WORDS}: Data integration; distributed inference; efficient score; surrogate likelihood; two-index asymptotics
\end{abstract}
The growth of availability and variety of clinical data has induced the trend of multicenter research \citep{sidransky2009multicenter}.  Multicenter research confers many distinct advantages over single-center studies, including the ability to study rare exposures/outcomes that require larger sample sizes, accelerating the discovery of more generalizable findings, and bringing together investigators who share and leverage resources, expertise, and ideas \citep{cheng2017conducting}. 
Since individual-level information is often protected by privacy regularities and rules, directly pooling data across multiple clinical sites is less feasible or requires large amount of operational efforts \citep{barrows1996privacy}.  As a consequence, healthcare systems need more effective tools for evidence synthesis across clinical sites.

Distributed algorithms, also known as ``divide-and-conquer" procedures, have been applied to multicenter studies. In the classical divide-and-conquer framework, the entire data set is split into multiple subsets and the final estimator is obtained by averaging the local estimators computed using the data from each subset \citep{li2013statistical,chen2014split,lee2017communication, tian2016communication,zhao2016partially,  lian2017divide, battey2018distributed,wang2019distributed}.  The class of methods adopts the same principle as meta-analysis in the area of evidence synthesis and systematic review, where the local estimates are combined through a fixed effect or random effects model \citep{dersimonian1986meta}. When the number of research sites is relatively small, these averaging type of methods are able to perform equally well as the combined analysis using data from all the sites \citep{hedges1983combining,olkin1998comparison,battey2018distributed}. When the number of research sites is large, as we will demonstrate in the simulation studies, these averaging methods may not be as good as the combined analysis. More importantly, when studying rare conditions, some clinical sites do not have enough number of cases to achieve the asymptotic properties.  In such cases, the averaging methods can be suboptimal. 

Recently, \cite{wang2017efficient} and \cite{jordan2018communication} proposed a novel surrogate likelihood approach, which approximates the higher order derivatives of the global likelihood by using the likelihood function in a local site. 
This method has low communication cost and improves the performance of the average method especially when the number of sites is large, see \cite{duan2019odal} for a real data application to pharamcoepidemiology. From the practical perspective, the surrogate likelihood approach endowed a highly feasible framework for sharing sensitive data in a collaborative environment, especially in biomedical sciences, where the lead investigators often have access to the individual-level data in their home institute, and the collaborative investigators from other sites are willing to share summary statistics but not individual-level information.

Most of the aforementioned distributed algorithms assumed that the data at different sites are independently and identically distributed. However, a prominent concern in multi-center analysis is that there may exist a non-negligible degree of heterogeneity across sites because the samples collected in different sites may come from different sub-populations and environments. One concrete example is the Observational Health Data Sciences and Informatics consortium, which contains over 82 clinical databases from over 20 countries around the world \citep{hripcsak2015observational}. The amount of heterogeneity cannot be ignored when implementing distributed algorithms in such healthcare networks.




To the best of our knowledge, \cite{zhao2016partially} is the only work in this area that considers a similar heterogeneous setting. They generalized the divide-and-conquer approach by averaging all the local estimators and studied theoretical properties under the partially linear model. Different from this work, we propose to account for the heterogeneous distributions via a general parametric likelihood framework by allowing site-specific nuisance parameters. In particular, we extend the surrogate likelihood function approach to a surrogate estimating  equation approach, and propose a density-ratio tilted surrogate efficient score function which only requires the individual-level data from a local site and summary statistics from the other sites. To reduce the influence of estimation of the site-specific nuisance parameters, we propose to use the efficient score function for distributed inference rather than the score function as in \cite{jordan2018communication}. We further adjust for the degree of heterogeneity by applying a novel density ratio tilting method to the efficient score function. We refer the resulting score function to the surrogate efficient score function. The estimator is defined as the root of this function. We show that the communication cost of the proposed algorithm is of the same order as \cite{jordan2018communication} assuming no heterogeneity and therefore is communication-efficient. Theoretically, we show that our estimator approximates the global maximum likelihood estimator with a faster rate than the average approach in the two-index asymptotic setting; see Remarks 3 and 4. From the inference perspective, our estimator attains the Cram\'er-Rao lower bound whereas the average approach has larger asymptotic variance and is not efficient when the number of sites is less than the sample size at each site;  see Remark 6.  We show that the proposed estimator outperforms the average approach in numerical studies.


\section{The surrogate likelihood approach for homogeneous Distributions}
In this section, we briefly review the surrogate likelihood approach for distributed inference by \cite{wang2017efficient} and \cite{jordan2018communication}. Consider a general parametric likelihood framework, where the random variable $Y$ follows the density function $f(y; \theta)$ indexed by a finite dimensional unknown parameter $\theta$. In the distributed inference problem, we suppose there are $K$ different sites. Denote $\{Y_{ij}\}$ to be the $i$-th observation in the $j$-th site. For notation simplicity, we assume that each site has equal sample size $n$.  The existing works on distributed inference such as \cite{wang2017efficient} and \cite{jordan2018communication} further assume that all the observations are independently and identically distributed across sites, $Y_{ij} \sim f(y; \theta)$. Under this assumption, the combined log likelihood function can be written as 
\begin{equation*}\label{global}
L (\theta) = \frac{1}{Kn}\sum_{j=1}^{K}\sum_{i=1}^{n}\log f(y_{ij} ; \theta) := \frac{1}{K}\sum_{j=1}^{K}L_j (\theta),
\end{equation*}
where $L_j (\theta) = \sum_{i=1}^{n}\log f(y_{ij} ; \theta)/n$ is the log-likelihood function obtained at each site. Due to the communication constraint and privacy concerns, one cannot directly combine data across multiple sites to compute the maximum likelihood estimator. Motivated by the following Taylor expansion of the combined likelihood function around some initial value $\bar \theta$,
\begin{equation}\label{global2}
L(\theta) = L(\bar\theta) + \nabla L(\bar\theta)^\T (\theta - \bar\theta) + \sum_{k = 2}^{\infty}\frac{1}{k!}\nabla^k L(\bar\theta) (\theta - \bar\theta)^{\otimes k},
\end{equation}
\cite{wang2017efficient} and \cite{jordan2018communication} proposed to construct a surrogate likelihood function by approximating all the higher-order derivatives in equation~(\ref{global2}) using the individual-level data in one of the $K$ sites (such as the first site). When the data are identically and independently distributed across sites, it holds that $\nabla^k L_1(\bar\theta) - \nabla^k L(\bar\theta)=o_P(1)$ for any $k \ge 0$, where  $L_1 (\theta)$  is the log-likelihood at the first site. Thus, $\nabla^k L_1(\bar\theta)$ is an asymptotically unbiased surrogate of $\nabla^k L(\bar\theta)$. However, in a distributed framework, communicating $\nabla L_j(\theta)$ from site $j$ to site $1$ requires to transfer only $O(d)$ numbers where $d$ is the dimension of $\theta$, whereas communicating higher order derivatives can be very costly. Replacing $\nabla^k L(\bar\theta)$ with $\nabla^k L_1(\bar\theta)$, the communication of the higher-order derivatives across sites can be avoided. Hence, by replacing $\sum_{k = 2}^{\infty}\nabla^k L(\bar\theta) (\theta - \bar\theta)^{\otimes k}/{k!}$ with $\sum_{k = 2}^{\infty}\nabla^k L_1(\bar\theta) (\theta - \bar\theta)^{\otimes k}/{k!}$, which also equals to $L_1(\theta) - \nabla L_1(\bar\theta)^\T(\theta - \bar\theta)$ and dropping the terms independent of $\theta$, the surrogate likelihood is defined as 
\begin{equation}\label{sur}
\tilde L(\theta) := L_1(\theta) + \{\nabla L(\bar\theta)-\nabla L_1(\bar\theta)\}^\T \theta.
\end{equation}
From the perspective of estimating equations, the surrogate likelihood approach is equivalent to a surrogate score approach which approximates the combined score function $\nabla L(\theta)$ by
\begin{equation*}
\tilde S(\theta) := \nabla L(\bar\theta)+\nabla L_1(\theta) -\nabla L_1(\bar\theta).
\end{equation*}
The theoretical properties of the estimator obtained by maximizing the surrogate likelihood function (or solving the surrogate score function) have been thoroughly studied; see \cite{wang2017efficient} and \cite{jordan2018communication} for details. 


\section{Surrogate efficient score method for heterogeneous distributions}
We consider a heterogeneous setting by assuming the $i$-th observation in the $j$-th site satisfies
\begin{equation*}
Y_{ij} \sim f(y; \theta_j), \quad \text{for  } i \in \{1, \dots, n\} \text{  and  } j \in \{1, \dots, K\},
\end{equation*}
where the unknown parameter $\theta_j$ can be decomposed into $\theta_j = (\beta, \gamma_j)\in R^d$. In this partition, $\beta$ is a $p$-dimensional parameter of interest assumed to be common in every site, which is the main motivation of evidence synthesis, and the $(d-p)$-dimensional nuisance parameter $\gamma_j$ is allowed to be different across sites. The true value of $\theta_j$ is denoted by $\theta_j^*$. 

If all patient-level data could be pooled together,  the combined log-likelihood function is 
\begin{equation*}
L_N (\beta, \Gamma) = \frac{1}{Kn}\sum_{j=1}^{K}\sum_{i=1}^{n}\log f(y_{ij}; \beta, \gamma_j):= \frac{1}{K}\sum_{j=1}^{K}L_j (\theta_j) ,
\end{equation*}
where $L_j (\theta) = \sum_{i=1}^{n}\log f(y_{ij} ; \theta_j)/n$ and $\Gamma = \{\gamma_j\}_{j \in \{1, \dots, K\}}\in R^{(d-p)K}$. 
In a distributed setting, the method reviewed in Section 2 is not directly applicable due to the following two reasons: the higher order derivatives of the log likelihood function in any site is a biased surrogate of the corresponding higher order derivatives of $L_N (\beta, \Gamma)$, and the total number of nuisance parameters $\textrm{dim}(\Gamma)=(d-p)K$ increases with sample size $n$ if we allow $K$ to increase with $n$. 

With the site-specific nuisance parameters, we propose to approximate the efficient score function instead of the score function. Motivated from theories of semiparametric models, the efficient score function is a way of reducing the influence of the less accurate estimation of the site-specific $\gamma_j$ which is essentially a projection of the score function of $\beta$ on the space that is orthogonal to the space spanned by the score function of nuisance parameter $\gamma_j$  \citep{van2000asymptotic}. In our setting, it is defined as 
\begin{equation*}\label{eif_each}
s_j(y;\beta,\gamma_j) = \nabla_\beta \log f(y;\beta,\gamma_j) - I^{(j)}_{\beta\gamma}{I^{(j)}_{\gamma\gamma}}^{-1} \nabla_\gamma \log f(y;\beta,\gamma_j)
\end{equation*}
where $I_{\gamma\gamma}^{(j)}$ and $I_{\gamma\beta}^{(j)}$ are the corresponding submatrices of the information matrix in the $j$-th site, i.e., 
$
I^{(j)} = \E\{-\nabla^2L_j(\theta_j^*)\}
$.  In parametric models, the estimator of $\beta$ obtained from solving the efficient score function defined above has the asymptotic variance reaching the Cramer-Rao lower bound, which is considered as an efficient estimator. In addition, it satisfies that $E\{\nabla_\gamma s_j(y;\beta,\gamma_j) \}=0$, which shows it is less sensitive to small perturbations of the nuisance parameter $\gamma_j$.  We then define 
\begin{equation*}\label{eif}
S (\beta, \Gamma) = \frac{1}{Kn}\sum_{j=1}^{K}\sum_{i=1}^{n}s_j(y_{ij};\beta,\gamma_j).
\end{equation*}
Treating the above combined efficient score function as a target function, we aim to construct a surrogate efficient score equation to approximate the target function using individual-level data from the first site and summary-level data from the other sites. To explain the origin of our estimator, we first consider an ideal situation where we know the true parameter value  $\gamma_j^*$. Using the key idea of the surrogate likelihood approach, we aim to construct a function $g^*(y; \beta)$  in the first site such that
\begin{equation}\label{goal}
E_{\theta_1^*} \{\nabla_\beta^k g^*(Y_{i1}; \beta)\} = E \{\nabla_\beta^k S (\beta, \Gamma^*)\},
\end{equation}
holds for any $k\geq 1$, where we use $E_{\theta_j^*}(\cdot)$ to denote the expectation with respect to the distribution $f(y,\beta^*,\gamma_j^*)$, $E(\cdot)$ to denote the expectation with respect to the joint distribution of the full data, and $\Gamma^*$ to denote the true value of $\Gamma$.  The right hand side of equation~(\ref{goal}) can be written as 
\[
E \{\nabla_\beta^k S (\beta,\Gamma^*)\} = \frac{1}{K}\sum_{j=1}^{K}E_{\theta_j^*} \{\nabla_\beta^k s_j(Y_{ij};\beta, \gamma_j^*)\}.
\]
However, the function $g^*(Y_{i1}; \beta)$ only involves samples in the first local site, which follows the distribution $f(y,\beta^*,\gamma_1^*)$ different from $f(y,\beta^*,\gamma_j^*)$ for $j\neq 1$. To achieve equation~(\ref{goal}),  we propose to construct $g^*(y; \beta)$ by using the density ratio tilting method
\begin{equation*}\label{gstar}
g^*(y; \beta) = \frac{1}{K}\sum_{j=1}^{K} \frac{f(y; \beta^*, \gamma_j^*)}{f(y; \beta^*, \gamma_1^*)}s_j(y_{ij};\beta,\gamma_j^*),
\end{equation*}
where the density ratio $f(y; \beta^*, \gamma_j^*)/f(y; \beta^*, \gamma_1^*)$ is the adjustment that accounts for the heterogeneity of the distributions.
It can be shown that $E_{\theta_1^*} \{\nabla_\beta^k g^*(Y_{i1}; \beta)\}=E \{\nabla_\beta^k S (\beta, \Gamma^*)\}$ holds for any $k\ge 0$ and observation $Y_{i1}$ in the first local site (see Supplementary Material for details). 

The map $g^*(y;\beta)$ cannot be computed in practice as it depends on the unknown parameters $\beta^*, \gamma_j^*$, and the information matrix $I^{(j)}$. Nevertheless, a natural surrogate can be used instead, by plugging in some initial estimators $\bar\beta$ and $\bar\gamma_j$, and replacing the matrix $I^{(j)}$ by its density ratio
tilting estimator $\tilde{H}^{(1,j)}$, defined as
\begin{equation*}\label{estI}
\tilde H^{(1,j)}= -\frac{1}{n}\sum_{i=1}^{n}\nabla^2\log f(y_{i1};\bar\beta,\bar\gamma_j)\frac{f(y_{i1};\bar \beta, \bar \gamma_j)}{f(y_{i1}; \bar \beta, \bar \gamma_1)}.
\end{equation*}
We then have 
\begin{equation*}\label{g}
g(y; \beta, \bar\beta, \bar \Gamma) = \frac{1}{K}\sum_{j=1}^{K} \left[\frac{f(y; \bar\beta, \bar\gamma_j)}{f(y; \bar\beta, \bar\gamma_1)}\left\{\nabla_\beta\log f(y; \beta, \bar\gamma_j)-\tilde H_{\beta\gamma}^{(1,j)}\{ \tilde H_{\gamma\gamma}^{(1,j)}\}^{-1}\nabla_\gamma\log f(y; \beta, \bar\gamma_j) \right\}\right].
\end{equation*}
We denote $U_1 (\beta; \bar\beta, \bar \Gamma) = \sum_{i=1}^{n} g(y_{i1};  \beta, \bar\beta, \bar \Gamma)/n$, and define the surrogate efficient score function as 
\begin{equation*}\label{msurrogate}
\tilde U(\beta;\bar\beta, \bar \Gamma) =  U_1( \beta; \bar\beta, \bar \Gamma) + \frac{1}{K}\sum_{j=1}^{K}\{\nabla_\beta L_j (\bar\beta, \bar\gamma_j)- \bar H^{(j)}_{\beta\gamma} ({{{{\bar H}}_{\gamma\gamma}}^{(j)}})^{-1}  \nabla_\gamma L_j (\bar\beta, \bar\gamma_j)\}-   U_1( \bar\beta; \bar\beta, \bar \Gamma) , 
\end{equation*}
where $\bar H^{(j)}_{\beta\gamma} = \nabla_{\beta\gamma} L_j(\bar\beta, \bar\gamma_j)$ and $\bar H^{(j)}_{\gamma\gamma} = \nabla_{\gamma\gamma} L_j(\bar\beta, \bar\gamma_j)$. Recall that $U_1 (\beta; \bar\beta, \bar \Gamma)$ is constructed based on the samples in Site 1. Thus the surrogate efficient score only requires to transfer a $p$-dimensional score vector  $S_j(\bar\beta, \bar\gamma_j) = \nabla_\beta L_j (\bar\beta, \bar\gamma_j)- \bar H^{(j)}_{\beta\gamma} ({{{{\bar H}}_{\gamma\gamma}}^{(j)}})^{-1}  \nabla_\gamma L_j (\bar\beta, \bar\gamma_j)$ from each site together with some initial estimators. The surrogate efficient score estimator $\tilde\beta$ is obtained by solving the following equation for $\beta$ within Site 1,
\begin{equation}\label{estimator}
\tilde U(\beta;\bar\beta, \bar \Gamma) = 0. 
\end{equation}
In Section 4, we show that the estimation accuracy of the above estimator $\tilde\beta$  can be further improved by iterating the above surrogate efficient score procedures. The method is summarized in the following algorithm. The estimator $\tilde \beta$ defined in equation~(\ref{estimator}) is equivalent to the estimator with $T=1$ in the following algorithm, which is also known as a oneshot procedure.  

\begin{algorithm}[h]
	\caption{Algorithm for the proposed surrogate efficient score estimator}
	\begin{algorithmic}[1]
		\STATE Set the number of iterations $T$
		\STATE In Site $j=1$ to $j=K$ \textbf{do}
		\STATE \hspace{4 mm}  Obtain and broadcast $(\bar\beta_j, \bar\gamma_j) = \arg\max_{\beta,\gamma_j}L_j(\beta, \gamma_j)$;
		\STATE \hspace{4 mm} Choose a proper weight $w_j$ and obtain $\bar \beta = \sum_{j=1}^{K}w_j\bar{\beta}_j/\{\sum_{j=1}^{K}w_j\}$; 
		\STATE \hspace{4 mm} Calculate and transfer $S_j(\bar\beta, \bar\gamma_j)$ to Site 1;
		\STATE \textbf{end} 
		\STATE In Site 1
		\STATE \hspace{4 mm} Construct $\tilde U(\beta;\bar\beta, \bar \Gamma)$ using $\bar \beta$, $\{\bar{\gamma}_j\}$, and $\{S_j(\bar\beta, \bar\gamma_j)\}$;
		\STATE \hspace{4 mm} Obtain $\tilde \beta^{(1)}$ by solving $\tilde U(\beta;\bar\beta, \bar \Gamma) = 0$;
		\STATE If $T = 1$, output $\tilde \beta^{(1)}$
		\STATE If $T \ge 2$, for $t = 2$ to $t = T$ \textbf{do}
		\STATE \hspace{4 mm} Broadcast $\bar{\beta}^{(t)} = \tilde \beta^{(t-1)}$;
		\STATE \hspace{4 mm} In Site $j=1$ to $j=K$ \textbf{do}
		\STATE \hspace{8 mm} Obtain and transfer $\bar{\gamma}_j^{(t)}= \arg\max_{\gamma_j} L_j(\bar{\beta}^{(t)}, \gamma_j)$ and  $S_j(\bar\beta^{(t)}, \bar\gamma_j^{(t)})$ to Site 1;
		\STATE \hspace{4 mm} \textbf{end}
		\STATE \hspace{4 mm} In Site 1
		\STATE \hspace{8 mm} Construct $\tilde U(\beta;\bar\beta^{(t)}, \bar \Gamma^{(t)})$ using $\bar \beta^{(t)}$, $\{\bar{\gamma}_j^{(t)}\}$ and $\{S_j(\bar\beta^{(t)}, \bar\gamma_j^{(t)})\}$;
		\STATE \hspace{8 mm} Obtain $\tilde \beta^{(t)}$ by solving $\tilde U(\beta;\bar\beta^{(t)}, \bar \Gamma^{(t)}) = 0$;
		\STATE \textbf{end}
		\STATE Output $\tilde \beta^{(T)}$
	\end{algorithmic}
\end{algorithm}

\begin{remark}
	The broadcast step (line 2) in the above algorithm can be done by transferring $\bar\theta_j$ from each site to Site 1, and Site 1 returns the initial estimator $\bar\beta$ to all the sites. It can also be done by uploading all $\bar\theta_j$ to a shared repository and obtaining $\bar\beta$ at each site. The initial estimator $\bar\beta$ is chosen as a weighted average of the local estimators $\bar\beta_j$. When $w_j = 1$ for all $j$, $\bar\beta=\sum_{j=1}^K \bar{\beta}_j/K$ is the  average estimator \citep{zhao2016partially}. We can also choose $w_j$ to be the sample size of each site in the unbalanced design. When $w_j$ is chosen as the inverse of the estimated variance of $\bar\beta_j$, the resulting estimator $\bar\beta$ is referred to as the fixed effect meta-analysis estimator. In this paper, we simply choose $w_j = 1$. The total communication cost per iteration is to transfer $O(Kd)$ numbers across all sites, where $d$ is the dimension of $\theta_j = (\beta,\gamma_j)$. Comparing to the homogeneous setting, the communication cost is of the same order as \cite{jordan2018communication}, and is communication-efficient. 
\end{remark}

\begin{remark}
	To further reduce the computational complexity of  solving the surrogate efficient score function, we can approximate the combined efficient score function $S(\beta, \Gamma)$ via one-step Taylor expansion,
	\[
	S(\beta, \Gamma)\approx S(\bar \beta, \bar\Gamma)+\nabla_\beta S(\bar \beta, \bar\Gamma) (\beta-\bar{\beta})+\nabla_\Gamma S(\bar \beta, \bar\Gamma) (\Gamma-\bar{\Gamma}).
	\]
	First, the property of the efficient score implies $\nabla_\Gamma S(\bar \beta, \bar\Gamma)\approx 0$ so that  the last term can be neglected. Next, we replace the Hessian matrix  $\nabla_\beta S(\bar \beta, \bar\Gamma)$ computed by pooling over all the samples with the local surrogate $\nabla_\beta\check U_1(\bar \beta)$, where $\check U_1(\beta)=U_1(\beta; \bar\beta, \bar \Gamma)$. The resulting linear approximation $S(\bar \beta, \bar\Gamma)+\nabla_\beta\check U_1(\bar \beta) (\beta-\bar{\beta})$ as an estimating function of $\beta$ defines the following estimator 
	\[
	\tilde{\beta}^O = \bar\beta - \{\nabla_\beta\check U_1(\bar \beta)\}^{-1}S(\bar \beta, \bar\Gamma). 
	\]
	If we treat $\bar\beta$ as an initial estimator, the above estimator $\tilde{\beta}^O$ can be also viewed as a one-step estimator with a local surrogate of the Hessian matrix. Hereafter, our discussion will be focused on the estimator from Algorithm 1, as we show in Lemma S12 in  Supplementary Materials that this one-step estimator shares the same theoretical properties as the estimator from Algorithm 1.  When  calculating  the inverse of $({{{{\bar H}}_{\gamma\gamma}}^{(j)}})^{-1}$ becomes a bottleneck of computation, we proposed a modified algorithm in Appendix A of  Supplementary Material. 
\end{remark}

\section{Main Results}
In this section, we study the theoretical properties of the surrogate efficient score estimator $\tilde \beta^{(T)}$ obtained from Algorithm 1.  For convenience, we use $C, C_1$ and $C_2$ to denote positive constants which can vary from place to place. For sequence $\{a_n\}$ and $\{b_n\}$, we write $a_n \lesssim b_n$ ($a_n \gtrsim b_n$) if there exist a constant $C$ such that $a_n \le Cb_n$ ($a_n \ge Cb_n$) for all $n$. We first introduce the following assumptions.
\begin{assumption}
	The parameter space of $\beta$, denoted by $\mathcal{B}$, is a compact and convex subset of $\mathbb{R}^p$. The true value $\beta^*$ is an interior point of $\mathcal{B}$.
\end{assumption}

\begin{assumption}[Local Strong Convexity]
	Define the expected second-order derivative of the negative log likelihood function to be $I^{(j)}(\theta_j) = E_{\theta_j^*} \{-\nabla^2\log f(Y_{ij};\theta_j)\}$. There exist positive constants $(\mu_-, \mu_+)$, such that for any $j \in \{1, \dots, K\}$, the population Hessian matrix $I^{(j)}(\theta_j^*)$  satisfies
	\[
	\mu_-I_d \preceq I^{(j)}(\theta_j^*) \preceq \mu_+I_d,
	\]	
	where $I_d$ is the $d$ dimensional identity matrix. Here, we use the notation that $A\preceq B$ for two matrices $A$ and $B$ if $A-B$ is positive semi-definite. 
\end{assumption}


\begin{assumption} 
	For all $j \in \{1, \dots, K\}$ and $i \in \{1, \dots , n\}$, all components in $\nabla \log f(Y_{ij},\theta_j^*)$ and $\nabla^2 \log f(Y_{ij},\theta_j^*)$ are sub-exponential random variables.
\end{assumption}

\begin{assumption}[Identifiability]
	For any $j \in \{1, \dots, K\}$, we denote $F_j(\beta,\gamma_j) =E_{\theta_j^*} \{\log f(Y_{ij};\beta,\gamma_j)\}$.  The parameter $(\beta^*, \gamma_j^*)$ is the unique maximizer of $F_j(\beta,\gamma_j)$.
\end{assumption}

\begin{assumption}[Smoothness]
	For each $j \in \{1, \dots, K\}$, let $\bar\eta_j = (\bar{\beta},\bar{\gamma}_1,\bar{\gamma}_j)$, and define 
	\[
	H(\theta_j;y) = \nabla^2\log f(y;\beta, \gamma_j),
	\]
	and 	
	\[
	\tilde H(\beta, \bar\eta_j; y) = \nabla^2\log f(y;\beta, \bar\gamma_j)\frac{f(y; \bar\beta, \bar\gamma_j)}{f(y; \bar\beta, \bar\gamma_1)}.
	\]
	Define $U_\theta (\rho) = \{\theta;\|\theta-\theta\|_2\le\rho\}$ for some radius $\rho>0$. There exist some function $m_1 (y)$ and  $m_2(y)$, where $m_1 (Y_{ij})$ and  $m_2(Y_{ij})$  are  sub-exponentially distributed for all $j \in \{1, \dots, K\}$ and $i \in \{1, \dots , n\}$, such that for any $\theta_j$ and $\theta_j'\in U_{\theta_j}(\rho)$, we have
	\[
	\|H(\theta_j; y) -H(\theta_j'; y)\|_2 \le m_1(y)\|\theta-\theta_j'\|_2.
	\]
	And for any $\beta$, $\beta' \in U_{\beta}(\rho)$, $\bar\eta_j$, $\bar\eta_j'\in U_{\eta_j}(\rho)$, we have
	\[
	\|\tilde H(\beta, \bar\eta_j; y) -\tilde H(\beta', \bar\eta_j'; y)\|_2 \le m_2(y)\{\|\beta-\beta'\|_2+\|\bar\eta_j-\bar\eta_j'\|_2\}.
	\]
\end{assumption}

Assumptions 1, 2, 4, and 5 are standard assumptions in the distributed inference literature; see  \cite{jordan2018communication}.   Assumption 3 is a general distributional requirements of the data, which covers a wide range of parametric models.


When all individual-level data can be pooled together, the global maximum likelihood estimator $(\hat\beta,\hat \Gamma)=\arg\max_{\beta,\Gamma} L_N (\beta, \Gamma)$ is considered as the gold standard in practice. The asymptotic property of the global estimator has been studied in \cite{li2003efficiency} under the asymptotic regime $K/n \rightarrow c \in (0,\infty)$.  Our first result characterizes a non-asymptotic bound for the distance between the global maximum likelihood estimator $\hat \beta$ and the true parameter value $\beta^*$. 

\begin{lemma}
	Under Assumptions 1-5, the global maximum likelihood estimator $\hat \beta$ satisfies
	\[
	E\|\hat \beta - \beta^*\|_2 \le \frac{C_1}{(Kn)^{1/2}} +\frac{C_2}{n}
	\] 
	for some positive constants $C_1 $ and $C_2$ not related to $n$ and $K$.
\end{lemma}

To the best of our knowledge, this is one of the first nonasymptotic results on the rate of convergence of the maximum likelihood estimator in the presence of site-specific nuisance parameters. Under the classical two-index asymptotics setting, we allow the number of sites $K$ to grow with the sample size $n$. As a result, the dimension of nuisance parameters $\Gamma$ also increases with $n$. Let $N=Kn$ denote the total sample size. This lemma implies that the convergence rate of $\hat \beta$ is of order $O_p(N^{-1/2})$ when $K/n=O(1)$, which attains the optimal rate of convergence with known nuisance parameters. However, when $K/n\rightarrow\infty$, the estimator has a slower rate $O_p({1}/{n})$. In particular if $n$ is fixed, the global maximum likelihood estimator is no longer consistent, which is known as the Neyman-Scott problem \citep{neyman1948consistent}.

In the following, we characterize the difference between the proposed estimator and the maximum likelihood estimator $\hat\beta$. We first focus on the estimator $\tilde \beta$ defined in equation~(\ref{estimator}), which is identical to $\tilde\beta^{(1)}$ in algorithm 1 with the number of iterations $T=1$. 

\begin{theorem}
	Suppose Assumptions 1-5 hold. In Algorithm 1, if the number of iterations $T=1$, assuming $n\gtrsim\log K$, we have
	\begin{align*}
	\E\|\tilde\beta^{(1)} - \hat\beta\|_2\le& \frac{C}{n}.
	\end{align*}
	where $C$ is a positive constant not related to $n$ and $K$.
\end{theorem}

The above theorem shows that the proposed estimator with only one iteration converges to the global estimator $\hat{\beta}$ with a rate only depending on $n$. Together with Lemma 1, we obtain $\E\|\tilde\beta^{(1)} - \beta^*\|_2\lesssim {1}/{(Kn)^{1/2}}+{1}/{n}$. In other words, the estimator has the same rate of convergence as the global maximum likelihood estimator. 

\begin{remark}
	When $K/n\rightarrow 0$,  we showed in Lemma S.10 of  Supplementary Material that the  average estimator $\bar\beta^{(1)}=\sum_{j=1}^K \bar{\beta}^{(1)}_j/K$ defined in algorithm 1 satisfies $\E\|\bar\beta^{(1)}-\hat\beta\|_2 \gtrsim {1}/{(Kn)^{1/2}}$. By comparing with the bound in Theorem 1, we have
	\begin{equation}\label{eq_barbeta}
	\E\|\tilde{\beta}^{(1)}-\hat{\beta}\|_2\le C\Big(\frac{K}{n}\Big)^{1/2}\E\|\bar\beta^{(1)}-\hat\beta\|_2. 
	\end{equation} 
	Thus our estimator $\tilde\beta^{(1)}$ is closer to the global maximum likelihood estimator than the  average estimator under the condition $K/n\rightarrow 0$.
\end{remark}

Our next result shows that after at least one iteration the estimator $\tilde\beta^{(T)}$ in Algorithm 1 with $T\ge 2$ has a tighter bound than $\tilde\beta^{(1)}$ in Theorem 1. 

\begin{theorem}
	Suppose all the assumptions in Theorem 1 hold. In Algorithm 1, if the number of iterations $T\ge 2$, we have
	\begin{align*}
	\E\|\tilde\beta^{(T)} - \hat\beta\|_2\le& \frac{C_1}{(K)^{1/2}n}+\frac{C_2}{n^{3/2}},
	\end{align*}
	where $C_1$ and $C_2$ are positive constants not related to $n$ and $K$.
\end{theorem}

\begin{remark}
	The above theorem implies that when $K/n\rightarrow 0$, for any $T\ge 2$
	\begin{equation}\label{eqrate}
	\E\|\tilde{\beta}^{(T)}-\hat{\beta}\|_2\le Cn^{-1/2}\E\|\bar\beta^{(1)}-\hat\beta\|_2, 
	\end{equation}
	which improves the result in equation (\ref{eq_barbeta}).  When $K$ is relatively small, our estimator $\tilde{\beta}^{(T)}$ with $T\ge 2$ is closer to the global maximum likelihood estimator by a factor of $n^{-1/2}$ than the  average estimator. We also see an interesting fact that the dimension of the nuisance parameters has no effect on the relative error $\E\|\tilde{\beta}^{(T)}-\hat{\beta}\|_2/\E\|\bar\beta^{(1)}-\hat\beta\|_2$. This dimension-free phenomenon provides an explanation of why the proposed estimator consistently outperforms the  average method in our simulation studies in Section 6.
\end{remark}

Our next theorem establishes the asymptotic normality of the proposed estimator.

\begin{theorem} Suppose all the assumptions in Theorem 1 hold. 
	Define
	$
	I_{\beta|\gamma} = \sum_{j=1}^{K}I_{\beta|\gamma}^{(j)}/K,
	$
	where $I_{\beta|\gamma}^{(j)}$ is the partial information matrix of $\beta$ defined as $I_{\beta|\gamma}^{(j)} = I_{\beta\beta}^{(j)} - I_{\beta\gamma}^{(j)}(I_{\gamma\gamma}^{(j)})^{-1}I_{\gamma\beta}^{(j)}$.  Assuming $K = C n^r$ for some fixed $r \in [0,1)$, we have for any $T \ge 1$, as $n \rightarrow \infty$, 
	\[
	Kn(\tilde{\beta}^{(T)}-\beta^*)^\T I_{\beta|\gamma}(\tilde{\beta}^{(T)}-\beta^*) \rightarrow \chi^2_p.
	\]
\end{theorem}

To obtain the $\surd{(Kn)}$-asymptotic normality of the proposed estimator $\tilde{\beta}^{(T)}$, we have to restrict to the setting $K = C n^r$ for some $r \in [0,1)$. In particular, when $K / n  \rightarrow C \in (0, \infty)$ or equivalently $r=1$, \cite{li2003efficiency} showed that the maximum likelihood estimator $\hat \beta$ is asymptotically biased, that is ${(Kn)}^{1/2}I_{\beta|\gamma}^{1/2}(\hat \beta-\beta^*)\rightarrow N(b, I_p)$, for some $b\neq 0$. Since the proposed estimator $\tilde{\beta}^{(T)}$ (with $T\ge 2$) satisfies $\tilde{\beta}^{(T)}-\hat{\beta}=O_p({K}^{-1/2}n^{-1}+{n}^{-3/2})$ by Theorem 2, it implies that the same asymptotic distribution holds for $\tilde\beta^{(T)}$, ${(Kn)}^{1/2}I_{\beta|\gamma}^{1/2}(\tilde \beta^{(T)}-\beta^*)\rightarrow_d N(b, I_p)$ for the same $b\neq 0$ if $r=1$. The same limiting distribution also holds for $T=1$. This leads to a phase transition of the limiting distribution of $\tilde\beta^{(T)}$ at $r=1$. As a result, the condition $r \in [0,1)$ is essential for the asymptotic unbiasedness of $\tilde \beta^{(T)}$ and cannot be further relaxed. 

\begin{remark}
	The choice of the initial value $\bar\theta_j^{(1)}$ in line 3 of Algorithm 1 is not necessarily restricted to the local maximum likelihood estimator. Due to the use of the efficient score, the impact of the initial estimators of the nuisance parameters is alleviated. We can show that the conclusions of Theorem 1-3 still hold if $\bar\theta_j^{(1)}$ is replaced with any $\surd{n}$-consistent estimator. 
\end{remark}

It is well known that the  average estimator is fully efficient under the homogeneous setting; see e.g., \cite{battey2018distributed} and \cite{jordan2018communication}. However, the following proposition shows that this  estimator is no longer efficient under the considered heterogeneous setting.

\begin{proposition}
	Recall that the  average estimator is $\bar\beta = \sum_{j=1}^{K}\bar\beta_j/K$, where $(\bar\beta_j, \bar\gamma_j) = \arg\max_{\beta,\gamma_j}L_j(\beta, \gamma_j)$. Suppose all the conditions in Theorem 3 hold. We have as $n \rightarrow \infty$, 
	\[
	Kn(\bar{\beta}-\beta^*)^T \left\{\frac{1}{K}\sum_{j=1}^{K}{I^{(j)}_{\beta|\gamma}}^{-1}\right\}^{-1}(\bar{\beta}-\beta^*) \rightarrow \chi^2_p.
	\]
\end{proposition}

\begin{remark}
	In this remark, we compare the asymptotic variance of $\tilde\beta^{(T)}$ in Theorem 3 and $\bar{\beta}$ in Proposition 1.  Our proposed estimator $\tilde{\beta}^{(T)}$ is efficient in the sense that its asymptotic variance is equal to the Cram\'er-Rao lower bound, i.e., $\lim_{K\rightarrow\infty}I_{\beta|\gamma}= \lim_{K\rightarrow\infty}\{\sum_{j=1}^{K}{I^{(j)}_{\beta|\gamma}}/K\}^{-1}$, for any $T\ge 1$. On the other hand, the  average estimator is not efficient as  $\lim_{K\rightarrow\infty}\sum_{j=1}^{K}{I^{(j)}_{\beta|\gamma}}^{-1}/K\succeq \lim_{K\rightarrow\infty}I_{\beta|\gamma}^{-1}$.
\end{remark}

Finally, to construct the confidence interval of $\beta^*$, we need to provide a consistent estimator for the averaged partial information matrix $I_{\beta|\gamma}$. In the following theorem, we apply the density ratio tilting approach to estimate the variance using data only from the first local site. 

\begin{theorem}
	Suppose all the assumptions in Theorem 3 hold. Define
	\[
	\tilde I^{(j)} = -\frac{1}{Kn}\sum_{j=1}^{K}\sum_{i=1}^{n} \frac{f(y_{i1},\tilde{\beta}^{(T)}, \bar{\gamma}_j^{(T)})}{f(y_{i1},\tilde{\beta}^{(T)}, \bar{\gamma}_1^{(T)})}\nabla^2\log f (y_{i1}; \tilde{\beta}^{(T)}, \bar{\gamma}_j^{(T)}),
	\]
	and  
	$\tilde I_{\beta|\gamma}^{(j)} = \tilde I_{\beta\beta}^{(j)} - I_{\beta\gamma}^{(j)}({\tilde I_{\gamma\gamma}^{(j)}})^{-1}\tilde I_{\gamma\beta}^{(j)}$.  We have as $n \rightarrow \infty$, 
	\[
	Kn (\tilde{\beta}^{(T)}-\beta^*)^\T{{\tilde{ I}_{\beta|\gamma}^{(j)}}} (\tilde{\beta}^{(T)}-\beta^*)  \rightarrow \chi^2_p.
	\] 
\end{theorem}

\section{Reduce the influence of the local site}

In a practical collaborative research network, each site can act as the local site and obtain an estimate using Algorithm 1. To further reduce the impact of the choice of the local site and improve the stability of the algorithm, we can combine these estimates in Algorithm 1 by an  average approach. At the $j$-th site, we define the site-specific surrogate score function to be 
\begin{equation*}
\tilde U_j(\beta;\bar\beta, \bar \Gamma) =  U_j( \beta; \bar\beta, \bar \Gamma) + \frac{1}{K}\sum_{k=1}^{K}\{\nabla_\beta L_k (\bar\beta, \bar\gamma_k)- \bar H^{(k)}_{\beta\gamma} ({{{{\bar H}}_{\gamma\gamma}}^{(k)}})^{-1}  \nabla_\gamma L_k (\bar\beta, \bar\gamma_j)\}-   U_j(\bar \beta; \bar\beta, \bar \Gamma), 
\end{equation*}
where 
\begin{align*}
&U_j( \beta; \bar\beta, \bar \Gamma) =\frac{1}{Kn}\sum_{k=1}^{K}\sum_{i=1}^{n}\ \frac{f(y_{ij}; \bar\beta, \bar\gamma_k)}{f(y_{ij}; \bar\beta, \gamma_j)}\left\{\nabla_\beta\log f(y_{ij}; \beta, \bar\gamma_k)-\tilde H_{\beta\gamma}^{(j,k)}\{\tilde H_{\gamma\gamma}^{(j,k)}\}^{-1}\nabla_\gamma\log f(y_{ij}; \beta, \bar\gamma_k) \right\}
\end{align*}
and 
\begin{equation*}
\tilde H^{(j,k)}= -\frac{1}{n}\sum_{i=1}^{n}\nabla^2\log f(y_{ij};\bar\beta,\bar\gamma_k)\frac{f(y_{ij};\bar \beta, \bar \gamma_k)}{f(y_{ij}; \bar \beta, \bar \gamma_j)}.
\end{equation*}
The surrogate score function $\tilde U_j(\beta;\bar\beta, \bar \Gamma)$ is obtained using the individual-level data in the $j$-th site and summary-level data from the other $K-1$ sites. In this case, each site can obtain a surrogate efficient score estimator $\tilde\beta_j$ by solving $\tilde U_j(\beta;\bar\beta, \bar \Gamma) = 0$, and 
%
we further combine these estimators by
$
\tilde{\beta}_{all} = \sum_{j=1}^{K}\tilde\beta_j/K.
$
The algorithm is summarized below.

\begin{algorithm}[h]
	\caption{Algorithm for the proposed surrogate efficient score estimator}
	\begin{algorithmic}[1]
		\STATE In Site $j=1$ to $j=K$ \textbf{do}
		\STATE \hspace{4 mm}  Obtain an initial estimator $\bar\theta_j = (\bar \beta_j, \bar \gamma_j)$ for the parameter $\beta $ and $\gamma_j$;
		\STATE \hspace{4 mm} Broadcast $\bar \theta_j$; 
		\STATE \hspace{4 mm} Choose $w_j^{(1)}$ and obtain $\bar \beta^{(1)} = \sum_{j=1}^{K}w_j^{(1)}\bar{\beta}_j/\{\sum_{j=1}^{K}w_j^{(1)}\}$;
		\STATE \hspace{4 mm} Obtain and broadcast $ S_j (\bar\beta,\bar\gamma_j)$;
		\STATE \hspace{4 mm} Construct $\tilde U_j(\beta;\bar\beta, \bar \Gamma)$ using $\bar \beta$, $\{\bar{\gamma}_j\}$ and $\{S_k (\bar\beta,\bar\gamma_k)\}_{1 \le k\le K}$;
		\STATE \hspace{4 mm} Obtain $\tilde \beta_j$ by solving $\tilde U_j(\beta;\bar\beta, \bar \Gamma) = 0$;
		\STATE \hspace{4 mm} Broadcast $\tilde \beta_j$;
		\STATE \textbf{end}
		\STATE Obtain  $
		\tilde{\beta}_{all} = \sum_{j=1}^{K}\tilde\beta_j/K
		$
		\STATE Output $
		\tilde{\beta}_{all}$
	\end{algorithmic}
\end{algorithm}


\section{Simulation study}
We consider a logistic regression between a binary outcome $Y$ and a binary exposure $X$, controlling for a confounding variable $Z$. It is assumed that for data in the $k$-th site, we have
\[
\text{logit}\{Pr(Y=1\mid X,Z)\} = \gamma_{0k} +\beta X+\gamma_{1k}Z.
\]
We set the true value of $\beta = -1$ for all the $K$ sites. The nuisance parameters $\gamma_{0k}$ and $\gamma_{1k}$ are generated from the uniform distribution $\text{U}  (a-1, a+1)$ and the uniform distribution $\text{U} (-2, 2)$, respectively. The binary exposure $X$ is generated from a Bernoulli distribution with probability $b$, and the confounder variable is generated by $Z\sim N(X-0.3,1)$. Under each setting, we set the sample size $n$ to be $100$ and the number of sites $K$ to be $10$ or $50$. We compare the performance of five different methods:
(1) The estimator from averaging all local maximum likelihood estimators (Average); (2) The surrogate likelihood method in \cite{jordan2018communication} assuming the homogeneous model (Homo); (3) The proposed estimator in Algorithm 1 with $T=1$ (i.e., the oneshot algorithm) (M1); (4) The proposed estimator in Algorithm 1 with $T=2$ (M2), and (5) The proposed estimator in Algorithm 2 (M3). 

We first investigate how the prevalence of binary events in a regression model influences the performance of the compared methods.  We vary the value of $a$ and the probability $b$ which control the prevalence of the exposure and the outcome, and consider the following four scenarios: (1) both the outcome and the exposure are common; (2) the outcome is rare and the exposure is common;
(3) the outcome is common and the exposure is rare; (4) both the outcome and the exposure are rare. The parameter values for $a$ and $b$ are presented in Table S1 of  Supplementary Material. We observed from Figure 1 that the surrogate likelihood method which ignores the heterogeneity has substantial bias in all settings. When both the outcome and exposure are common, all the proposed estimators and the average estimator perform well. The average estimator starts to show large bias when either the outcome or the exposure is rare. Our oneshot estimator (M1) reduces the bias of the average estimator in all settings. However, when both outcome and exposure are rare, the oneshot algorithm illustrates non-negligible bias and relatively large variation. Through only one more iteration, our estimator (M2) has sizeably improved performance and becomes more stable in the rare outcome or exposure setting. The estimator in Algorithm 2 (M3) also outperforms  the oneshot method in terms of reducing the bias and variance, especially when both the outcome and the exposure are rare. When we increase the number of sites, the bias of the average method remains the same while the variation is smaller. Our proposed algorithms, however, are able to take advantage of the increased total sample size and provide estimates with smaller bias.

We then investigate whether the level of heterogeneity  influences  the performance of the compared methods. To alter the level of heterogeneity, we  generate the nuisance parameters $\gamma_{0k}$ from the uniform distribution $\text{U}  (-v, v)$,  and $\gamma_{1k}$ from the uniform distribution $\text{U} (-2v, 2v)$. We increase $v$ from $0.1$ to $4$. When $v$ is $0.1$, the heterogeneity of the values of nuisance parameters is small across sites. We observe from Figure 2 that all methods work comparably well, including the surrogate likelihood method which ignores the heterogeneity. As $v$ increases from $0.1$ to $4$, the bias of the surrogate likelihood method is increasing. The average approach, in theory,  is a consistent estimator, but is shown to have larger bias and variation when $v$ is increasing. The proposed three methods have similar performance under all settings. We can observe a slightly increasing variation and more outlying points of the proposed methods when $v$ is large. Overall, the proposed methods are more robust to the change of  $v$ compared to the other two methods.

Finally, we investigate how the dimension of the nuisance parameters affects the performance of the compared methods. We  generate $Z$ from $N(0, I_{q})$, and the corresponding coefficient vector $\gamma_{1k}$ is generated from $q$ independent uniform distributions  $\text{U}  (-1, 1)$. Including the intercept, the total dimension of the nuisance parameters is denoted as $d_\gamma = q+1$.  We increase $d_\gamma$ from $2$ to $14$. From Figrue 3, we see as $d_\gamma$ increases, the estimation errors of all compared methods become larger.  M2 has slightly better performance compared to M1 and M3, implying that  iterations might help reduce the bias. The average approach, however, has the worst performance compared to the other approaches when $d_\gamma$ is large.

In sum, the increase of the rareness of the disease or exposure, the level of heterogeneity and the dimension of the model can increase the estimation errors of all compared methods. Ignoring heterogeneity in a distributed setting can lead to substantial amount of bias, and our proposed methods can greatly  improve the estimation accuracy based on the commonly used average approach.

\begin{figure}[h]
	\caption{Boxplots of the estimates (subtracted by the true parameter value)  under the four parameter settings in Table 1. Each site has a sample size $100$, and each setting is replicated $1000$ times.}
	\centering
	\includegraphics[scale = 0.24]{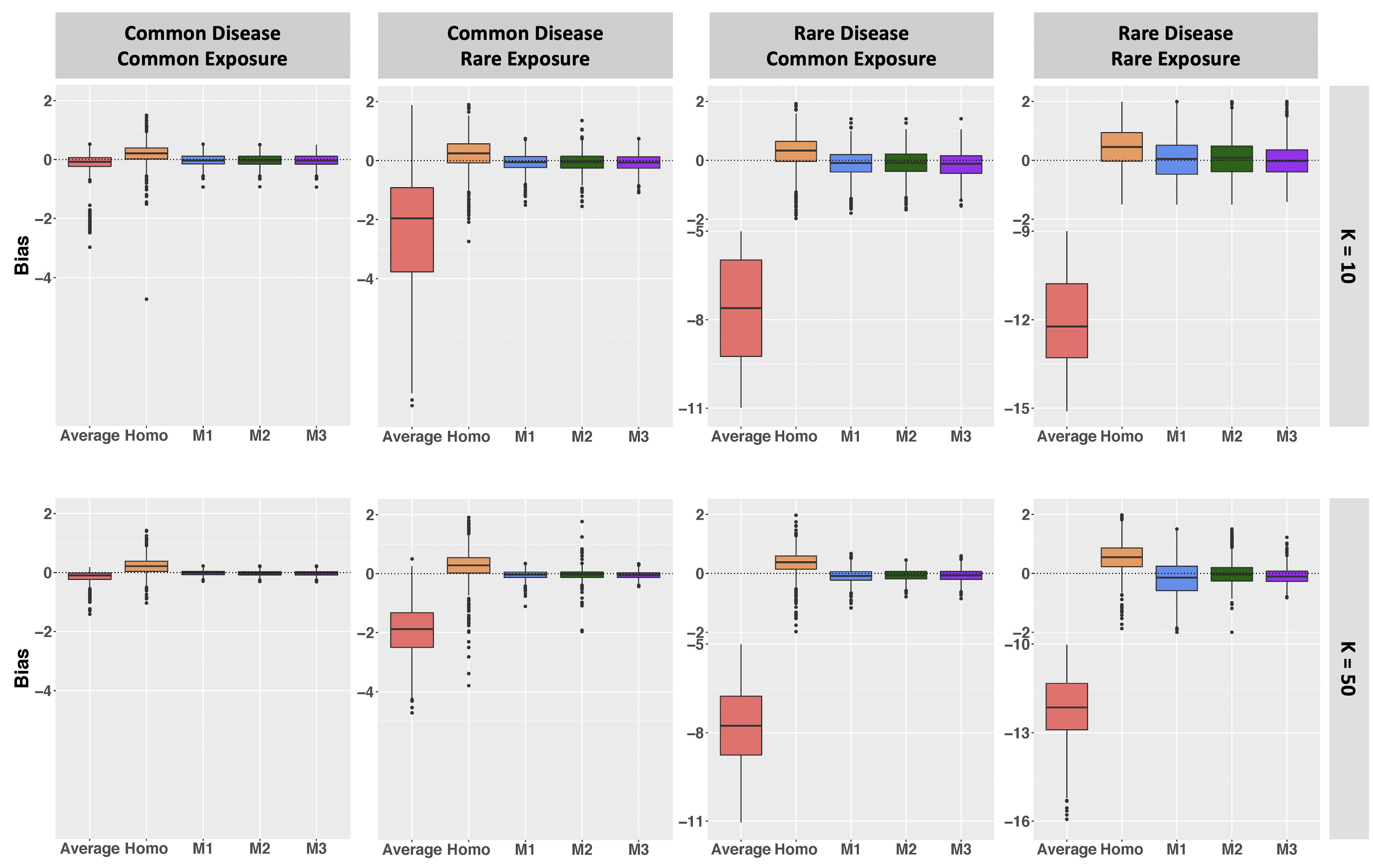}
\end{figure}

\begin{figure}[h]
	\caption{Boxplots of the estimates (subtracted by the true parameter value)  when $v$ takes values in $(0.1,1,2,4)$, and $K$ varies from $10$ to $50$.  Each site has a sample size $100$, and each setting is replicated $1000$ times.}
	\centering
	\includegraphics[width= 5.8 in]{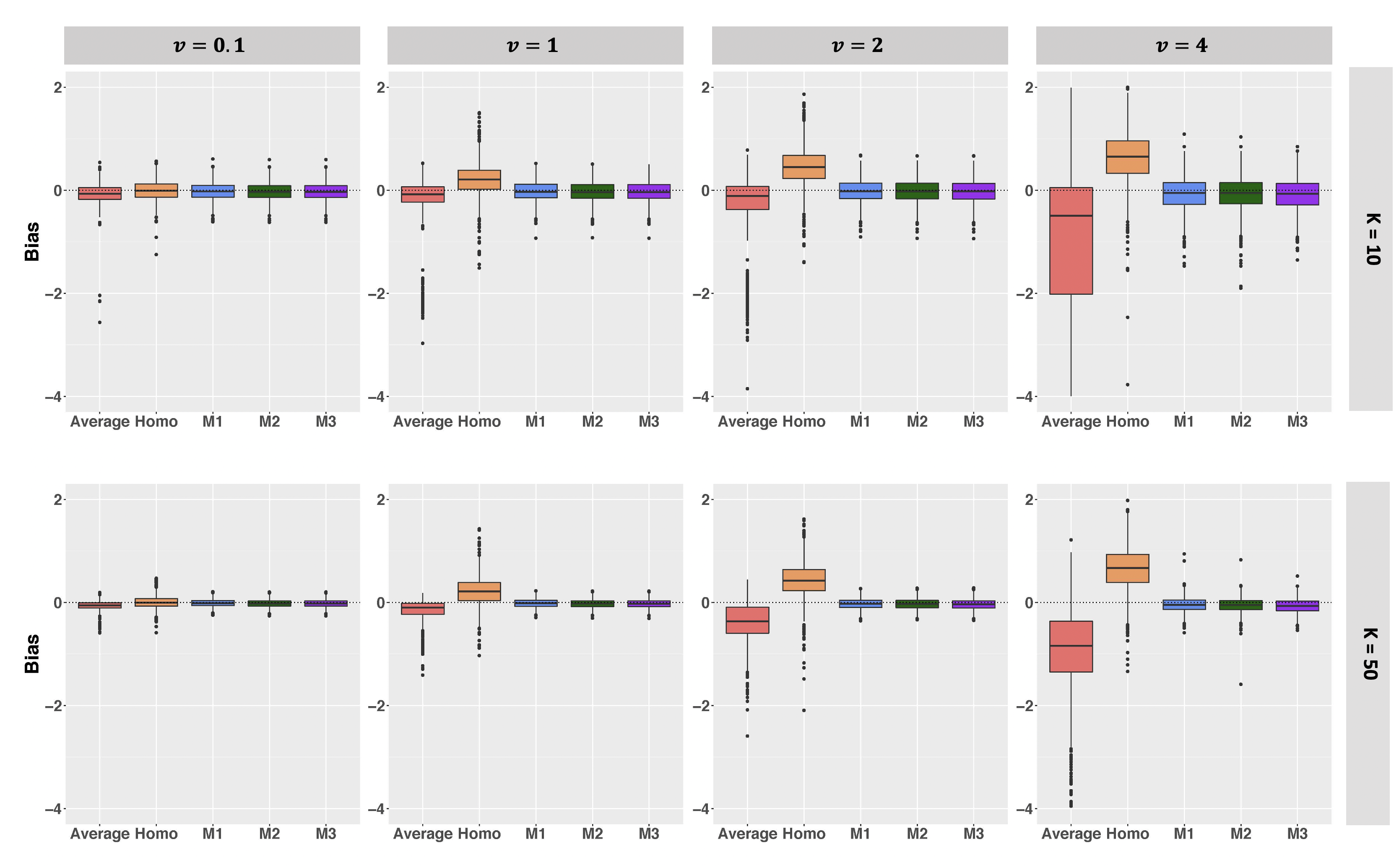}
\end{figure}

\begin{figure}[h]
	\caption{Boxplots of the estimates (subtracted by the true parameter value) when $d_\gamma$ takes values in $(2,6,10,14)$, and $K$ varies from $10$ to $50$.  Each site has a sample size $100$, and each setting is replicated $1000$ times.}
	\centering
	\includegraphics[width= 5.8 in]{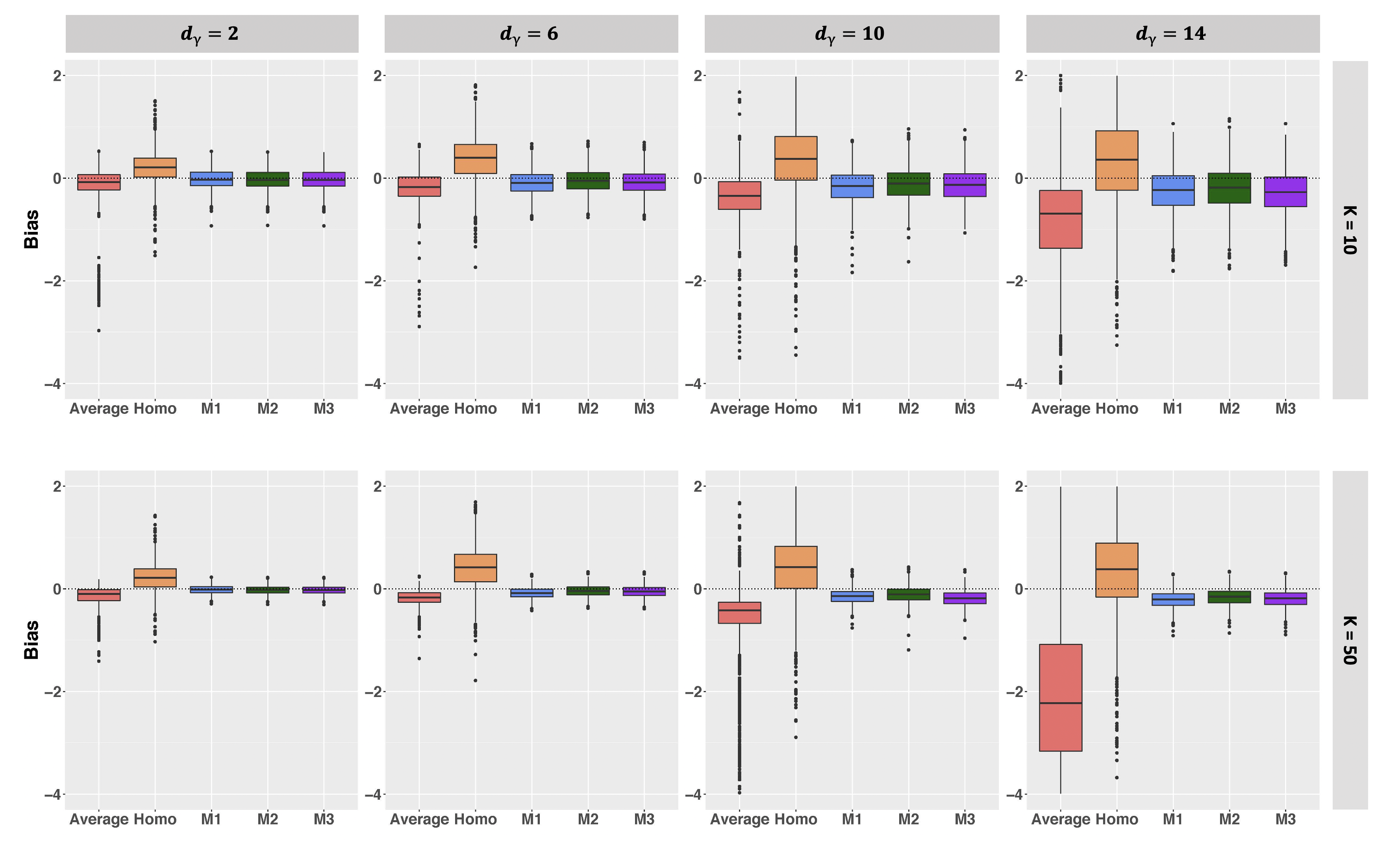}
\end{figure}

\section{Real data application}
We applied the proposed algorithms to data from five sites within the OneFlorida Clinical Research Consortium to quantify the association between mental disorders, including major depression and anxiety, with the risk of opioid use disorder using a logistic regression model.  Each participating site extracted electronic health records  between 01/01/2012 and 03/01/2019 for patients who had opioid prescription (including Codeine, Fentanyl, Hydromorphone, Meperidine, Methadone, Morphine, Oxycodone, Tramadol, Hydrocodone, Buprenorphine), and no cancer or  diagnosis of opioid use disorder before their first prescription. Among these patients who were exposed to opioid, a case of opioid use disorder is defined as having first diagnosis of opioid use disorder within 12 months after their first prescription and a control is defined as having no diagnosis of opioid use disorder in the entire time window. We obtained in total 1458 cases from the five clinical sites, and we randomly selected 2908 controls to maintain a case-control ratio of 1:2.  In addition to the two risk factors of interest (i.e., major depression and anxiety), we  requested a list of relevant covariate variables to be adjusted in the regression model, including age, gender, race (non-Hispanic White vs others), alcohol-related disorders, pain, cannabis-related disorder, cocaine-related disorder, nicotine-related disorder, smoking status (ever-smoker, non-smoker, and unknown), as well as the Charlson comorbidity index \citep{quan2005coding}.  Records with missing values were removed, resulting in sample sizes of $680$, $1311$,  $920$,  $270$, and $1106$ from Site 1 to Site 5, respectively; see Appendix C of  Supplementary Material for more information about data processing. 

In the logistic regression model, we treated the coefficients of major depression and anxiety to be common parameters across sites, and all the other coefficients including the intercept were assumed to be site-specific. We chose Site 1 as the local site and applied our methods M1, M2, M3, and the average approach. The estimated log odds ratios with their 95\% confidence intervals are shown in Figure 4. We  observed consistent results from the three proposed methods  for both anxiety and depression, suggesting one round of communication already led to stable estimation results. Anxiety was identified to be statistically significantly associated with opioid use disorder by all the methods. The relative difference based on the point estimates is about 18.4\% comparing the average approach to M1. All methods failed to identify significant association between depression and opioid use disorder, possibly due to the relatively low prevalence (9\%) of depression in the overall sample and the limited sample size. We observed opposite signs of the point estimates obtained from the proposed methods and the average approach, leading to a large relative difference of 164.3\%. Since it has been shown in many studies that depression is associated with increased risk of developing opioid use disorder \citep{martins2012mood,sullivan2018depression}, the negative association estimated by the average approach can be less reliable. More details of model fitting results can be found in the Appendix C of Supplementary Material. 

This real data application demonstrated the feasibility of implementing the proposed distributed algorithms in real-world distributed research networks. Although the average approach is easy to implement in practice, the proposed methods provide more reliable parameter estimates with only one extra step of sharing aggregate data.

\begin{figure}[h]
	\caption{Estimated log odds ratios of anxiety and depression, with 95\% confidence intervals from the four methods.}
	\centering
	\includegraphics[width= 5.8 in]{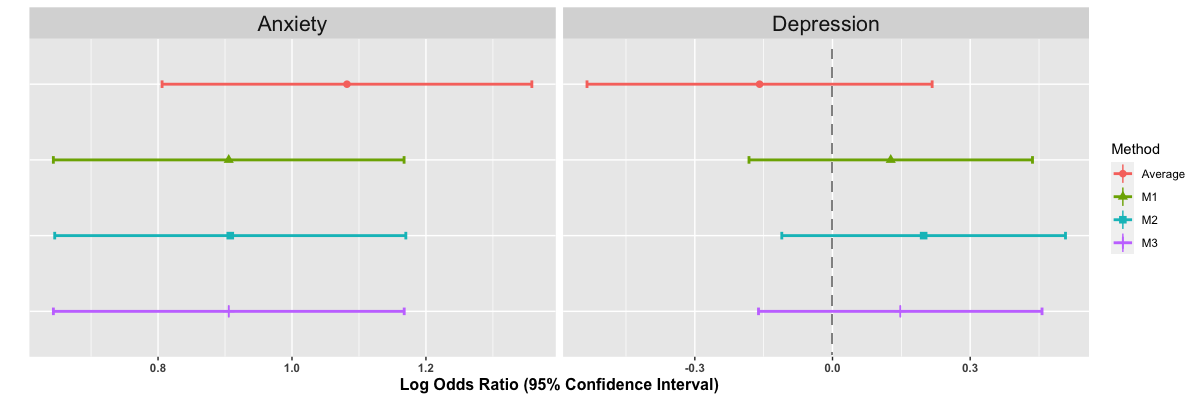}
\end{figure} 

\section{Discussion}
Motivated from a practical consideration that the data stored at different clinical sites are often heterogeneously distributed, we propose a surrogate efficient score approach for distributed inference. Our approach provides  flexibility to allow site-specific nuisance parameters, and bridges the gap in the current research on healthcare distributed research networks. There are several future research directions. To account for the large number of clinical,  environmental and genetic related variables in the modern healthcare datasets, it will be interesting to extend our method to the high-dimensional settings where either the dimension of $\beta$ or the dimension of the nuisance parameters is  larger than the sample size. Moreover, to extend the scope of the proposed framework, it would be of interest to relax the parametric assumption by using methods such as the generalized estimating equations \citep{liang1986longitudinal} and the generalized methods of moments \citep{hansen1982large}. However, as the density ratio tilting relies on the  distributional assumption, it may require new methodological development to adjust for the heterogeneity under these new settings.  Another practical challenge is that some sites only have a subset of all covariates. Recent work including \cite{kundu2019generalized} and \cite{zhang2020generalized} proposed novel methods to integrate summary statistics from external datasets with different covariate information. It is of interest to develop distributed inference that can handle heterogeneity and account for  incomplete covariate information across sites. These topics are currently under investigation and will be reported in the future.

\setstretch{1.24}
\bibliographystyle{chicago}
\bibliography{bib1}

\begin{thebibliography}{}

\bibitem[\protect\citeauthoryear{Barrows~Jr and Clayton}{Barrows~Jr and
  Clayton}{1996}]{barrows1996privacy}
Barrows~Jr, R.~C. and P.~D. Clayton (1996).
\newblock Privacy, confidentiality, and electronic medical records.
\newblock {\em Journal of the American Medical Informatics Association\/}~{\em
  3\/}(2), 139--148.

\bibitem[\protect\citeauthoryear{Battey, Fan, Liu, Lu, and Zhu}{Battey
  et~al.}{2018}]{battey2018distributed}
Battey, H., J.~Fan, H.~Liu, J.~Lu, and Z.~Zhu (2018).
\newblock Distributed testing and estimation under sparse high dimensional
  models.
\newblock {\em Annals of Statistics\/}~{\em 46\/}(3), 1352--1382.

\bibitem[\protect\citeauthoryear{Chen and Xie}{Chen and
  Xie}{2014}]{chen2014split}
Chen, X. and M.-g. Xie (2014).
\newblock A split-and-conquer approach for analysis of extraordinarily large
  data.
\newblock {\em Statistica Sinica\/}, 1655--1684.

\bibitem[\protect\citeauthoryear{Cheng, Kessler, Mackinnon, Chang, Nadkarni,
  Hunt, Duval-Arnould, Lin, Pusic, and Auerbach}{Cheng
  et~al.}{2017}]{cheng2017conducting}
Cheng, A., D.~Kessler, R.~Mackinnon, T.~P. Chang, V.~M. Nadkarni, E.~A. Hunt,
  J.~Duval-Arnould, Y.~Lin, M.~Pusic, and M.~Auerbach (2017).
\newblock Conducting multicenter research in healthcare simulation: Lessons
  learned from the inspire network.
\newblock {\em Advances in Simulation\/}~{\em 2\/}(1), 6.

\bibitem[\protect\citeauthoryear{DerSimonian and Laird}{DerSimonian and
  Laird}{1986}]{dersimonian1986meta}
DerSimonian, R. and N.~Laird (1986).
\newblock Meta-analysis in clinical trials.
\newblock {\em Controlled Clinical Trials\/}~{\em 7\/}(3), 177--188.

\bibitem[\protect\citeauthoryear{Duan, Boland, Moore, and Chen}{Duan
  et~al.}{2019}]{duan2019odal}
Duan, R., M.~R. Boland, J.~H. Moore, and Y.~Chen (2019).
\newblock {ODAL}: A one-shot distributed algorithm to perform logistic
  regressions on electronic health records data from multiple clinical sites.
\newblock {\em Pacific Symposium on Biocomputing\/}, 30--41.

\bibitem[\protect\citeauthoryear{Hansen}{Hansen}{1982}]{hansen1982large}
Hansen, L.~P. (1982).
\newblock Large sample properties of generalized method of moments estimators.
\newblock {\em Econometrica: Journal of the Econometric Society\/}, 1029--1054.

\bibitem[\protect\citeauthoryear{Hedges}{Hedges}{1983}]{hedges1983combining}
Hedges, L.~V. (1983).
\newblock Combining independent estimators in research synthesis.
\newblock {\em British Journal of Mathematical and Statistical
  Psychology\/}~{\em 36\/}(1), 123--131.

\bibitem[\protect\citeauthoryear{Hripcsak, Duke, Shah, Reich, Huser, Schuemie,
  Suchard, Park, Wong, Rijnbeek, et~al.}{Hripcsak
  et~al.}{2015}]{hripcsak2015observational}
Hripcsak, G., J.~D. Duke, N.~H. Shah, C.~G. Reich, V.~Huser, M.~J. Schuemie,
  M.~A. Suchard, R.~W. Park, I.~C.~K. Wong, P.~R. Rijnbeek, et~al. (2015).
\newblock Observational health data sciences and informatics (ohdsi):
  opportunities for observational researchers.
\newblock {\em Studies in Health Technology and Informatics\/}~{\em 216},
  574--578.

\bibitem[\protect\citeauthoryear{Jordan, Lee, and Yang}{Jordan
  et~al.}{2018}]{jordan2018communication}
Jordan, M.~I., J.~D. Lee, and Y.~Yang (2018).
\newblock Communication-efficient distributed statistical inference.
\newblock {\em Journal of the American Statistical Association\/}, 1--14.

\bibitem[\protect\citeauthoryear{Kundu, Tang, and Chatterjee}{Kundu
  et~al.}{2019}]{kundu2019generalized}
Kundu, P., R.~Tang, and N.~Chatterjee (2019).
\newblock Generalized meta-analysis for multiple regression models across
  studies with disparate covariate information.
\newblock {\em Biometrika\/}~{\em 106\/}(3), 567--585.

\bibitem[\protect\citeauthoryear{Lee, Liu, Sun, and Taylor}{Lee
  et~al.}{2017}]{lee2017communication}
Lee, J.~D., Q.~Liu, Y.~Sun, and J.~E. Taylor (2017).
\newblock Communication-efficient sparse regression.
\newblock {\em The Journal of Machine Learning Research\/}~{\em 18\/}(1),
  115--144.

\bibitem[\protect\citeauthoryear{Li, Lindsay, and Waterman}{Li
  et~al.}{2003}]{li2003efficiency}
Li, H., B.~G. Lindsay, and R.~P. Waterman (2003).
\newblock Efficiency of projected score methods in rectangular array
  asymptotics.
\newblock {\em Journal of the Royal Statistical Society: Series B (Statistical
  Methodology)\/}~{\em 65\/}(1), 191--208.

\bibitem[\protect\citeauthoryear{Li, Lin, and Li}{Li
  et~al.}{2013}]{li2013statistical}
Li, R., D.~K. Lin, and B.~Li (2013).
\newblock Statistical inference in massive data sets.
\newblock {\em Applied Stochastic Models in Business and Industry\/}~{\em
  29\/}(5), 399--409.

\bibitem[\protect\citeauthoryear{Lian and Fan}{Lian and
  Fan}{2017}]{lian2017divide}
Lian, H. and Z.~Fan (2017).
\newblock Divide-and-conquer for debiased l 1-norm support vector machine in
  ultra-high dimensions.
\newblock {\em The Journal of Machine Learning Research\/}~{\em 18\/}(1),
  6691--6716.

\bibitem[\protect\citeauthoryear{Liang and Zeger}{Liang and
  Zeger}{1986}]{liang1986longitudinal}
Liang, K.-Y. and S.~L. Zeger (1986).
\newblock Longitudinal data analysis using generalized linear models.
\newblock {\em Biometrika\/}~{\em 73\/}(1), 13--22.

\bibitem[\protect\citeauthoryear{Martins, Fenton, Keyes, Blanco, Zhu, and
  Storr}{Martins et~al.}{2012}]{martins2012mood}
Martins, S.~S., M.~C. Fenton, K.~M. Keyes, C.~Blanco, H.~Zhu, and C.~L. Storr
  (2012).
\newblock Mood and anxiety disorders and their association with non-medical
  prescription opioid use and prescription opioid-use disorder: longitudinal
  evidence from the national epidemiologic study on alcohol and related
  conditions.
\newblock {\em Psychological medicine\/}~{\em 42\/}(6), 1261--1272.

\bibitem[\protect\citeauthoryear{Neyman, Scott, et~al.}{Neyman
  et~al.}{1948}]{neyman1948consistent}
Neyman, J., E.~L. Scott, et~al. (1948).
\newblock Consistent estimates based on partially consistent observations.
\newblock {\em Econometrica\/}~{\em 16\/}(1), 1--32.

\bibitem[\protect\citeauthoryear{Olkin and Sampson}{Olkin and
  Sampson}{1998}]{olkin1998comparison}
Olkin, I. and A.~Sampson (1998).
\newblock Comparison of meta-analysis versus analysis of variance of individual
  patient data.
\newblock {\em Biometrics\/}, 317--322.

\bibitem[\protect\citeauthoryear{Quan, Sundararajan, Halfon, Fong, Burnand,
  Luthi, Saunders, Beck, Feasby, and Ghali}{Quan et~al.}{2005}]{quan2005coding}
Quan, H., V.~Sundararajan, P.~Halfon, A.~Fong, B.~Burnand, J.-C. Luthi, L.~D.
  Saunders, C.~A. Beck, T.~E. Feasby, and W.~A. Ghali (2005).
\newblock Coding algorithms for defining comorbidities in icd-9-cm and icd-10
  administrative data.
\newblock {\em Medical care\/}, 1130--1139.

\bibitem[\protect\citeauthoryear{Sidransky, Nalls, Aasly, et~al.}{Sidransky
  et~al.}{2009}]{sidransky2009multicenter}
Sidransky, E., M.~A. Nalls, J.~O. Aasly, et~al. (2009).
\newblock Multicenter analysis of glucocerebrosidase mutations in parkinson's
  disease.
\newblock {\em New England Journal of Medicine\/}~{\em 361\/}(17), 1651--1661.

\bibitem[\protect\citeauthoryear{Sullivan}{Sullivan}{2018}]{sullivan2018depression}
Sullivan, M.~D. (2018).
\newblock Depression effects on long-term prescription opioid use, abuse, and
  addiction.
\newblock {\em The Clinical journal of pain\/}~{\em 34\/}(9), 878--884.

\bibitem[\protect\citeauthoryear{Tian and Gu}{Tian and
  Gu}{2016}]{tian2016communication}
Tian, L. and Q.~Gu (2016).
\newblock Communication-efficient distributed sparse linear discriminant
  analysis.
\newblock {\em arXiv preprint arXiv:1610.04798\/}.

\bibitem[\protect\citeauthoryear{Van~der Vaart}{Van~der
  Vaart}{2000}]{van2000asymptotic}
Van~der Vaart, A.~W. (2000).
\newblock {\em Asymptotic statistics}, Volume~3.
\newblock Cambridge university press.

\bibitem[\protect\citeauthoryear{Vershynin}{Vershynin}{2010}]{vershynin2010introduction}
Vershynin, R. (2010).
\newblock Introduction to the non-asymptotic analysis of random matrices.
\newblock {\em arXiv preprint arXiv:1011.3027\/}.

\bibitem[\protect\citeauthoryear{Wang, Kolar, Srebro, and Zhang}{Wang
  et~al.}{2017}]{wang2017efficient}
Wang, J., M.~Kolar, N.~Srebro, and T.~Zhang (2017).
\newblock Efficient distributed learning with sparsity.
\newblock In {\em Proceedings of the 34th International Conference on Machine
  Learning-Volume 70}, pp.\  3636--3645. JMLR. org.

\bibitem[\protect\citeauthoryear{Wang, Yang, Chen, and Liu}{Wang
  et~al.}{2019}]{wang2019distributed}
Wang, X., Z.~Yang, X.~Chen, and W.~Liu (2019).
\newblock Distributed inference for linear support vector machine.
\newblock {\em Journal of Machine Learning Research\/}~{\em 20\/}(113), 1--41.

\bibitem[\protect\citeauthoryear{Zhang and Zhou}{Zhang and
  Zhou}{2018}]{zhang2018non}
Zhang, A. and Y.~Zhou (2018).
\newblock A non-asymptotic, sharp, and user-friendly reverse
  chernoff-cram$\backslash$er bound.
\newblock {\em arXiv preprint arXiv:1810.09006\/}.

\bibitem[\protect\citeauthoryear{Zhang, Deng, Schiffman, Qin, and Yu}{Zhang
  et~al.}{2020}]{zhang2020generalized}
Zhang, H., L.~Deng, M.~Schiffman, J.~Qin, and K.~Yu (2020).
\newblock Generalized integration model for improved statistical inference by
  leveraging external summary data.
\newblock {\em Biometrika\/}~{\em 107\/}(3), 689--703.

\bibitem[\protect\citeauthoryear{Zhang, Wainwright, and Duchi}{Zhang
  et~al.}{2012}]{zhang2012communication}
Zhang, Y., M.~J. Wainwright, and J.~C. Duchi (2012).
\newblock Communication-efficient algorithms for statistical optimization.
\newblock {\em Advances in Neural Information Processing Systems\/},
  1502--1510.

\bibitem[\protect\citeauthoryear{Zhao, Cheng, and Liu}{Zhao
  et~al.}{2016}]{zhao2016partially}
Zhao, T., G.~Cheng, and H.~Liu (2016).
\newblock A partially linear framework for massive heterogeneous data.
\newblock {\em Annals of Statistics\/}~{\em 44\/}(4), 1400--1437.

\end{thebibliography}
\newpage
\title{Supplementary Material to "Heterogeneity-aware and communication-efficient distributed statistical inference"}
\author{Rui Duan, Yang Ning and Yong Chen
}
\maketitle
\section*{Appendix A: a modified algorithm }

When the calculating the inverse of $({{{{\bar H}}_{\gamma\gamma}}^{(j)}})^{-1}$ becomes a bottleneck of computation, we proposed the following algorithm 
\begin{itemize}
	\item Set  initial values $\tilde{\beta}^{(0)} = \bar \beta$, $\tilde\Gamma^{(0)} = \bar\Gamma$, for $t = 1, \dots, T$,  do 
	\begin{enumerate}
		\item From  site $1$ to site $K$, calculate and transfer $
		\nabla_\beta L_j  (\tilde{\beta}^{(t-1)} , \tilde\gamma_j^{(t-1)} )$ to site 1.
		\item At site $1$, construct 
		$\tilde S (\beta) = \check S_1 (\beta) +\{\nabla_\beta L_N (\tilde{\beta}^{(t-1)}, \tilde\Gamma^{(t-1)} )-\check S_1 (\tilde{\beta}^{(t-1)})\}$
		where 
		\[
		\check S_1 (\beta) = \sum_{i=1}^{n}g'(y; \beta; \tilde{\beta}^{(t-1)}, \tilde\Gamma^{(t-1)} )
		\]
		with 
		\begin{equation*}
		g'(y; \beta; \beta'; \Gamma') = \frac{1}{K}\left\{\sum_{j=1}^{K} \nabla_\beta\log f(y; \beta,  \gamma_j') \frac{f(y; \beta',  \gamma_j')}{f(y;  \beta',  \gamma_1')}\right\}, 
		\end{equation*}
		\item Update $\tilde{\beta}^{(t)}$ by solving $\tilde S (\beta) =0$ and update $\tilde\gamma_j^{(t)}$ at each site.
	\end{enumerate}
	\item Use $\tilde{\beta}^{(T)}$ as initial value and run Algorithm 1 in the main paper once to obtain $\tilde{\beta}^{(T+1)}$.
	
\end{itemize}

From a computational perspective, this algorithm could potentially cost less time for each of the first $T$ iterations compared to the surrogate efficiency score approach, by avoiding the computation of the inverse of Fisher information matrix. When the dimension $d$ increases, the computational time saved by using this new algorithm might be more obvious compared to the original algorithm we proposed. 

\section*{ Appendix B: parameter settings for simulation study}
The parameter $a$ and $b$ are set to values  in Table 1 to adjust the prevalence of the binary outcome and exposure variables.
\begin{table}[h]
	\def~{\hphantom{0}}
	\caption{Parameter values for four simulation settings}{%
		\begin{tabular}{lll}
			& Common Outcome&  Rare Outcome\\
			\multirow{2}{*}{Common Exposure}
			& $a = 0$ (outcome prevalence: $0.43$) & $a = -3$ (outcome prevalence: $0.06$)  \\
			& $b=0.3$ (exposure prevalence: $0.3$)  & $b=0.3$ (exposure prevalence: $0.3$) \\
			\multirow{2}{*}{Rare Exposure}& $a = 0$ (outcome prevalence: $0.45$) & $a = -3$ (outcome prevalence: $0.07$)  \\
			& $b=0.1$ (exposure prevalence: $0.1$)  & $b=0.1$ (exposure prevalence: $0.1$) \\
	\end{tabular}}
\end{table}

\section*{Appendix C: additional information for Data Analysis}

Table 2 shows the definition of the risk factors included in the regression model.
\begin{table}[h]
	\def~{\hphantom{0}}
	\caption{Definition of variables.}
		\scalebox{0.9}{
		\begin{tabular}{ll}
			\hline
			\hline
			Variables	 &Definition\\
			\hline	                            	                            
			age                    &    age at 1st prescription  \\
			female                &    basic info in demographic table  \\
			alcohol related disorders  &  ICD-9 Code: 291, 303 ICD-10 code:F10  within 12 months before 1st prescription\\
			depression        &            ICD-9 Code: 311	ICD-10 code: F33, F32 within 12 months before 1st prescription  \\
			anxiety         &    ICD-9 Code:  300	 ICD-10 Code: F41   within 12 months before 1st prescription    \\
			pain                           &  ICD-9 Code: 338	 ICD-10 Code: G89, R52 within 12 months before 1st prescription \\
			cannabis related disorder    &   ICD-9 Code: 304.3, 305.2 ICD-10 Code: F12 within 12 months before 1st prescription\\
			cocaine related disorder      &  	 ICD-9 Code: 304.2, 305.6 ICD-10 Code: F14 within 12 months before 1st prescription\\
			Charlson comorbidity index  &   defined diagnosis within 12 months before 1st prescription \citep{quan2005coding}\\
			nicotine related disorder      &ICD-9 Code: 305.1 ICD-10 Code: F17 within 12 months before 1st prescription\\
			smoke1                      &  1: ever smoker; 0: otherwise \\
			smoke2                      &  1: unknown; 0: otherwise\\
			non-Hispanic White    & basic info in demographic table \\		
			\hline
	\end{tabular}}
\end{table}

Table 3 shows the model fitting results from all participating sites.

\begin{table}[h]
	\def~{\hphantom{0}}
	\caption{Estimated log odds ratios (standard errors) from five participating sites.}{
		\begin{tabular}{llllll}
			\hline
			\hline
			Variables	 &Site $1$ & Site $2$ &Site $3$&Site $4$& Site $5$\\
			\hline	                            
			(Intercept)       &              -2.57      (0.48)  &  -2.29       (0.42)  &  -1.93       (1.09)  &  -2.65      (0.68) &  -2.27      ( 0.21)\\
			age                    &        -0.03       (0.01)   &  0.05       (0.01)  &  -0.02       (0.01)&    -0.03       (0.01) &   -0.01       (0.00)\\
			female                &       -1.08       (0.22)   &  0.18       (0.15)  & -0.89       (0.25)  &  -1.19       (0.45)   &  0.00       (0.14)\\
			alcohol related disorders  &    -0.71      (0.89)   &  1.50       (0.80)     &1.25       (0.56)   & 0.42       (1.30)  &   0.55       (0.30)\\
			depression                   &  -1.19       (0.71)  &  -0.27       (0.26)  &   0.28       (0.48)  &  -0.28       (0.91)  &   0.67       (0.23)\\
			anxiety                       &  1.49       (0.43)  &   0.60       (0.22)  &   1.59       (0.36)   &  0.92       (0.63)   &  0.81       (0.22)\\
			pain                           & 1.38       (0.34)  &   0.86       (0.23)  &   1.65       (0.31)  &   3.23       (0.92)   &  0.88       (0.18)\\
			cannabis related disorder    &   1.05       (0.99)   &  0.71       (0.50)     &0.95       (0.55)   & -0.36       (1.84)  &   0.94       (0.42)\\
			cocaine related disorder      &  0.51       (1.02)    & 2.00       (1.11)   &  2.68       (0.69)  &   2.94       (1.94)  &  1.26       (0.48)\\
			Charlson comorbidity index  &                         -0.02       (0.10) &   -0.05       (0.05) &   -0.18       (0.12) &    0.06       (0.15)  &  -0.05       (0.05)\\
			nicotine related disorder      &1.09       (0.44) &    0.21       (0.46)  &   0.14       (0.52)  &   0.20       (0.67)&     0.42       (0.21)\\
			smoke1                      &   1.44       (0.54)   &  1.77       (0.57)   & 0.19       (1.14)  &   1.20       (0.57)  &   1.07       (0.22)\\
			smoke2                      &   1.16       (0.45)   &  1.35       (0.40)  &  -1.05       (1.06)  &  -0.32       (0.56)    & 1.03       (0.20)\\
			non-Hispanic White    &                     0.73       (0.22)   &  1.68       (0.14)   &  0.88       (0.26)    & 1.25       (0.52)   &  0.60      (0.16)\\
			\hline
			\label{tablelabel}
	\end{tabular}}
\end{table}

\section*{Appendix D: theoretical lemmas}
In this section we provide three lemmas and their proofs, and for convenience we use $C, C_1, \dots, $ to denote positive constants which can vary from place to place.

\begin{lem1}
	For $n$ centered independent sub-exponential random variables $X_1, X_2, \dots, X_n$, assume that there exists a constant $C_1$ such that $\sup_{p>1}p^{-1}\{\E |X_i|^p\}^{1/p}\le C_1$ for all $i$, then we have 
	\[
	\E\|\frac{1}{n}\sum_{i=1}^{n}X_i\|_2^{2k} \le \frac{C_2}{n^{k}},
	\]	
	for $k \leq 16$.
\end{lem1}

\begin{lem1}
	For any $j \in \{1, \dots, K\}$, let $\bar \theta_j = (\bar \beta_j,\bar\gamma_j) = \arg\max_{\beta, \gamma_j} L_j(\beta, \gamma_j)$, under Assumption 1-5, we have $\bar\theta_j$ satisfies
	\begin{equation*}
	\|\bar\theta_j- \theta_j^*\|_2 \le C_1\|\nabla L_j(\theta_j^*)\|_2
	\end{equation*}
	with probability at least $1-\exp(-C_2n)$. In addition
	\begin{equation}
	\bar\theta_j- \theta_j^*= {I^{(j)}(\theta_j^*)}^{-1}\nabla L_j (\theta_j^*) +\delta_j,
	\end{equation}
	where $\delta_j$ satisfies $\E\|\delta_j\|^{k}_2\lesssim1/n^{k}$ for $k \in \{1, \dots, 16\}$.  
\end{lem1}

\begin{lem1}
	Define $\hat\Theta = (\hat \beta, \hat\Gamma) = \arg\max_{\beta, \Gamma} L_N(\beta; \Gamma)$, where $\Gamma = (\gamma_1, \dots, \gamma_K)$. Under Assumption 1-5, we have  
	\[
	\E\|\hat{\Theta} - \Theta^*\|_2^2 \le C\frac{K}{n}
	\]
	for some positive constant $C$.  
\end{lem1}

\begin{lem1}
	Under Assumption 1-5, we have for $j \in \{1, \dots, K\}$,   
	\[
	\E\|\hat{\gamma_j} - \gamma_j^*\|_2^2 \le \frac{C}{n}
	\]
	for some positive constant $C$.  
\end{lem1}

\begin{lem1}
	Under Assumption 1-5, the global maximum likelihood estimator satisfies, 
	\[
	\hat{\Theta}-\Theta^* = I^{-1}S(\Theta^*) +\delta
	\] 
	where 
	\[
	I = \begin{pmatrix} 
	\sum_{j=1}^{K}I_{\beta\beta}^{(j)} & & I_{\beta\gamma}^{(1)}&\dots&I_{\beta\gamma}^{(K)} \\
	I_{\gamma\beta}^{(1)}& I_{\gamma\gamma}^{(1)} & 0&\dots &0\\
	\dots& 0&I_{\gamma\gamma}^{(2)}&\dots &0\\
	\dots&&\dots&&\dots
	\\
	I_{\beta\gamma}^{(K)}&0&\dots &0&I_{\gamma\gamma}^{(K)}
	\end{pmatrix},
	\]
	
	\[
	S = \begin{pmatrix} 
	\sum_{j=1}^{K}\nabla_{\beta}L_j(\theta_j^*)\\
	\nabla_{\gamma}L_1(\theta_1^*)\\
	\dots\\
	\nabla_{\gamma}L_K(\theta_K^*)
	\end{pmatrix},
	\]
	and each entry of $\delta$ satisfies  $\E|\delta_t|^2\lesssim 1/n^2$ for all $t$.
\end{lem1}

\begin{lem1}
	Define $\check{\beta}$ to be the solution of the following equation
	\[
	0 = \sum_{j=1}^{K}\{\nabla_{\beta}L_j(\beta, \bar{\gamma}_j) -\bar H^{(j)}_{\beta\gamma} ({{{{\bar H}}_{\gamma\gamma}}^{(j)}})^{-1}\nabla_{\beta}L_j(\beta, \bar{\gamma}_j)\},
	\] 
	we have under assumption 1-5, 
	\[
	\check{\beta} - \beta^* = \{\sum_{j=1}^{K}I^{(j)}_{\beta|\gamma}\}^{-1}\sum_{j=1}^{K}\{\nabla_\beta L_j (\beta^*, \gamma_j^*)-I^{(j)}_{\beta\gamma}{I^{(j)}_{\gamma\gamma}}^{-1}\nabla_{\gamma}L_j(\beta^*, \gamma_j^*)\}+\check\delta
	\]
	where the remaining term $\check\delta$ satisfies $\E\|\check\delta\|_2^8\lesssim 1/n^8$.
\end{lem1}

\begin{lem1}
	Under assumption 1-5, for $T=1$, the surrogate estimator $\tilde{\beta}^{(1)}$ satisfies that
	\[
	\E\{\|\tilde{\beta}^{(1)}- \check{\beta}\|_2^4\}\lesssim\frac{1}{K^2n^4}+\frac{1}{n^6}.
	\] 
\end{lem1}

\begin{lem1}
	Under assumption 1-5, for $T=2$, the updated initial estimator, which is obtained by
	\[
	\bar{\gamma}^{(2)} = \arg\max_{\gamma_j}L_j(\tilde{\beta}^{(1)}, \gamma_j),
	\]
	satisfies
	\begin{align*}
	\bar{\gamma}_j^{(2)} - \gamma_j^* =&-{I^{(j)}_{\gamma\gamma}}^{-1}I^{(j)}_{\gamma\beta}\{\sum_{j=1}^{K}I^{(j)}_{\beta|\gamma}\}^{-1} \sum_{j=1}^{K}\{\nabla_\beta L_j (\beta^*, \gamma_j^*)-I^{(j)}_{\beta\gamma}{I^{(j)}_{\gamma\gamma}}^{-1}\nabla_{\gamma}L_j(\beta^*, \gamma_j^*)\}\\&+{I^{(j)}_{\gamma\gamma}}^{-1}\nabla_{\gamma}L_j(\beta^*, \gamma_j^*)+\bar\delta^{(2)}
	\end{align*}
	where it satisfies that $\E\|\bar\delta^{(2)}\|_2^2 \lesssim1/{n^2}$.
\end{lem1}

\begin{lem1}
	Define  
	\begin{equation*}
	H^{(j)}(\beta, \gamma_j) = -\frac{1}{n}\sum_{i=1}^{n}\nabla^2\log f(y_{i1};\beta,\gamma_j),
	\end{equation*} 
	and 
	\begin{equation*}
	\tilde H^{(1,j)}(\beta, \gamma_j) = -\frac{1}{n}\sum_{i=1}^{n}\nabla^2\log f(y_{i1};\beta,\gamma_j)\frac{f(y_{i1};\bar \beta, \bar \gamma_j)}{f(y_{i1}; \bar \beta, \bar \gamma_1)}.
	\end{equation*}
	For some $\rho \in (0, 1)$, we have $\mu_1(1-\rho)\preceq H^{(j)}(\beta, \gamma_j) \preceq 2\mu_+$, and $\mu_1(1-\rho)\preceq\tilde H^{(1,j)}(\beta, \gamma_j) \preceq 2\mu_+$ for $\theta_j \in U(\delta_\rho)$ with probability at least $1-C\exp(-n)$, where $\delta_\rho \in (0, \rho\mu_-/(4M)).$
\end{lem1}
\begin{lem1}
	Suppose $(\bar\beta_j^{(1)}, \bar\gamma_j^{(1)}) = \arg\max_{\beta,\gamma_j}L_j(\beta, \gamma_j)$, and 
	$\bar\beta = \sum_{j=1}^{K}\bar\beta_j^{(1)}/K$. We have $\E\|\bar\beta-\hat\beta\|_2 \gtrsim {1}/{(Kn)^{1/2}}$ when $K/n \rightarrow 0$.
\end{lem1}

\begin{lem1}
	Assume $Y_{j}\sim f(y; \theta_j^*)$, we have for any function $g(y)$, we have 
	\[
	\E_{\theta_j^*}\{g(Y_j)\} = \E_{\theta_1^*}\{g(Y_1)\frac{f(y; \theta_j^*)}{f(y; \theta_1^*)}\}.
	\]
\end{lem1}

\begin{lem1}
	Under assumption 1-5,  the one-step estimator defined in Remarks 2 satisfies that
	\[
	\E\{\|\tilde{\beta}^{(1)}-\tilde{\beta}^{(O)}\|_2^4\}\lesssim\frac{1}{K^2n^4}+\frac{1}{n^6}.
	\] 
\end{lem1}

\section*{Appendix E: proofs of theorems, lemmas, and corollaries in the main paper}
\subsection*{Proof of Lemma 1} 
We first write $I$ as 
\begin{align*}
I = \begin{pmatrix} 
I_{\beta\beta} & I_{\beta\Gamma}\\
I_{\Gamma\beta} & I_{\Gamma\Gamma}
\\
\end{pmatrix},
\end{align*}
where $I_{\beta\beta} = \sum_{j=1}^{K}I_{\beta\beta}^{(j)}$, $I_{\beta\Gamma} = I_{\Gamma\beta}^\T = \left( I_{\beta\gamma}^{(1)},\dots, I_{\beta\gamma}^{(K)}\right)$, and $I_{\Gamma\Gamma} = \text{diag}\{I^{(1)}_{\gamma\gamma}, \dots,I^{(K)}_{\gamma\gamma} \}$, which is a block diagonal matrix. By Inversion of block matrix, we have 
\begin{align*}
I^{-1} &= \begin{pmatrix} 
I^{-1}_{\beta\beta} & I^{-1}_{\beta\Gamma}\\
I^{-1}_{\Gamma\beta} & I^{-1}_{\Gamma\Gamma}
\\
\end{pmatrix}\\&= \begin{pmatrix} 
(I_{\beta\beta}- I_{\beta\Gamma}I_{\Gamma\Gamma}^{-1}I_{\Gamma\beta})^ {-1} & & -(I_{\beta\beta}- I_{\beta\Gamma}I_{\Gamma\Gamma}^{-1}I_{\Gamma\beta})^{-1}I_{\Gamma\beta}I_{\Gamma\Gamma}^{-1}\\
-I_{\Gamma\Gamma}^{-1}I_{\Gamma\beta}(I_{\beta\beta}- I_{\beta\Gamma}I_{\Gamma\Gamma}^{-1}I_{\Gamma\beta})^ {-1}& & I_{\Gamma\Gamma}^{-1}+I_{\Gamma\Gamma}^{-1}I_{\Gamma\beta}(I_{\beta\beta}- I_{\beta\Gamma}I_{\Gamma\Gamma}^{-1}I_{\Gamma\beta})^ {-1}I_{\beta\Gamma}I_{\Gamma\Gamma}^{-1}
\\
\end{pmatrix}.
\end{align*}
Define the partial information matrix to be $ I^{-1}_{\beta\beta}=I_{\beta|\gamma}^{(j)} = I_{\beta\beta}^{(j)}-I^{(j)}_{\beta\gamma}{I^{(j)}_{\gamma\gamma}}^{-1}I^{(j)}_{\gamma\beta}$. We have
\[
(I_{\beta\beta}- I_{\beta\Gamma}I_{\Gamma\Gamma}^{-1}I_{\Gamma\beta})^ {-1} = \left(\sum_{j=1}^{K}I_{\beta|\gamma}^{(j)}\right)^{-1},
\]
and 
\[
I^{-1}_{\beta\Gamma}=-(I_{\beta\beta}- I_{\beta\Gamma}I_{\Gamma\Gamma}^{-1}I_{\Gamma\beta})^{-1}I_{\Gamma\beta}I_{\Gamma\Gamma}^{-1} = \{\sum_{j=1}^{K}I_{\beta|\gamma}^{(j)}\}^{-1}\left(I_{\beta\gamma}^{(1)}{I_{\gamma\gamma}^{(1)}}^{-1}, \dots, I_{\beta\gamma}^{(K)}{I_{\gamma\gamma}^{(K)}}^{-1}\right).
\]
From Lemma S.5, we have 
\begin{align*}
\hat\beta - \beta^* = \left(\sum_{j=1}^{K}I_{\beta|\gamma}^{(j)}\right)^{-1}\sum_{j=1}^{K}\nabla_{\beta}L_j(\theta_j^*)+\{\sum_{j=1}^{K}I_{\beta|\gamma}^{(j)}\}^{-1}\left(\sum_{j=1}^{K}I_{\beta\gamma}^{(j)}{I_{\gamma\gamma}^{(j)}}^{-1}\nabla_\gamma  L_j(\theta_j^*)\right)+\delta_\beta.
\end{align*}
By Assumption 2, we know that $\mu_-\text{I}_d\preceq I^{(j)}\preceq \mu_+\text{I}_d$, which implies $\mu_-\text{I}_p \preceq I^{(j)}_{\beta|\gamma}\preceq \mu_+\text{I}_p$. We have 
\begin{align*}
&\E\|\hat\beta - \beta^*\|_2\le \|\left(\frac{1}{K}\sum_{j=1}^{K}I_{\beta|\gamma}^{(j)}\right)^{-1}\|_2\E\|\frac{1}{K}\sum_{j=1}^{K}\nabla_{\beta}L_j(\theta_j^*)\|_2\\&+\|\frac{1}{K}\{\sum_{j=1}^{K}I_{\beta|\gamma}^{(j)}\}^{-1}\|_2\E\|\frac{1}{K}\sum_{j=1}^{K}I_{\beta\gamma}^{(j)}{I_{\gamma\gamma}^{(j)}}^{-1}\nabla_\gamma  L_j(\theta_j^*)\|_2+\E\|\delta_{\beta}\|_2\\
&\le \mu_-^{-1} \E\|\frac{1}{Kn}\sum_{j=1}^{K}\sum_{i=1}^{n}\nabla_{\beta}\log f(y_{ij};\theta_j^*)\|_2\\&+\mu_-^{-1} \E\|\frac{1}{Kn}\sum_{j=1}^{K}\sum_{i=1}^{n}I_{\beta\gamma}^{(j)}{I_{\gamma\gamma}^{(j)}}^{-1}\nabla_{\gamma}\log f(y_{ij};\theta_j^*)\|_2+\E\|\delta_{\beta}\|_2
\end{align*}
From Lemma S.1 and Lemma S.5, we obtain 
\begin{align*}
&\E\|\hat\beta - \beta^*\|_2\le \frac{C_1}{\sqrt{Kn}} +\frac{C_2}{n},
\end{align*}
which completes the proof.$\hfill\square$
\subsection*{Proof of Theorem 1}
From Lemma S.7 we have 
$\E\{\|\tilde{\beta}^{(1)}-\check\beta\|_2\}\lesssim 1/(K^{1/2}n)+1/n^{3/2}$. Therefore we only need to show that 
$\E\{\|\hat{\beta}-\check\beta\|_2\}\lesssim1/n$. 

From Lemma S.5, and S.6, we have 
\[
\hat{\beta} - \beta^* = \{\sum_{j=1}^{K}I^{(j)}_{\beta|\gamma}\}^{-1}\sum_{j=1}^{K}\{\nabla_\beta L_j (\beta^*, \gamma_j^*)-I^{(j)}_{\beta\gamma}{I^{(j)}_{\gamma\gamma}}^{-1}\nabla_{\gamma}L_j(\beta^*, \gamma_j^*)\}+\delta_\beta,
\]
and 
\[
\check{\beta} - \beta^* = \{\sum_{j=1}^{K}I^{(j)}_{\beta|\gamma}\}^{-1}\sum_{j=1}^{K}\{\nabla_\beta L_j (\beta^*, \gamma_j^*)-I^{(j)}_{\beta\gamma}{I^{(j)}_{\gamma\gamma}}^{-1}\nabla_{\gamma}L_j(\beta^*, \gamma_j^*)\}+\check\delta,
\]
where $\E\|\check\delta\|_2\le {1}/{n}$, and $\E\|\delta_\beta\|_2\le {1}/{n}$. Then we have 
\[
\E\|\hat{\beta} - \check\beta\|_2\le\E\|\delta_\beta\|_2+\E\|\check\delta\|_2\lesssim 1/n.
\]
$\hfill \square$

\subsection*{Proof of Theorem 2}
During the second iteration, the initial value becomes 
$\bar\beta = \tilde{\beta}^{(1)}$ and $\bar{\gamma}_j = \bar{\gamma}_j^{(2)}$. In the following proof, we only use $\bar\beta$ and $\bar{\gamma}_j$ for easier notation.

Since $\hat\beta$ and $\hat{\gamma}_j$ maximize the function $L_N(\beta, \Gamma)$, we have 
$\sum_{j=1}^{K}\nabla_{\beta}L_j(\hat{\beta}, \hat{\gamma}_j) = 0$ and $\nabla_{\gamma}L_j(\hat{\beta}, \hat{\gamma}_j) = 0$. Therefore, for the efficient score function construct as 
\[
S(\beta, \Gamma) = \frac{1}{K}\sum_{j=1}^{K}\{\nabla_{\beta}L_j({\beta}, {\gamma}_j) - \bar H^{(j)}_{\beta\gamma} ({{{{\bar H}}_{\gamma\gamma}}^{(j)}})^{-1}\nabla_{\gamma}L_j({\beta}, {\gamma}_j)\}
\]
we have $S(\hat\beta, \hat\Gamma) = 0$, where 
\[
\bar H^{(j)}_{\beta\gamma} = \nabla_{\beta\gamma} L_j({\bar\beta, \bar{\gamma}_j})
\]
and
\[
\bar H^{(j)}_{\gamma\gamma} = \nabla_{\gamma\gamma} L_j({\bar\beta, \bar{\gamma}_j}).
\]
And $\tilde{\beta}^{(2)}$ satisfies 
\[
\tilde U(\tilde{\beta}^{(2)}) =  U_1(\tilde{\beta}^{(2)}) + \{\frac{1}{K}\sum_{j=1}^{K}\{\nabla_\beta L_j (\bar\beta, \bar\gamma_j)- \bar H^{(j)}_{\beta\gamma} ({{{{\bar H}}_{\gamma\gamma}}^{(j)}})^{-1}  \nabla_\gamma L_j (\bar\beta, \bar\gamma_j)\}-   U_1(\bar\beta)\} = 0. 
\]
We define the following events:

\[
\mathcal{E}_{0j}: =\{ \frac{1}{n}\sum_{i=1}^{n}m_k(Y_{ij}) \le  2M, \text{ for } k = 1, 2\},
\]

\[
\mathcal{E}_1 : = \{\|\nabla_\beta\tilde U(\hat\beta)-\nabla_\beta S(\hat\beta, \hat\Gamma)\|_2\le C_1\},
\]
and 
\[
\mathcal{E}_2 : = \{\|\tilde U(\hat\beta)\|_2 \le C_2\}.
\] 
for some constants $M$, $C_1$ and $C_2$ which satisfy $\E\{ m_k (Y_{ij})\}< M$ for all $j \in \{1, \dots, K\}$, and $k = 1, 2$, $C_1\le \rho\mu_-/2$ and $C_2<(1-\rho)\rho\mu_-^2/8M$.
Define $\mathcal{E}_{0} = \cap_{1\le j\le K}\mathcal{E}_{0j}$. Applying Lemma 6 in \cite{zhang2012communication} we have under
event $\mathcal{E} = \cap_{i=0,1,2}\mathcal{E}_i$, 
\[
\|\tilde{\beta} -\hat{\beta}\|_2 \le C\|\tilde U(\hat\beta)\|_2.
\]
Now we control the term $\|\tilde U(\hat\beta)\|_2$. We have 
\begin{align*}
\tilde U(\hat\beta) =   U_1(\hat\beta) + S(\bar\beta, \bar\Gamma) -  U_1(\bar\beta).
\end{align*}
Since $S(\hat\beta, \hat\Gamma) = 0$, we have 
\begin{align*}
&\tilde U(\hat\beta) =   U_1(\hat\beta)-S(\hat\beta, \hat\Gamma) + S(\bar\beta, \bar\Gamma) -  U_1(\bar\beta)\\
& = \{\nabla_\beta  U_1(\beta') -\nabla_\beta S(\beta', \Gamma')\}(\bar{\beta}-\hat{\beta}) +\nabla_\gamma S(\beta', \Gamma') (\bar{\gamma}_j-\hat{\gamma}_j),
\end{align*}
where $\beta'$ and $\gamma_j'$ satisfy $\|\beta'-\hat\beta\|_2^2 \le\|\bar{\beta}-\hat\beta\|_2^2$, and $\|\gamma_j'-\hat\gamma_j\|_2^2 \le\|\bar{\gamma}_j-\hat\gamma_j\|_2^2$.  
Following the same proof as Lemma S.7 (See (\ref{ee4}) - (\ref{ee6})), we have
\textcolor{black}{
	\beq\label{e4}
	\E \|\{\nabla_\beta  U_1(\beta') -\nabla_\beta S(\beta', \Gamma')\}(\bar{\beta}-\hat{\beta})\|_2 = \frac{1}{K^{1/2}n}+\frac{1}{n^{3/2}}. 
	\eeq}
Now we control the term $\nabla_\gamma S(\beta', \Gamma') (\bar{\gamma}_j-\hat{\gamma}_j)$. 
Now we control the term $\nabla_\gamma S(\beta', \Gamma') (\bar{\gamma}_j-\hat{\gamma}_j)$. 
From Lemma S.5, and Equations (\ref{inv0})-(\ref{inv4}), we have 
\begin{align*}
\hat{\gamma}_j - {\gamma}_j^* = 
&-{I^{(j)}_{\gamma\gamma}}^{-1}I^{(j)}_{\gamma\beta}\{\sum_{j=1}^{K}I^{(j)}_{\beta|\gamma}\}^{-1} \sum_{j=1}^{K}\{\nabla_\beta L_j (\beta^*, \gamma_j^*)-I^{(j)}_{\beta\gamma}{I^{(j)}_{\gamma\gamma}}^{-1}\nabla_{\gamma}L_j(\beta^*, \gamma_j^*)\}\\&+{I^{(j)}_{\gamma\gamma}}^{-1}\nabla_{\gamma}L_j(\beta^*, \gamma_j^*)+\hat\delta,
\end{align*}
and by Lemma S.8, we have 
\begin{align*}
\bar{\gamma}_j^{(2)} - \gamma_j^* =&-{I^{(j)}_{\gamma\gamma}}^{-1}I^{(j)}_{\gamma\beta}\{\sum_{j=1}^{K}I^{(j)}_{\beta|\gamma}\}^{-1} \sum_{j=1}^{K}\{\nabla_\beta L_j (\beta^*, \gamma_j^*)-I^{(j)}_{\beta\gamma}{I^{(j)}_{\gamma\gamma}}^{-1}\nabla_{\gamma}L_j(\beta^*, \gamma_j^*)\}\\&+{I^{(j)}_{\gamma\gamma}}^{-1}\nabla_{\gamma}L_j(\beta^*, \gamma_j^*)+\bar\delta^{(2)},
\end{align*}
and therefore we have, 
\begin{align*}
\bar{\gamma}_j-\hat{\gamma}_j = \hat\delta-\bar\delta^{(2)},
\end{align*}
where it satisfies $\E\|\hat\delta\|^2_2\lesssim1/n^2$, and $\E \|\bar\delta^{(2)}\|_2^2\lesssim 1/n^2$, which implies 
$\E\|\bar{\gamma}_j-\hat{\gamma}_j\|^2_2\lesssim1/n^2$. 
In addition we have 
\begin{align}
\nabla_{\gamma}S(\beta', \Gamma')& = \frac{1}{K}\sum_{j=1}^{K}\{\nabla_{\beta\gamma}L_j({\beta}', {\gamma}_j') - \bar H^{(j)}_{\beta\gamma} ({{{{\bar H}}_{\gamma\gamma}}^{(j)}})^{-1}\nabla_{\gamma\gamma}L_j({\beta}', {\gamma}_j')\}\nonumber\\&
= \frac{1}{K}\sum_{j=1}^{K}\{\nabla_{\beta\gamma}L_j({\beta}^*, {\gamma}_j^*) - I^{(j)}_{\beta\gamma} ({{{{I}}_{\gamma\gamma}}^{(j)}})^{-1}\nabla_{\gamma\gamma}L_j({\beta}^*, {\gamma}_j^*)\}\nonumber\\&+
\frac{1}{K}\sum_{j=1}^{K}\{\nabla_{\beta\gamma}L_j({\beta}', {\gamma}_j')-\nabla_{\beta\gamma}L_j({\beta}^*, {\gamma}_j^*)\}\nonumber\\&
+ \frac{1}{K}\sum_{j=1}^{K}\{I^{(j)}_{\beta\gamma} ({{{{I}}_{\gamma\gamma}}^{(j)}})^{-1}\nabla_{\gamma\gamma}L_j({\beta}^*, {\gamma}_j^*)- \bar H^{(j)}_{\beta\gamma} ({{{{\bar H}}_{\gamma\gamma}}^{(j)}})^{-1}\nabla_{\gamma\gamma}L_j({\beta}', {\gamma}_j')\}.\label{e2}
\end{align}
By Lemma S.1, we have 
\begin{align*}
\|\frac{1}{K}\sum_{j=1}^{K}\nabla_{\beta\gamma}L_j({\beta}^*, {\gamma}_j^*) - I^{(j)}_{\beta\gamma} ({{{{I}}_{\gamma\gamma}}^{(j)}})^{-1}\nabla_{\gamma\gamma}L_j({\beta}^*, {\gamma}_j^*)\|_2^2 \lesssim \frac{1}{Kn}+\frac{1}{n^2}, 
\end{align*}
and by Assumption 5 and $\mathcal{E}_0$, we have 
\begin{align*}
&\E\|\frac{1}{K}\sum_{j=1}^{K}\{\nabla_{\beta\gamma}L_j({\beta}', {\gamma}_j')-\nabla_{\beta\gamma}L_j({\beta}^*, {\gamma}_j^*)\}\|_2^2\le 4M\E\{\|\bar\beta-\hat{\beta}\|_2^2+\|\hat\beta-{\beta}^*\|_2^2+\|\bar\gamma_j-\hat{\gamma}_j\|_2^2+\|\hat\gamma_j-{\gamma}_j^*\|_2^2\}\\&\lesssim\frac{1}{n} 
\end{align*}

For the term in (\ref{e2}), we have
\begin{align}
&\frac{1}{K}\sum_{j=1}^{K}\{I^{(j)}_{\beta\gamma} ({{{{I}}_{\gamma\gamma}}^{(j)}})^{-1}\nabla_{\gamma\gamma}L_j({\beta}^*, {\gamma}_j^*)- \bar H^{(j)}_{\beta\gamma} ({{{{\bar H}}_{\gamma\gamma}}^{(j)}})^{-1}\nabla_{\gamma\gamma}L_j({\beta}', {\gamma}_j')\}\nonumber\\ 
= &\frac{1}{K}\sum_{j=1}^{K}\{I^{(j)}_{\beta\gamma} ({{{{I}}_{\gamma\gamma}}^{(j)}})^{-1}\nabla_{\gamma\gamma}L_j({\beta}^*, {\gamma}_j^*)+I^{(j)}_{\beta\gamma} ({{{{I}}_{\gamma\gamma}}^{(j)}})^{-1}I_{\gamma\gamma}^{(j)}\}\nonumber\\&-\frac{1}{K}\sum_{j=1}^{K}\{I^{(j)}_{\beta\gamma}+ H^{(j)}_{\beta\gamma}\}+\frac{1}{K}\sum_{j=1}^{K}\{H^{(j)}_{\beta\gamma} - \bar H^{(j)}_{\beta\gamma}\}\nonumber\\&+\frac{1}{K}\sum_{j=1}^{K}\bar H^{(j)}_{\beta\gamma} ({{{{\bar H}}_{\gamma\gamma}}^{(j)}})^{-1}\{\nabla_{\gamma\gamma}L_j(\bar{\beta}, \bar{\gamma}_j)-\nabla_{\gamma\gamma}L_j({\beta}', {\gamma}_j')\}\label{e3}
\end{align}
where $H_{\beta\gamma}^{(j)} = \nabla_{\beta\gamma}L_j(\beta^*,\gamma_j^*)$, and $H_{\gamma\gamma}^{(j)} = \nabla_{\gamma\gamma}L_j(\beta^*,\gamma_j^*)$. By Lemma S.1, we have 
\[
\E\|\frac{1}{K}\sum_{j=1}^{K}\{I^{(j)}_{\beta\gamma} ({{{{I}}_{\gamma\gamma}}^{(j)}})^{-1}\nabla_{\gamma\gamma}L_j({\beta}^*, {\gamma}_j^*)+I^{(j)}_{\beta\gamma} ({{{{I}}_{\gamma\gamma}}^{(j)}})^{-1}I_{\gamma\gamma}^{(j)}\}\|_2^2\lesssim \frac{1}{Kn},
\]
and 
\[
\E\|\frac{1}{K}\sum_{j=1}^{K}\{I^{(j)}_{\beta\gamma}+ H^{(j)}_{\beta\gamma}\}\|_2^2\lesssim\frac{1}{Kn}.
\]
Under Assumption 5, and $\mathcal{E}_0$, we have
\begin{align*}
&\E\|\frac{1}{K}\sum_{j=1}^{K}\{H^{(j)}_{\beta\gamma} - \bar H^{(j)}_{\beta\gamma}\}\|_2 =\E\| \frac{1}{K}\sum_{j=1}^{K}\{\nabla_{\beta\gamma}L_j(\beta^*,\gamma_j^*) - \nabla_{\beta\gamma}L_j(\bar\beta,\bar\gamma_j)\}\|^2_2\\
&\le 4M^2\frac{1}{K}\sum_{j=1}^{K}\E\{\|\bar\beta-\hat{\beta}\|_2^2+\|\hat\beta-{\beta}^*\|_2^2+\|\bar\gamma_j-\hat{\gamma}_j\|_2^2+\|\hat\gamma_j-{\gamma}_j^*\|_2^2\}\le \frac{1}{n}.
\end{align*}

To control the term in (\ref{e3}), by Lemma S.2 we have $\bar H^{(j)}\succeq (1-\rho)\mu_- \text{I}_d$ with probability at least $1-\exp(-Cn)$. Thus, 
\begin{align*}
&\E\|\frac{1}{K}\sum_{j=1}^{K}\bar H^{(j)}_{\beta\gamma} ({{{{\bar H}}_{\gamma\gamma}}^{(j)}})^{-1}\{\nabla_{\gamma\gamma}L_j(\bar{\beta}, \bar{\gamma}_j)-\nabla_{\gamma\gamma}L_j({\beta}', {\gamma}_j')\}\|_2^2\\&\le\frac{1}{K}\sum_{j=1}^{K}\E\|\bar H^{(j)}_{\beta\gamma} ({{{{\bar H}}_{\gamma\gamma}}^{(j)}})^{-1}\{\nabla_{\gamma\gamma}L_j(\bar{\beta}, \bar{\gamma}_j)-\nabla_{\gamma\gamma}L_j({\beta}', {\gamma}_j')\}\|_2^2\\&\le \frac{4\mu_+}{(1-\rho)\mu_-}\frac{2M}{K}\sum_{j=1}^{K}\E\{\|\bar\beta-\hat{\beta}\|_2^2+\|\hat\beta-{\beta}^*\|_2^2+\|\bar\gamma_j-\hat{\gamma}_j\|_2^2+\|\hat\gamma_j-{\gamma}_j^*\|_2^2\}\lesssim\frac{1}{n}. 
\end{align*}
Thus, we have under $\mathcal{E}_0$,
\begin{align*}
\E\|\frac{1}{K}\sum_{j=1}^{K}\{I^{(j)}_{\beta\gamma} ({{{{I}}_{\gamma\gamma}}^{(j)}})^{-1}\nabla_{\gamma\gamma}L_j({\beta}^*, {\gamma}_j^*)- \bar H^{(j)}_{\beta\gamma} ({{{{\bar H}}_{\gamma\gamma}}^{(j)}})^{-1}\nabla_{\gamma\gamma}L_j({\beta}', {\gamma}_j')\}\|_2^2\lesssim \frac{1}{n}. 
\end{align*}
Combining all we have 
\beq\label{e5}
\E\{\|\nabla_\gamma S(\beta', \Gamma') (\bar{\gamma}_j-\hat{\gamma}_j)\|_2I(\mathcal{E}_0)\}\lesssim \frac{1}{n^{3/2}}. 
\eeq
Combining (\ref{e4}) and (\ref{e5}), we have $\E\{\|\tilde U(\hat\beta)\|_2I(\mathcal{E})\}\lesssim 1/(K^{1/2}n)+1/n^{3/2}$. \textcolor{black}{Following the same argument as Lemma S.7 (See (\ref{ee10}) to (\ref{ee9}))}, we have $\p(\mathcal{E}^c)\lesssim 1/(K^{1/2}n)+1/n^{3/2}$. Thus, $\E\|\tilde{\beta} -\hat{\beta}\|_2\lesssim 1/(K^{1/2}n)+1/n^{3/2}$.
$\hfill\square$
\subsection*{Proof of Theorem 3}
From Theorem 1 and 2, we have $\E\{\|\tilde{\beta}^{(T)}-\beta^*\|_2^2\}\lesssim 1/n^2$ for $T\ge 1$. Therefore we have 
\begin{align*}
{(Kn)}^{1/2}(\tilde\beta-\beta^*) = {(Kn)}^{1/2}(\tilde\beta-\hat\beta) +{(Kn)}^{1/2}(\hat\beta-\beta^*). 
\end{align*}
Since we assume $K/n \rightarrow 0$, we have  $\E {(Kn)}^{1/2}\|\tilde\beta-\hat\beta\|_2\}\rightarrow 0$, which implies ${(Kn)}^{1/2}(\tilde\beta-\hat\beta) = o_P(1)$.
From Lemma S.5, we have 
\begin{align*}
{(Kn)}^{1/2}(\hat\beta-\beta^*) = {(Kn)}^{1/2}\{\sum_{j=1}^{K}I^{(j)}_{\beta|\gamma}\}^{-1}\sum_{j=1}^{K}\{\nabla_\beta L_j (\beta^*, \gamma_j^*)-I^{(j)}_{\beta\gamma}{I^{(j)}_{\gamma\gamma}}^{-1}\nabla_{\gamma}L_j(\beta^*, \gamma_j^*)\}+{(Kn)}^{1/2}\delta_{\beta}
\end{align*}
where $\E\|{(Kn)}^{1/2}\delta_{\beta}\|_2 \rightarrow 0$. Thus, 
\begin{align*}
{(Kn)}^{1/2}(\tilde\beta^{(T)}-\beta^*) = {(Kn)}^{1/2}\{\sum_{j=1}^{K}I^{(j)}_{\beta|\gamma}\}^{-1}\sum_{j=1}^{K}\{\nabla_\beta L_j (\beta^*, \gamma_j^*)-I^{(j)}_{\beta\gamma}{I^{(j)}_{\gamma\gamma}}^{-1}\nabla_{\gamma}L_j(\beta^*, \gamma_j^*)\}+o_P(1),
\end{align*}
which implies 
$$
Kn(\tilde\beta^{(T)}-\beta^*)^\T I_{\beta|\gamma}(\tilde\beta^{(T)}-\beta^*) \rightarrow \chi^2_p.$$ 
$\hfill\square$

\subsection*{Proof of Theorem 4}
From Theorem 3, we know that ${(Kn)}^{1/2}(\tilde\beta-\beta^*)$ converge in distribution to $N(0,I_{\beta|\gamma}^{-1})$, which implies
\[
Kn(\tilde\beta-\beta^*)^\T I_{\beta|\gamma}(\tilde\beta-\beta^*) \rightarrow \chi^2_p.
\]
Since $I_{\beta|\gamma} = \sum_{j=1}^{K}I^{(j)}_{\beta|\gamma}/K$ and $\tilde I_{\beta|\gamma} = \sum_{j=1}^{K}\tilde I^{(j)}_{\beta|\gamma}/K$.  We only need to prove that  $\tilde I^{(j)}$ is a consistent estimator of  $I^{(j)}$.
We have 
\begin{align*}
\|\tilde I^{(j)} - I^{(j)}\|_2 &\le \|\frac{1}{n}\sum_{i=1}^{n} \nabla_{\beta\beta}\log f(y_{i1},\tilde\beta, \bar\gamma_j)\frac{f(y_{i1}\tilde\beta, \bar\gamma_j)}{f(y_{i1}\tilde\beta, \bar\gamma_1)}-\log f(y_{i1},\beta^*, \gamma_j^*)\frac{f(y_{i1},\beta^*, \gamma_j^*)}{f(y_{i1},\beta^*, \gamma_1^*)}\|_2 \\
&+\|\frac{1}{n}\sum_{i=1}^{n}\log f(y_{i1},\beta^*, \gamma_j^*)\frac{f(y_{i1},\beta^*, \gamma_j^*)}{f(y_{i1},\beta^*, \gamma_1^*)}-I^{(j)}\|_2\\
&\le\{\frac{1}{n}\sum_{i=1}^{n} m_2(y_{i1})\}\{\|\tilde{\beta}-\beta^*\|_2+\|\bar \gamma_1 - \gamma_1^*\|_2+\|\bar \gamma_j - \gamma_j^*\|_2\} +o_p(1)= o_p(1)
\end{align*}
Thus $\tilde I_{\beta|\gamma}^{(j)}$ is a consistent estimator of $I^{(j)}_{\beta|\gamma}$, which implies $ \tilde I_{\beta|\gamma}- I_{\beta|\gamma} \rightarrow o_P(1)$.
\[
{Kn}(\tilde{\beta}-\beta^*)^\T\tilde I_{\beta|\gamma}(\tilde{\beta}-\beta^*)  \rightarrow \chi^2_p.
\] 
$\hfill\square$
\subsection*{Proof of Proposition 1}
By Lemma S.2, we have 
\begin{align*}
\bar{\beta}-\beta^* = \frac{1}{K}\sum_{j=1}^{K}\bar\beta_j-\beta^* = \frac{1}{K}\sum_{j=1}^{K}\{(I^{(j)}_{\beta|\gamma})^{-1}\nabla_{\beta}L_j(\theta_j^*)-(I^{(j)}_{\beta|\gamma})^{-1}I^{(j)}_{\beta\gamma}(I^{(j)}_{\gamma\gamma})^{-1}\nabla_{\gamma}L_j(\theta_j^*)\} +\frac{1}{K}\sum_{j=1}^{K}\delta_{\beta,j}
\end{align*}
where $\delta_{\beta,j}$ is the subvector of $\delta_j$ defined in Lemma S.2, which satisfies $\E\|\delta_j\|_2 \lesssim 1/n$. Then we have 
\begin{align*}
(Kn)^{1/2}(\bar{\beta}-\beta^*) =& \frac{1}{(Kn)^{1/2}}\sum_{j=1}^{K}\sum_{i=1}^{n}\{(I^{(j)}_{\beta|\gamma})^{-1}\nabla_{\beta}\log f(y_{ij};\theta_j^*)-(I^{(j)}_{\beta|\gamma})^{-1}I^{(j)}_{\beta\gamma}(I^{(j)}_{\gamma\gamma})^{-1}\nabla_{\gamma}\log f(y_{ij};\theta_j^*)\}\\& +\frac{1}{K}\sum_{j=1}^{K}(Kn)^{1/2}\delta_{\beta,j}.
\end{align*}

Assuming $K/n \rightarrow 0$, we have 
$\E\|\frac{1}{K}\sum_{j=1}^{K}(Kn)^{1/2}\delta_{\beta,j}\|_2 = {K^{1/2}/n^{1/2}}\rightarrow 0$. 
Thus, $\frac{1}{K}\sum_{j=1}^{K}(Kn)^{1/2}\delta_{\beta,j} = o_p(1)$. 
Therefore, let $\phi_{ij} = \{(I^{(j)}_{\beta|\gamma})^{-1}\nabla_{\beta}\log f(y_{ij};\theta_j^*)-(I^{(j)}_{\beta|\gamma})^{-1}I^{(j)}_{\beta\gamma}(I^{(j)}_{\gamma\gamma})^{-1}\nabla_{\gamma}\log f(y_{ij};\theta_j^*)\}$, we have $Kn(\bar{\beta}-\beta^*)^\T V^{-1}(\bar{\beta}-\beta^*) \rightarrow \chi^2_p$, where 
$$V = \frac{1}{K}\sum_{j=1}^{K}\sum_{i=1}^{n}\E\phi_{ij}\phi_{ij}^\T = \frac{1}{K}\sum_{j=1}^{K}(I^{(j)}_{\beta|\gamma})^{-1}.$$

$\hfill\square$
\section*{Appendix F: proofs of lemmas}
\subsection*{Proof of Lemma S.1.} From Proposition 5.16 in \cite{vershynin2010introduction}, we have 
\[
P(\|\frac{1}{n}\sum_{i=1}^{n}X_i\|_2^2>t^2) \le 2\exp(-C\min\{\frac{nt^2}{C_1^2}, \frac{nt}{C_1}\}).
\]
Let $t^2 = s$, we have 
\[
P(\|\frac{1}{n}\sum_{i=1}^{n}X_i\|_2^2>s) \le 2\exp(-C\min\{\frac{ns}{C_1^2}, \frac{n{s}^{1/2}}{C_1}\}).
\]
We have
\begin{align*}
\E\|\frac{1}{n}\sum_{i=1}^{n}X_i\|_2^2  &= \int_0^{\infty}P(\|\frac{1}{n}\sum_{i=1}^{n}X_i\|_2^2>s)ds\\
&\le\int_0^{C_1^2}2\exp(-C\frac{n{s}}{C_1^2})ds+\int_{C_1^2}^\infty2\exp(-C\frac{n{s}^{1/2}}{C_1})ds\\
&= \frac{2C_1^2}{Cn}(1-e^{-Cn})+4\frac{C_1^2-C_1}{Cn}C_1e^{-Cn}\\
&\lesssim \frac{1}{n}.
\end{align*}
Following similar procedure, we can show that the conclusion holds for $k = 1, 2, 4$ and $8$.
$\hfill\square$

\subsection*{Proof of Lemma S.2.}
For a given $j$, define the following events:

\[
\mathcal{E}_0 :=\{ \frac{1}{n}\sum_{i=1}^{n}m_1(Y_{ij}) \le  2M\},
\]

\[
\mathcal{E}_1 := \{\|\nabla^2 L_j(\theta_j^*) + I^{(j)}(\theta_j^*)\|_2 \le C_3\},
\]

and 

\[
\mathcal{E}_2 := \{\|\nabla L_j(\theta_j^*)\|_2 \le C_4\}.
\]
for some constants $M$, $C_1$ and $C_2$ which satisfy $\E\{ m_k (Y_{ij})\}< M$ for all $j \in \{1, \dots, K\}$, and $k = 1, 2$, $C_1\le \rho\mu_-/2$ and $C_2<(1-\rho)\rho\mu_-^2/8M$.  By replacing $F_1(\theta)$, $F_0(\theta)$ by $L_j(\theta_j)$ and $F_j(\theta_j)$ to Lemma 6 in \cite{zhang2012communication}, we obtain that under event $\mathcal{E} = \cap_{i=0,1,2}\mathcal{E}_i$, we have
\[
\|\bar{\theta}_j -\theta_j^*\|_2 \le C_1\|\nabla L_j(\theta_j^*)\|_2.
\] 
Next we calculate $\text{pr}(\mathcal{E}^{c})$. We have 
\[
P(\mathcal{E}^c) = P(\mathcal{E}_0^c\cup\mathcal{E}_1^c\cup\mathcal{E}_2^c) \le \sum_{i=1}^{3} P(\mathcal{E}_i^c).
\]
For $\mathcal{E}_0$, denote $m = \E\{\sum_{i=1}^{n}m_1(Y_{ij})/n\}$, we have
\begin{align*}
&P\left\{\frac{1}{n}\sum_{i=1}^{n}m_1(Y_{ij})>2M\right\} =P\left\{\frac{1}{n}\sum_{i=1}^{n}m_1(Y_{ij})-m>2M-m\right\}\\
&\le P\left\{|\frac{1}{n}\sum_{i=1}^{n}m_1(Y_{ij})-m|>2M-m\right\}.
\end{align*}
Since  $2M-m > 0$, and by Proposition 5.16 in \cite{vershynin2010introduction}, we have
\[
P\left\{\frac{1}{Kn}\sum_{j=1}^{K}\sum_{i=1}^{n}m_1(Y_{ij})>2M\right\} \lesssim \exp(-n).
\]
Therefore $P(\mathcal{E}_0^c)\lesssim \exp(-n)$. 
For $\mathcal{E}_1$, since $\E\{\nabla L_j(\theta_j)\} =\nabla F_j(\theta_j)$, by Proposition 5.16 in \cite{vershynin2010introduction}
\begin{align*}
&\text{pr}\{\|\nabla^2 L_j(\theta_j^*) - \nabla^2F_j(\theta_j^*)\|_2 >C_3\} \lesssim \exp(-n).
\end{align*}
Similarly we have  
\begin{align*}
\text{pr}\{\|\nabla L_j(\theta_j^*)\|_2 >C_4\} \lesssim \exp(-n).
\end{align*}
Thus, we have
\[
P(\mathcal{E}^c) = P(\mathcal{E}_0^c\cup\mathcal{E}_1^c\cup\mathcal{E}_2^c) \le \sum_{i=1}^{3} P(\mathcal{E}_i^c)\lesssim \exp(-n).
\]
In summary, we have 
\[
\|\bar{\theta}_j -\theta_j^*\|_2 \le  C_1\|\nabla L_j(\theta_j^*)\|_2
\]
with probability at least $1- \exp(-C_2n)$, which proves the first condition.

Since $\bar{\theta}_j$ is the maximizer of $L_j({\theta}_j)$, we have
\begin{align*}
0 = \nabla L_j(\bar{\theta}_j)  = 
\nabla L_j({\theta}_j^*)+\nabla^2 L_j({\theta}_j')(\bar{\theta}_j-\theta^*_j)
\end{align*}
where ${\theta}_j'$ satisfies $\|{\theta}_j'-{\theta}_j^*\|_2\le\|\bar{\theta}_j-{\theta}_j^*\|_2$.  And we have 
\begin{align*}
\bar{\theta}_j-\theta_j ^*=  
{I^{(j)}}^{-1}\nabla L_j({\theta}_j^*)+{I^{(j)}}^{-1}\{\nabla^2 L_j({\theta}_j')+{I^{(j)}}\}(\bar{\theta}_j-\theta^*_j)
\end{align*}
Let $\delta_j = {I^{(j)}}^{-1}\{\nabla^2 L_j({\theta}_j')+{I^{(j)}}\}(\bar{\theta}_j-\theta^*_j)$, we have 
\begin{align*}
\|{I^{(j)}}^{-1}\{\nabla^2 L_j({\theta}_j')+{I^{(j)}}\}(\bar{\theta}_j-\theta^*_j)\|_2\le\frac{1}{\mu_-}\|\nabla^2 L_j({\theta}_j')+{I^{(j)}}\|_2\|\bar{\theta}_j-\theta^*_j\|_2. 
\end{align*}
By Assumption 5 and event $\mathcal{E}$, we have
\begin{align*}
\|\nabla^2 L_j({\theta}_j')+{I^{(j)}}\|_2\le&\|\nabla^2 L_j({\theta}_j')-\nabla^2 L_j({\theta}_j^*)\|+\|\nabla^2 L_j({\theta}_j^*)+{I^{(j)}}\|_2\\
&\le 2M\|\bar{\theta}_j - \theta_j^*\|_2+\|\nabla^2 L_j({\theta}_j^*)+{I^{(j)}}\|_2.
\end{align*}
Thus, under event $\mathcal{E}$, we have
\begin{align*}
\|\delta_j\|_2 &=\|{I^{(j)}}^{-1}\{\nabla^2 L_j({\theta}_j')+{I^{(j)}}\}(\bar{\theta}_j-\theta^*_j)\|_2\\
&\le \frac{M}{\mu_-}\|\bar{\theta}_j - \theta_j^*\|^2_2+\frac{1}{\mu_-}\|\nabla^2 L_j({\theta}_j^*)+{I^{(j)}}\|_2\|\bar{\theta}_j - \theta_j^*\|_2\\
&\le \frac{MC_1^2}{\mu_-}\|\nabla L_j(\theta_j^*)\|_2^2 +\frac{C_1}{\mu_-}\|\nabla L_j(\theta_j^*)\|_2\|\nabla^2 L_j({\theta}_j^*)+{I^{(j)}}\|_2.
\end{align*}
Therefore, we have 
\begin{align*}
\E\|\delta_j\|^k_2 &\le C_5\E\|\nabla L_j(\theta_j^*)\|_2^{2k} +C_6\{\E\|\nabla L_j(\theta_j^*)\|_2^{2k}\E\|\nabla^2 L_j({\theta}_j^*)+{I^{(j)}}\|_2^{2k}\}^{1/2}.\\
\end{align*}
By Lemma S.1, we have 
\begin{align*}
\E\|\delta_j\|^k_2\lesssim \frac{1}{n^{k}}
\end{align*}
for $k= 1, \dots, 16.$
$\hfill\square$

\subsection*{Proof of Lemma S.3.} 
We start by defining the following events:
\[
\mathcal{E}_0 :=\{ \frac{1}{Kn}\sum_{j=1}^{K}\sum_{i=1}^{n}m_1(Y_{ij}) \le  2M\},
\]

\[
\mathcal{E}_1 := \left\{\|K\nabla^2 L_N(\beta^*, \Gamma^*) - K\E \{\nabla^2 L_N(\beta^*, \Gamma^*)\} \|_2 \le C_1\right\},
\]
and 
\[
\mathcal{E}_2 := \{\|K\nabla  L_N(\beta^*, \Gamma^*)\|_2 \le C_2\},
\]
for some constants $M$, $C_1$ and $C_2$ which satisfy $\E\{ m_k (Y_{ij})\}< M$ for all $j \in \{1, \dots, K\}$, and $k = 1, 2$, $C_1\le \rho\mu_-/2$ and $C_2<(1-\rho)\rho\mu_-^2/8M$. By replacing $F_1(\theta)$, $F_0(\theta)$ by $KL_N(\beta, \Gamma)$ and $K\E\{L_N(\beta^*, \Gamma^*)\}$, we apply Lemma 6 in \cite{zhang2012communication}, and obtain that under event $\mathcal{E} = \cap_{i=0,1,2}\mathcal{E}_i$, we have
\[
\|\hat{\Theta} -{\Theta}^*\|_2 \le C\|K\nabla L_N(\beta^*, \Gamma^*)\|_2,
\]
which implies 
\[
\|\hat{\Theta} -{\Theta}^*\|_2^2 \le C^2\|K\nabla L_N(\beta^*, \Gamma^*)\|_2^2.
\]
Then we have 
\[
\E\{\|\hat{\Theta} -{\Theta}^*\|_2^2I(\mathcal{E})\} \le C^2\E\{\|K\nabla L_N(\beta^*, \Gamma^*)\|^{2}_2I(\mathcal{E})\} \le C^2\E\| K\nabla L_N(\beta^*, \Gamma^*)\|^{2}_2.
\]
Since $\E\nabla L_N(\beta^*, \Gamma^*) = 0$, for the subvector corresponding to $\beta$ we have, 
\[
\E\|\nabla_\beta KL_N(\beta^*, \Gamma^*)\|^{2}_2\lesssim \frac{K}{n}
\]
And for each $\gamma_j$, we have
\[
\E\|\nabla_{\gamma_j} KL_N(\beta^*, \Gamma^*)\|^{2}_2 = \E\|\frac{1}{n}\sum_{i=1}^{n}\nabla_{\gamma_j} \log f(y_{ij}; \beta^*, \gamma_j^*)\|^2_2\lesssim \frac{1}{n}.
\]
Therefore we have 
\[
\E\| K\nabla L_N(\beta^*, \Gamma^*)\|^2_2 \lesssim \frac{K}{n},
\]
which leads to 
\[
\E\{\|\bar{\theta}_j -\theta_j^*\|^2_2I(\mathcal{E})\} \lesssim \frac{K}{n}.
\]

Next we calculate $\text{pr}(\mathcal{E}^{c})$. We have 
\[
P(\mathcal{E}^c) = P(\mathcal{E}_0^c\cup\mathcal{E}_1^c\cup\mathcal{E}_2^c) \le \sum_{i=1}^{3} P(\mathcal{E}_i^c).
\]
For $\mathcal{E}_0$, denote $m = \E\{\sum_{j=1}^{K}\sum_{i=1}^{n}m_1(Y_{ij})/(Kn)\}$, we have
\begin{align*}
&P\left\{\frac{1}{Kn}\sum_{j=1}^{K}\sum_{i=1}^{n}m_1(Y_{ij})>2M\right\} =P\left\{\frac{1}{Kn}\sum_{j=1}^{K}\sum_{i=1}^{n}m_1(Y_{ij})-m>2M-m\right\}\\
&\le P\left\{|\frac{1}{Kn}\sum_{i=1}^{n}m_1(Y_{ij})-m|>2M-m\right\}.
\end{align*}
Since  $2M-m > 0$, and by Proposition 5.16 in \cite{vershynin2010introduction}, we have
\[
P\left\{\frac{1}{Kn}\sum_{j=1}^{K}\sum_{i=1}^{n}m_1(Y_{ij})>2M\right\} \lesssim \exp(-n).
\] Therefore $P(\mathcal{E}_0^c)\le \exp(-C_1n)$. 
For $\mathcal{E}_1$, since the number of non-zero entry of matrix $\nabla^2 L_N(\beta^*, \Gamma^*)$ is $2K-1$, by Proposition 5.16 in \cite{vershynin2010introduction} we have
\begin{align*}
&\text{pr}\{\|K\nabla^2 L_N(\beta^*, \Gamma^*) - K\E \nabla^2L_N(\beta^*, \Gamma^*)\|_2 >C_1\} \\=&  \text{pr}\{\|\nabla^2 L_N(\beta^*, \Gamma^*) - \E \nabla^2L_N(\beta^*, \Gamma^*)\|_2 >C_1/K\} \\
\le&\text{pr}\{\|\nabla^2 L_N(\beta^*, \Gamma^*) - \E \nabla^2L_N(\beta^*, \Gamma^*)\|_\infty >C_1/2K^2\} \lesssim \exp(-n/K). 
\end{align*}
Since $\exp(-x)<1/x$ for all $x>0$, we have $\exp(-n/K) \le K/n$. 
Similarly 
\begin{align*}
\text{pr}\{\|K\nabla L_N(\beta^*, \Gamma^*)\|_2 >C_2\} \lesssim K/n.
\end{align*}
Thus, $\text{pr}(\mathcal{E}^C) \lesssim {K}/{n}$, and we have
\[
\E\|\hat{\Theta} -{\Theta}^*\|_2^2 \le \E\{\|\hat{\Theta} -{\Theta}^*\|_2^2I(\mathcal{E})\}+\text{pr}(\mathcal{E}^c) \lesssim  {K}/{n}.
\]
$\hfill\square$


\subsection*{Proof of Lemma S.4.} 
The proof of Lemma S.4 is consist of two steps. In Step 1, we show that the global maximum likelihood estimator $\hat\beta$ has a risk bound of $ E\{\|\hat\beta-\beta^*\|_2^2\}\lesssim 1/n$ using the previous results obtained in Lemma S.3; In the second step, we show the 
$E\{\|\hat\gamma_j-\gamma_j^*\|_2^2\}\lesssim 1/n$. Both steps are based on constructing the proper likelihood functions and using Lemma 6 in \cite{zhang2012communication}. 

\textit{Step 1:} Define the following events
\[
\mathcal{E}_0 :=\{ \frac{1}{Kn}\sum_{j=1}^{K}\sum_{i=1}^{n}m_1(Y_{ij}) \le  2M\},
\]

\[
\mathcal{E}_1 := \{\|\nabla_{\beta\beta} L_N(\beta^*, \hat \Theta) - \E\{\nabla_{\beta\beta} L_N(\beta^*, \Theta^*)\} \|_2\le C_1\}
\]
and
\[
\mathcal{E}_2 := \{\|\nabla_\beta L_N(\beta^*, \hat \Theta)\|_2 \le C_2\},
\]
for some constants $M$, $C_1$ and $C_2$ which satisfy $\E\{ m_k (Y_{ij})\}< M$ for all $j \in \{1, \dots, K\}$, and $k = 1, 2$, $C_1\le \rho\mu_-/2$ and $C_2<(1-\rho)\rho\mu_-^2/8M$. By replacing $F_1(\theta)$, $F_0(\theta)$ by $L_N(\beta, \hat\Gamma)$ and $\E\{L_N(\beta, \Gamma)\}$, we apply Lemma 6 in \cite{zhang2012communication}, and obtain that under event $\mathcal{E} = \{\cap_{i=0,1,2}\mathcal{E}_i\}$, we have
\[
\|\hat{\beta} -{\beta}^*\|_2 \le C\|\nabla L_N(\beta^*, \hat\Gamma)\|_2,
\]
which implies 
\[
\|\hat{\beta} -{\beta}^*\|^2_2 \le C\|\nabla L_N(\beta^*, \hat\Gamma)\|^2_2. 
\]
Then we have 
\[
\E\{\|\hat{\beta} -{\beta}^*\|^2_2I(\mathcal{E})\} \le \E\{\|\nabla L_N(\beta^*, \hat\Gamma)\|^2_2\}.
\]
Now we control the term $\E\{\|\nabla L_N(\beta^*, \hat\Gamma)\|^2_2\}$. We have
\[
\nabla_\beta L_N(\beta^*, \hat\Gamma) =  \nabla_\beta L_N(\beta^*, \Gamma^*) +\frac{1}{K}\sum_{j=1}^{K}\nabla_{\beta\gamma} L_j(\beta^*,\gamma_j')(\hat\gamma_j-\gamma_j^*),
\]
where $\gamma_j'$ satisfies $\|\gamma_j'-\gamma_j^*\|_2\le\|\hat\gamma_j-\gamma_j^*\|_2$.
For the last term we have
\begin{align*}
\frac{1}{K}\sum_{j=1}^{K}\nabla_{\beta\gamma} L_j(\beta^*,\gamma_j')(\hat\gamma_j-\gamma_j^*) = & \frac{1}{K}\sum_{j=1}^{K} \{\nabla_{\beta\gamma}L_j(\beta^*,\gamma_j')-\nabla_{\beta\gamma}L_j(\beta^*,\gamma_j^*)\}(\hat\gamma_j-\gamma_j^*)\\& + \frac{1}{K}\sum_{j=1}^{K} \{\nabla_{\beta\gamma}L_j(\beta^*,\gamma_j^*)\}(\hat\gamma_j-\gamma_j^*).
\end{align*}
By Assumption 1, and the definition of $\mathcal{E}_{0}$ and Lemma S.9, we know that 
\begin{align*}
&\|\frac{1}{K}\sum_{j=1}^{K}\nabla_{\beta\gamma} L_j(\beta^*,\gamma_j')(\hat\gamma_j-\gamma_j^*)\|_2^2 \\
&\le 2\|\frac{1}{K}\sum_{j=1}^{K} \{\nabla_{\beta\gamma}L_j(\beta^*,\gamma_j')-\nabla_{\beta\gamma}L_j(\beta^*,\gamma_j^*)\}(\hat\gamma_j-\gamma_j^*)\|_2^2 +2\|\frac{1}{K}\sum_{j=1}^{K} \{\nabla_{\beta\gamma}L_j(\beta^*,\gamma_j^*)\}(\hat\gamma_j-\gamma_j^*)\|_2^2\\
&\le \frac{8M^2}{K}\sum_{j=1}^{K}\|\hat\gamma_j-\gamma_j^*\|_2^2 +\frac{8\mu_+^2}{K}\sum_{j=1}^{K}\|\hat\gamma_j-\gamma_j^*\|_2^2. 
\end{align*}
And from Lemma S.3, we have 
\[
\E\{\sum_{j=1}^{K}\|\hat\gamma_j-\gamma_j^*\|_2^2\} \le \E\{\|\hat \Theta-\Theta^*\|_2^2\}\lesssim \frac{K}{n}. 
\]
Combine all, we have 
\begin{align*}
\E\{\|\nabla L_N(\beta^*, \hat\Gamma)\|^2_2\} \le 2\E\{\|L_N(\beta^*, \Gamma^*)\|^2_2\}+2\E\{\|\frac{1}{K}\sum_{j=1}^{K}\nabla_{\beta\gamma} L_j(\beta^*,\gamma_j')(\hat\gamma_j-\gamma_j^*)\|^2_2\}
\le \frac{C}{n}.
\end{align*}

Next we calculate $\text{pr}(\mathcal{E}^{c})$. 
For $\mathcal{E}_{0}^c$, denote $m = \E\{\sum_{j=1}^{K}\sum_{i=1}^{n}m_1(Y_{ij})/(Kn)\}$, we have
\begin{align*}
&P\left\{\sum_{j=1}^{K}\sum_{i=1}^{n}m_1(Y_{ij})/(Kn)>2M\right\} =P\left\{\sum_{j=1}^{K}\sum_{i=1}^{n}m_1(Y_{ij})/(Kn)-m>2M-m\right\}\\
&\le P\left\{|\sum_{j=1}^{K}\sum_{i=1}^{n}m_1(Y_{ij})/(Kn)-m|>2M-m\right\}.
\end{align*}
Since  $2M-m > 0$, and by Proposition 5.16 in \cite{vershynin2010introduction}, we have
\[
P\left\{\frac{1}{n}\sum_{i=1}^{n}m_1(Y_{ij})>2M\right\} \lesssim \exp(-Kn).
\]
Therefore $P(\mathcal{E}_{0}^c)\lesssim \exp(-Kn)$. \\For $\mathcal{E}_{1}^c$, we have 
\begin{align*}
&\text{pr}\{\|\nabla_{\beta\beta} L_N(\beta^*, \hat \Theta) - \E\{\nabla_{\beta\beta} L_N(\beta^*, \Theta^*)\} \|_2\le C_1\}
\\&\le \text{pr}\{\|\nabla_{\beta\beta} L_N(\beta^*, \hat \Theta) - \nabla_{\beta\beta} L_N(\beta^*, \Theta^*)\|_2 >C_1/2\}  \\&+\text{pr}\{\|\nabla_{\beta\beta} L_N(\beta^*, \Theta^*)-\E\nabla_{\beta\beta} L_N(\beta^*, \Theta^*)\|_2 >C_1/2\}
\end{align*}
Under $\mathcal{E}_{0}$ we have 
\begin{align*}
&\text{pr}\{\|\nabla_{\beta\beta} L_N(\beta^*, \hat \Theta) - \nabla_{\beta\beta} L_N(\beta^*, \Theta^*)\|_2 >C_1/2\} \\&\le \text{pr}\{ \frac{1}{K}\sum_{j=1}^{K}\|\nabla_{\beta\beta} L_j(\beta^*, \hat \gamma_j) - \nabla_{\beta\beta} L_j(\beta^*, \gamma_j^*)\|_2 >C_1/2\} \\ &\le  
\text{pr}\{ \frac{M}{K}\sum_{j=1}^{K}\|\hat{\gamma}_j-\gamma_j^*\|_2 >C_1/2\} = \text{pr}\{ \frac{1}{K}\sum_{j=1}^{K}\|\hat{\gamma}_j-\gamma_j^*\|^2_2 >(C_1/(2M))^2\}\\&\le \frac{\E\{\frac{1}{K}\sum_{j=1}^{K}\|\hat{\gamma}_j-\gamma_j^*\|^2_2\}}{(C_1/(2M))^2}\lesssim \frac{1}{n}.
\end{align*}
and 
\[
\text{pr}\{\|\nabla_{\beta\beta} L_N(\beta^*, \Theta^*)-\E\nabla_{\beta\beta} L_N(\beta^*, \Theta^*)\|_2 >C/2\} \le \exp(-CKn).
\]
For $\mathcal{E}_{2}^c$ we have 
\begin{align*}
&\text{pr}\{\|\nabla_\beta L_N(\beta^*, \hat \Theta)\|_2 > C_2\} \\&\le \text{pr}\{\|\nabla_\beta L_N(\beta^*, \Theta^*)\|_2 > C_2/3\} +  \text{pr}\{\|\frac{1}{K}\sum_{j=1}^{K}\nabla_{\beta\gamma} L_j(\beta^*, \gamma_j^*)(\hat{\gamma}_j-\gamma_j^*)\|_2 > C_2/3\}\\
& +\text{pr}\{\|\frac{1}{K}\sum_{j=1}^{K}\{\nabla_{\beta\gamma} L_j(\beta, \gamma_j')-\nabla_{\beta\gamma} L_j(\beta^*, \gamma_j^*)\}(\hat{\gamma}_j-{\gamma}_j^*)\|_2 > C_2/3\}.
\end{align*}
where $\gamma_j'$ satisfies  $\|\hat\gamma_j-\gamma_j^*\|_2\le\|\gamma_j'-\gamma_j^*\|_2$.
We have 
\[
\text{pr}\{\|\nabla_\beta L_N(\beta^*, \Theta^*)\|_2 > C_2/3\}\lesssim \exp(-Kn).
\]
Under $\mathcal{E}_{0}$ and Lemma S.9, we have 
\begin{align*}
&\text{pr}\{\|\frac{1}{K}\sum_{j=1}^{K}\nabla_{\beta\gamma} L_j(\beta^*, \gamma_j^*)(\hat{\gamma}_j-\gamma_j^*)\|_2 > C_2/3\}\le   \text{pr}\{2\mu_+\frac{1}{K}\sum_{j=1}^{K}\|\hat{\gamma}_j-\gamma_j^*\|_2 > C_2/3\}\\& = \text{pr}\{\frac{1}{K}\sum_{j=1}^{K}\|\hat{\gamma}_j-\gamma_j^*\|^2_2 > (C_2/(6\mu_+))^2\}\le \frac{\E\{\frac{1}{K}\sum_{j=1}^{K}\|\hat{\gamma}_j-\gamma_j^*\|^2_2\}}{\{C_2/(6\mu_+)\}^2} \lesssim \frac{1}{n},
\end{align*}
and 
\begin{align*}
&\text{pr}\{\|\frac{1}{K}\sum_{j=1}^{K}\{\nabla_{\beta\gamma} L_j(\beta, \gamma_j')-\nabla_{\beta\gamma} L_j(\beta^*, \gamma_j^*)\}(\hat{\gamma}_j-{\gamma}_j^*)\|_2 > C_2/3\}\le   \text{pr}\{\frac{2M}{K}\sum_{j=1}^{K}\|\hat{\gamma}_j-{\gamma}_j^*\|_2 > C_2/3\}\\
&\le\text{pr}[\frac{1}{K}\sum_{j=1}^{K}\|\hat{\gamma}_j-{\gamma}_j^*\|_2^2 > \{C_2/(6M)\}^2]\le \frac{\E\{\frac{1}{K}\sum_{j=1}^{K}\|\hat{\gamma}_j-{\gamma}_j^*\|^2_2\}}{\{C_2/(6M)\}^2} \lesssim \frac{1}{n}.
\end{align*}
Combine all, we have 
\[
\text{pr} (\mathcal{E}^c)\le \text{pr} (\mathcal{E}_{0}^c) +\text{pr} (\mathcal{E}_{0}\cap\mathcal{E}_{1}) +\text{pr} (\mathcal{E}_{0}\cap\mathcal{E}_{2}^c)\le \frac{C}{n}. 
\]
Therefore we have 
\begin{align*}
&\E\{\|\hat\beta-\beta^*\|_2^2\}\le\E\{\|\hat\beta-\beta^*\|_2^2I(\mathcal{E})\} +P(\mathcal{E}^c)
\le\frac{C}{n}.
\end{align*}

\textit{Step 2:} In this step, we prove the risk bound for $\hat{\gamma}_j$. For each site $j$,  we define three more events
\[
\mathcal{E}'_{0j} :=\{ \frac{1}{n}\sum_{i=1}^{n}m_1(Y_{ij}) \le  2M\},
\]
\[
\mathcal{E}'_{1j} := \{\|\nabla_{\gamma\gamma} L_j(\hat\beta,  \gamma_j^*) - \nabla_{\gamma\gamma}F_j(\beta^*, \gamma_j^*)\} \|_2\le C_1\},
\]
and
\[
\mathcal{E}'_{2j} := \{\|\nabla_\gamma L_j(\hat\beta, \gamma_j^*)\|_2 \le C_2\}.
\]

By replacing $F_1(\theta)$, $F_0(\theta)$ by $L_j(\hat\beta, \gamma_j)$ and $F_j(\beta^*, \gamma_j)$, we apply Lemma 6 in \cite{zhang2012communication}, and obtain that under event $\mathcal{E}'_j = \cap_{i=0,1,2}\mathcal{E}'_{ji}$, we have
\[
\|\hat{\gamma}_j -{{\gamma}_j}^*\|^2 \le C\|\nabla_\gamma L_j(\hat\beta, \gamma_j^*)\|^2. 
\]
which implies 
\[
\|\hat{\gamma}_j -{{\gamma}_j}^*\|^2_2 \le C^2\|\nabla_\gamma L_j(\hat\beta, \gamma_j^*)\|^2_2. 
\]
Then we have 
\[
\E\{\|\hat{\gamma}_j -{{\gamma}_j}^*\|^2_2I(\mathcal{E})\} \le C^2\E\{\|\nabla_\gamma L_j(\hat\beta, \gamma_j^*)\|^2_2\}.
\]
Now we control the term $\E\{\|\nabla_\gamma L_j(\hat\beta, \gamma_j^*)\|^2_2\}$. We have

\[
\nabla_\gamma L_j(\hat\beta, \gamma_j^*) =  \nabla_{\gamma} L_j(\beta^*, \gamma_j^*) +\nabla_{\beta\gamma} L_j(\beta',\gamma_j^*)(\hat\beta-{\beta}^*),
\]
where $\beta'$ satisfies $\|\beta'-\beta^*\|_2\le\|\hat\beta-\beta^*\|_2$.
For the last term we have
\begin{align*}
\nabla_{\beta\gamma} L_j(\beta',\gamma_j^*)(\hat\beta-{\beta}^*)=& \nabla_{\beta\gamma}L_j(\beta',\gamma_j^*)(\hat\beta-{\beta}^*)-\nabla_{\beta\gamma}L_j(\beta^*,\gamma_j^*)(\hat\beta-{\beta}^*)+\nabla_{\beta\gamma}L_j(\beta^*,\gamma_j^*)(\hat\beta-{\beta}^*)
\end{align*}
By Assumption 1, Lemma S.9, and the definition of $\mathcal{E}_{0j}$, we know that 
\begin{align*}
&\|\nabla_{\beta\gamma}L_j(\beta',\gamma_j^*)(\hat\beta-{\beta}^*)\|_2^2 \\
&\le 2\|\nabla_{\beta\gamma}L_j(\beta',\gamma_j^*)(\hat\beta-{\beta}^*)-\nabla_{\beta\gamma}L_j(\beta^*,\gamma_j^*)(\hat\beta-{\beta}^*)\|_2^2 +2\|\nabla_{\beta\gamma}L_j(\beta^*,\gamma_j^*)(\hat\beta-{\beta}^*)\|_2^2\\
&\le 8(M^2+\mu_+^2)\|\hat\beta-\beta^*\|_2^2.
\end{align*}
Therefore we have
\[
\E\|\nabla_\gamma L_j(\hat\beta, \gamma_j^*)\|_2^2\le 2\E\|\nabla_{\gamma} L_j(\beta^*, \gamma_j^*)\|_2^2 +16(M^2+\mu_+^2)\E\|\hat\beta-\beta^*\|_2^2 \lesssim \frac{1}{n}.
\]
Next we calculate $\text{pr}(\mathcal{E}_j'^{c})$. 
For $\mathcal{E}_{0j}'^c$, denote $m = \E\{\sum_{i=1}^{n}m_1(Y_{ij})/n\}$, we have
\begin{align*}
&P\left\{\frac{1}{n}\sum_{i=1}^{n}m_1(Y_{ij})>2M\right\} =P\left\{\frac{1}{n}\sum_{i=1}^{n}m_1(Y_{ij})-m>2M-m\right\}\\
&\le P\left\{|\frac{1}{n}\sum_{i=1}^{n}m_1(Y_{ij})-m|>2M-m\right\}.
\end{align*}
Since  $2M-m > 0$, and by Proposition 5.16 in \cite{vershynin2010introduction}, we have
\[
P\left\{\frac{1}{n}\sum_{i=1}^{n}m_1(Y_{ij})>2M\right\} \lesssim \exp(-n).
\]
Therefore $P(\mathcal{E}_{0j}'^c)\lesssim \exp(-n)$. 
For $\mathcal{E}_{1j}'^c$, we have 
\begin{align*}
&\text{pr}\{\|\nabla_{\gamma\gamma} L_j(\hat\beta,  \gamma_j^*) - \nabla_{\gamma\gamma}F_j(\beta^*, \gamma_j^*) \|_2 >C_1\}  \le \text{pr}\{\|\nabla_{\gamma\gamma} L_j(\hat\beta,  \gamma_j^*) - \nabla_{\gamma\gamma} L_j(\beta^*,  \gamma_j^*) \|_2 >C_1/2\}  \\&+\text{pr}\{\|\nabla_{\gamma\gamma} L_j(\beta^*,  \gamma_j^*) -\nabla_{\gamma\gamma}F_j(\beta^*, \gamma_j^*)\|_2 >C_1/2\}.
\end{align*}
Under $\mathcal{E}_{0j}$ we have 
\begin{align*}
&\text{pr}\{\|\nabla_{\gamma\gamma} L_j(\hat\beta,  \gamma_j^*) - \nabla_{\gamma\gamma} L_j(\beta^*,  \gamma_j^*) \|_2 >C_1/2\} =  \text{pr}\{2M\|\hat\beta-\beta\|_2 >C_1/2\} \\ = & 
\text{pr}\{\|\hat\beta-\beta\|^2_2 >(C_1/4M)^2\} \le \frac{\E\|\hat\beta-\beta\|^2_2}{\{C_1/(4M)\}^2} \lesssim \frac{1}{n}.
\end{align*}
and 
\[
\text{pr}\{\|\nabla_{\gamma\gamma} L_j(\beta^*,  \gamma_j^*) -\nabla_{\gamma\gamma}F_j(\beta^*, \gamma_j^*)\|_2 >C_1/2\} \lesssim \exp(-n).
\]
For $\mathcal{E}_{2j}'^c$ we have 
\begin{align*}
&\text{pr}\{\|\nabla_\gamma L_j(\hat\beta, \gamma_j^*)\|_2 > C_2\} \le \text{pr}\{\|\nabla_\gamma L_j(\beta^*, \gamma_j^*)\|_2 > C_2/3\} +  \text{pr}\{\|\nabla_{\gamma\beta} L_j(\beta^*, \gamma_j^*)(\hat{\beta}-\beta)\|_2 > C_2/3\}\\
& +\text{pr}\{\|\{\nabla_{\gamma\beta} L_j(\beta', \gamma_j^*)-\nabla_{\gamma\beta} L_j(\beta^*, \gamma_j^*)\}(\hat{\beta}-\beta)\|_2 > C_2/3\}
\end{align*}
Under $\mathcal{E}_{0j}'$, we have 
\begin{align*}
&\text{pr}\{\|\nabla_{\gamma\beta} L_j(\beta^*, \gamma_j^*)(\hat{\beta}-\beta)\|_2 > C_2/3\}\le   \text{pr}\{2\mu_+\|\hat{\beta}-\beta\|_2 > C_2/3\}\\
&\le\text{pr}[\|\hat{\beta}-\beta\|^2_2 > \{C_2/(6\mu_+)\}^2]\le \frac{\E\|\hat\beta-\beta\|^2_2}{\{C_2/(6\mu_+)\}^2} \lesssim \frac{1}{n},
\end{align*}
and 
\begin{align*}
&\text{pr}\{\|\{\nabla_{\gamma\beta} L_j(\beta', \gamma_j^*)-\nabla_{\gamma\beta} L_j(\beta^*, \gamma_j^*)\}(\hat{\beta}-\beta)\|_2 > C_2/3\}\le   \text{pr}\{2M\|\hat{\beta}-\beta\|_2 > C_2/3\}\\
&\le\text{pr}[\|\hat{\beta}-\beta\|^2_2 > \{C_2/(6M)\}^2]\le \frac{\E\|\hat\beta-\beta\|^2_2}{\{C_2/(6M)\}^2} \lesssim \frac{1}{n}.
\end{align*}
And we have 
\[
\text{pr}\{\|\nabla_\gamma L_j(\beta^*, \gamma_j^*)\|_2 > C_2/3\}\lesssim \exp(-n).
\]

Combine all, we have 
\[
\text{pr} (\mathcal{E}_{j}'^c)\le \text{pr} (\mathcal{E}_{0j}'^c) +\text{pr} (\mathcal{E}_{0j}'\cap\mathcal{E}_{1j}') +\text{pr} (\mathcal{E}_{0j}'\cap\mathcal{E}_{2j}'^c)\le \frac{C}{n}. 
\]
Therefore we have 
\begin{align*}
&\E\{\|\hat\gamma_j-\gamma_j^*\|_2^2\}\le\E\{\|\hat\gamma_j-\gamma_j^*\|_2^2I(\mathcal{E}_j')\} +P(\mathcal{E}_j'^c)
\le\frac{C}{n}.
\end{align*}
$\hfill\square$\\

\subsection*{Proof of Lemma S.5} 

Since $\hat\Theta$ is the maximizer of $L_N(\Theta)$, we have 
\begin{align*}
0 = \nabla L_N(\hat\Theta) = \nabla L_N(\Theta^*)+\nabla^2 L_N(\Theta^*) (\hat\Theta-\Theta^*) +\{\nabla^2 L_N(\Theta')-\nabla^2 L_N(\Theta^*)\} (\hat\Theta-\Theta^*),
\end{align*}
where $\Theta' = (\beta',\gamma_1', \dots,\gamma_K')$ satisfies $\|\beta'-\beta^*\|_2 \le \|\hat\beta-\beta^*\|_2$.  Multiplying $K$ to the above equation we obtain  
\begin{align*}
&0 = K\nabla L_N(\Theta^*)+K\nabla^2 L_N(\Theta^*) (\hat\Theta-\Theta^*) +\{K\nabla^2 L_N(\Theta')-K\nabla^2 L_N(\Theta^*)\} (\hat\Theta-\Theta^*)\\
&= K\nabla L_N(\Theta^*)-I(\hat\Theta-\Theta^*) +\{K\nabla^2 L_N(\Theta^*)+I\} (\hat\Theta-\Theta^*) +\{K\nabla^2 L_N(\Theta')-K\nabla^2 L_N(\Theta^*)\}(\hat\Theta-\Theta^*)\\
&:=K\nabla L_N(\Theta^*)-I(\hat\Theta-\Theta^*) +d_1+d_2.
\end{align*}
We can then solve that 
\[
\hat{\Theta}-\Theta = I^{-1}\{K\nabla L_N(\Theta^*)\} +I^{-1} d_1 +I^{-1} d_2.
\]
Now we only need to show that each element in $\delta = I^{-1} d_1 +I^{-1} d_2$ satisfies $\E|\delta_t|_2 \le C/n$ for all $t$, where $\delta_t$ denotes the $t$-th entry.

For $d_1$, we have
\begin{align*}
K\nabla^2 L_N(\Theta^*)+I = \begin{pmatrix} 
A & B\\
B^\T & D
\\
\end{pmatrix},
\end{align*}
where $A = \sum_{j=1}^{K}\{\nabla_{\beta\beta}L_j(\theta_j^*)+I_{\beta\beta}^{(j)}\}$, $B = \left( \{\nabla_{\beta\gamma}L_1(\theta_1^*) +  I_{\beta\gamma}^{(1)}\},\dots,\{\nabla_{\beta\gamma}L_K(\theta_K^*) +  I_{\beta\gamma}^{(K)}\}\right)$, and 
\[
D = \begin{pmatrix} 
\{\nabla_{\gamma\gamma}L_1(\theta_1^*)+I_{\gamma\gamma}^{(1)}\} & 0&\dots &0\\
0&\{\nabla_{\gamma\gamma}L_2(\theta_2^*)+I_{\gamma\gamma}^{(2)}\}&\dots &0\\
\dots&&\dots&\dots
\\
0&\dots &0&\{\nabla_{\gamma\gamma}L_K(\theta_K^*)+I_{\gamma\gamma}^{(K)}\}
\end{pmatrix}.
\]
So we have
\begin{align*}
d_1 = \begin{pmatrix} 
\sum_{j=1}^{K}\Big\{\{\nabla_{\beta\beta}L_j(\theta_j^*)+I_{\beta\beta}^{(j)}\}(\hat{\beta}-\beta^*)+\{\nabla_{\beta\gamma}L_j(\theta_j^*)+I_{\beta\gamma}^{(j)}\}(\hat{\gamma}_j-\gamma_j^*)\Big\} \\
\{\nabla_{\gamma\beta}L_1(\theta_1^*)+I_{\gamma\beta}^{(1)}\}(\hat{\beta}-\beta^*)+\{\nabla_{\gamma\gamma}L_1(\theta_1^*)+I_{\gamma\gamma}^{(1)}\}(\hat{\gamma}_1-\gamma_1^*)\\
\dots\\
\{\nabla_{\gamma\beta}L_K(\theta_K^*)+I_{\gamma\beta}^{(K)}\}(\hat{\beta}-\beta^*)+\{\nabla_{\gamma\gamma}L_j(\theta_K^*)+I_{\gamma\gamma}^{(K)}\}(\hat{\gamma}_K-\gamma_K^*)
\end{pmatrix}.
\end{align*}
For the subvector corresponding to $\beta$ we have
\begin{align*}
&\E\|d_{1\beta}\|_2 \le \sum_{j=1}^{K}\left[\E\|\{\nabla_{\beta\beta}L_j(\theta_j^*)+I_{\beta\beta}^{(j)}\}(\hat{\beta}-\beta^*)\|_2+\E\|\{\nabla_{\beta\gamma}L_j(\theta_j^*)+I_{\beta\gamma}^{(j)}\}(\hat{\gamma}_j-\gamma_j^*)\|_2\right]\\
&\le 
\sum_{j=1}^{K}\left[\{\E\|\nabla_{\beta\beta}L_j(\theta_j^*)+I_{\beta\beta}^{(j)}\|^2_2\E\|\hat{\beta}-\beta^*\|^2_2\}^{1/2}+\{\E\|\nabla_{\beta\gamma}L_j(\theta_j^*)+I_{\beta\gamma}^{(j)}\|^2_2\E\|(\hat{\gamma}_j-\gamma_j^*)\|^2_2\}^{1/2}\right].
\end{align*}
From the proof of Lemma S.4, we have $\|\hat{\beta}-\beta^*\|^2_2\lesssim 1/n$ and $\|\hat{\gamma}_j-\gamma_j^*\|^2_2\lesssim 1/n$ for all $j$. By Lemma S.1 we have $\E\|\nabla^2 L_j(\theta_j^*)+I^{(j)}\|^2_2\lesssim 1/{n}$. Thus we have
\[
\E\|d_{1\beta}\|_2\lesssim \frac{K}{n}.
\]
And for subvector corresponding to $\gamma_j$, we have 
\begin{align*}
\E\|d_{1\gamma_j}\|_2 &\le \E\|\{\nabla_{\gamma\beta}L_1(\theta_1^*)+I_{\gamma\beta}^{(1)}\}(\hat{\beta}-\beta^*)\|_2+\E\|\{\nabla_{\gamma\gamma}L_1(\theta_1^*)+I_{\gamma\gamma}^{(1)}\}(\hat{\gamma}_1-\gamma_1^*)\|_2\\
&\le \{\E\|\nabla_{\gamma\beta}L_1(\theta_1^*)+I_{\gamma\beta}^{(1)}\|_2^2\E\|\hat{\beta}-\beta^*\|^2_2\}^{1/2}+\{\E\|\nabla_{\gamma\gamma}L_1(\theta_1^*)+I_{\gamma\gamma}^{(1)}\|_2^2\E\|\hat{\gamma}_1-\gamma_1^*\|_2^2\}^{1/2} \\
&\lesssim \frac{1}{n}
\end{align*}
Similarly, we have 

\begin{align*}
d_2 = \begin{pmatrix} 
\sum_{j=1}^{K}\Big\{\{\nabla_{\beta\beta}L_j(\theta_j')-\nabla_{\beta\beta}L_j(\theta_j^*)\}(\hat{\beta}-\beta^*)+\{\nabla_{\beta\gamma}L_j(\theta_j')-\nabla_{\beta\gamma}L_j(\theta_j^*)\}(\hat{\gamma}_j-\gamma_j^*)\Big\} \\
\{\nabla_{\gamma\beta}L_1(\theta_1')-\nabla_{\gamma\beta}L_1(\theta_1^*)\}(\hat{\beta}-\beta^*)+\{\nabla_{\gamma\gamma}L_1(\theta_1')-\nabla_{\gamma\gamma}L_1(\theta_1^*)\}(\hat{\gamma}_1-\gamma_1^*)\\
\dots\\
\{\nabla_{\gamma\beta}L_K(\theta_K')-\nabla_{\gamma\beta}L_K(\theta_K^*)\}(\hat{\beta}-\beta^*)+\{\nabla_{\gamma\gamma}L_j(\theta_K')-\nabla_{\gamma\gamma}L_j(\theta_K^*)\}(\hat{\gamma}_K-\gamma_K^*)
\end{pmatrix}.
\end{align*}
We have for the subvector corresponding to $\beta$
\begin{align*}
\E\|d_{2\beta}\|_2 &\le \sum_{j=1}^{K}\Big\{\E\{ \|\nabla_{\beta\beta}L_j(\theta_j')-\nabla_{\beta\beta}L_j(\theta_j^*)\}(\hat{\beta}-\beta^*)\|_2+\E\|\{\nabla_{\beta\gamma}L_j(\theta_j')-\nabla_{\beta\gamma}L_j(\theta_j^*)\}(\hat{\gamma}_j-\gamma_j^*)\|_2\Big\} \\
&\le \sum_{j=1}^{K}\Big\{\{\E \|\nabla_{\beta\beta}L_j(\theta_j')-\nabla_{\beta\beta}L_j(\theta_j^*)\|^2_2\E\|\hat{\beta}-\beta^*\|^2_2\}^{1/2}+\{\E\|\nabla_{\beta\gamma}L_j(\theta_j')-\nabla_{\beta\gamma}L_j(\theta_j^*)\|_2^2\E\|\hat{\gamma}_j-\gamma_j^*\|^2_2\}^{1/2}\Big\}\\
&\le M\sum_{j=1}^{K}\Big\{\E\|\hat{\beta}-\beta^*\|^2_2+\E\|\hat{\gamma}_j-\gamma_j^*\|^2_2\Big\}\lesssim \frac{K}{n}.
\end{align*}
And for the subvector corresponding to $\gamma_j$, we have 
\begin{align*}
\E\|d_{2\gamma_j}\|_2 &\le \E\|\{\nabla_{\gamma\beta}L_1(\theta_1')-\nabla_{\gamma\beta}L_1(\theta_1^*)\}(\hat{\beta}-\beta^*)\|_2+\E\|\{\nabla_{\gamma\gamma}L_1(\theta_1')-\nabla_{\gamma\gamma}L_1(\theta_1^*)\}(\hat{\gamma}_1-\gamma_1^*)\|_2\\
&\le
\{\E \|\nabla_{\gamma\beta}L_1(\theta_1')-\nabla_{\gamma\beta}L_1(\theta_1^*)\|^2_2\E\|\hat{\beta}-\beta^*\|^2_2\}^{1/2}+\{\E\|\nabla_{\gamma\gamma}L_1(\theta_1')-\nabla_{\gamma\gamma}L_1(\theta_1^*)\|_2^2\E\|\hat{\gamma}_j-\gamma_j^*\|^2_2\}^{1/2}\\
&\le M\{\E\|\hat{\beta}-\beta^*\|^2_2+\E\|\hat{\gamma}_j-\gamma_j^*\|^2_2\}\lesssim \frac{1}{n}.
\end{align*}

Then we write $I$ as 
\begin{align*}
I = \begin{pmatrix} 
I_{\beta\beta} & I_{\beta\Gamma}\\
I_{\Gamma\beta} & I_{\Gamma\Gamma}
\\
\end{pmatrix},
\end{align*}
where $I_{\beta\beta} = \sum_{j=1}^{K}I_{\beta\beta}^{(j)}$, $I_{\beta\Gamma} = I_{\Gamma\beta}^\T = \left( I_{\beta\gamma}^{(1)},\dots, I_{\beta\gamma}^{(K)}\right)$, and $I_{\Gamma\Gamma} = \text{diag}\{I^{(1)}_{\gamma\gamma}, \dots,I^{(K)}_{\gamma\gamma} \}$, which is a block diagonal matrix. By Inversion of block matrix, we have 
\begin{align}\label{inv0}
I^{-1} &= \begin{pmatrix} 
(I_{\beta\beta}- I_{\beta\Gamma}I_{\Gamma\Gamma}^{-1}I_{\Gamma\beta})^ {-1} & & -(I_{\beta\beta}- I_{\beta\Gamma}I_{\Gamma\Gamma}^{-1}I_{\Gamma\beta})^{-1}I_{\Gamma\beta}I_{\Gamma\Gamma}^{-1}\\
-I_{\Gamma\Gamma}^{-1}I_{\Gamma\beta}(I_{\beta\beta}- I_{\beta\Gamma}I_{\Gamma\Gamma}^{-1}I_{\Gamma\beta})^ {-1}& & I_{\Gamma\Gamma}^{-1}+I_{\Gamma\Gamma}^{-1}I_{\Gamma\beta}(I_{\beta\beta}- I_{\beta\Gamma}I_{\Gamma\Gamma}^{-1}I_{\Gamma\beta})^ {-1}I_{\beta\Gamma}I_{\Gamma\Gamma}^{-1}
\\
\end{pmatrix}.
\end{align}
Define the partial information matrix to be $ I_{\beta|\gamma}^{(j)} = I_{\beta\beta}^{(j)}-I^{(j)}_{\beta\gamma}{I^{(j)}_{\gamma\gamma}}^{-1}I^{(j)}_{\gamma\beta}$. We have
\beq\label{inv1}
(I_{\beta\beta}- I_{\beta\Gamma}I_{\Gamma\Gamma}^{-1}I_{\Gamma\beta})^ {-1} = \left(\sum_{j=1}^{K}I_{\beta|\gamma}^{(j)}\right)^{-1},
\eeq
\beq\label{inv2}
-(I_{\beta\beta}- I_{\beta\Gamma}I_{\Gamma\Gamma}^{-1}I_{\Gamma\beta})^{-1}I_{\Gamma\beta}I_{\Gamma\Gamma}^{-1} = \{\sum_{j=1}^{K}I_{\beta|\gamma}^{(j)}\}^{-1}\left(I_{\beta\gamma}^{(1)}{I_{\gamma\gamma}^{(1)}}^{-1}, \dots, I_{\beta\gamma}^{(K)}{I_{\gamma\gamma}^{(K)}}^{-1}\right),
\eeq
and 
\begin{align*}
&I_{\Gamma\Gamma}^{-1}+I_{\Gamma\Gamma}^{-1}I_{\Gamma\beta}(I_{\beta\beta}- I_{\beta\Gamma}I_{\Gamma\Gamma}^{-1}I_{\Gamma\beta})^ {-1}I_{\beta\Gamma}I_{\Gamma\Gamma}^{-1} = \begin{pmatrix} 
A_{11}&A_{12}&\dots& A_{1K}\\
\dots&\dots&\dots& \dots\\
A_{K1}&A_{K2}&\dots& A_{KK}\\
\end{pmatrix}, 
\end{align*}
where 
\beq\label{inv3}
A_{jj} = {I_{\gamma\gamma}^{(j)}}^{-1}+ {I_{\gamma\gamma}^{(j)}}^{-1}I_{\gamma\beta}^{(j)}\{\sum_{i=1}^{K}I_{\beta|\gamma}^{(i)}\}^{-1}I_{\beta\gamma}^{(j)}{I_{\gamma\gamma}^{(j)}}^{-1} ,
\eeq
and 
\beq\label{inv4}
A_{jk} = {I_{\gamma\gamma}^{(j)}}^{-1}I_{\gamma\beta}^{(j)}\{\sum_{i=1}^{K}I_{\beta|\gamma}^{(i)}\}^{-1}I_{\beta\gamma}^{(k)}{I_{\gamma\gamma}^{(k)}}^{-1},
\eeq
for $j, k \in \{1, \dots,K\}$ and $j\ne k$.

Now we control 
$
\delta = I^{-1}(d_1+d_2)$. By Assumption 2, we know that $\mu_-\text{I}_d\preceq I^{(j)}\preceq \mu_+\text{I}_d$. This implies $\mu_-\text{I}_p \preceq I^{(j)}_{\beta|\gamma}\preceq \mu_+\text{I}_p$.  For the sub-vector of $\delta$ that corresponding to $\beta$ we have

\begin{align*}
\E\|\delta_\beta\|_2 &=\E \| (\sum_{i=1}^{K}I_{\beta|\gamma}^{(i)})^{-1}(d_{1\beta}+d_{2\beta}) + (\sum_{i=1}^{K}I_{\beta|\gamma}^{(i)})^{-1}\sum_{j=1}^{K}I_{\beta\gamma}^{(j)}{I_{\gamma\gamma}^{(j)}}^{-1}(d_{1\gamma_j}+d_{2\gamma_j})\|_2\\
& = \E\| \frac{1}{K}(\frac{1}{K}\sum_{i=1}^{K}I_{\beta|\gamma}^{(i)})^{-1}(d_{1\beta}+d_{2\beta}) + \frac{1}{K}(\frac{1}{K}\sum_{i=1}^{K}I_{\beta|\gamma}^{(i)})^{-1}\sum_{j=1}^{K}I_{\beta\gamma}^{(j)}{I_{\gamma\gamma}^{(j)}}^{-1}(d_{1\gamma_j}+d_{2\gamma_j})\|_2\\
& \leq \|(\frac{1}{K}\sum_{i=1}^{K}I_{\beta|\gamma}^{(i)})^{-1}\|_2\frac{1}{K}(\E\|d_{1\beta}\|_2+\E\|d_{2\beta}\|_2)\\&+\frac{1}{K}\|(\frac{1}{K}\sum_{i=1}^{K}I_{\beta|\gamma}^{(i)})^{-1}\|_2\sum_{j=1}^{K}\|I_{\beta\gamma}^{(j)}{I_{\gamma\gamma}^{(j)}}^{-1}\|_2\{\E\|d_{1\gamma_j}\|_2+\E\|d_{2\gamma_j}\|_2\}\\
&\le \frac{1}{K\mu_-}\frac{CK}{n} +\frac{1}{K\mu_-}K\frac{\mu_+C}{\mu_-n}\lesssim \frac{1}{n}.
\end{align*}

And for the sub-vector corresponding to each $\gamma_j$, we have

\begin{align*}
\E\|\delta_{\gamma_j}\|_2 &= \E\|(\sum_{i=1}^{K}I_{\beta|\gamma}^{(i)})^{-1} I_{\beta\gamma}^{(j)}{I_{\gamma\gamma}^{(j)}}^{-1}(d_{1\beta}+d_{2\beta})+{I_{\gamma\gamma}^{(j)}}^{-1}(d_{1\gamma_j}+d_{2\gamma_j})\\&+\sum_{i=1}^{K} {I_{\gamma\gamma}^{(i)}}^{-1}I_{\gamma\beta}^{(i)}\{\sum_{i=1}^{K}I_{\beta|\gamma}^{(i)}\}^{-1}I_{\beta\gamma}^{(j)}{I_{\gamma\gamma}^{(j)}}^{-1}(d_{1\gamma_i}+d_{2\gamma_i})\|_2\\
& \le \frac{1}{K}\|(\frac{1}{K}\sum_{i=1}^{K}I_{\beta|\gamma}^{(i)})^{-1} I_{\beta\gamma}^{(j)}{I_{\gamma\gamma}^{(j)}}^{-1}\|_2(\E\|d_{1\beta}\|_2+\E \|d_{2\beta}\|_2) + \|{I_{\gamma\gamma}^{(j)}}^{-1}\|_2(\E\|d_{1\gamma_j}\|_2+\E\|d_{2\gamma_j}\|_2)\\
&+ \frac{1}{K}\sum_{i=1}^{K}\| {I_{\gamma\gamma}^{(i)}}^{-1}I_{\gamma\beta}^{(i)}\{\frac{1}{K}\sum_{i=1}^{K}I_{\beta|\gamma}^{(i)}\}^{-1}I_{\beta\gamma}^{(j)}{I_{\gamma\gamma}^{(j)}}^{-1}\|_2(\E\|d_{1\gamma_i}\|_2+\E\|d_{2\gamma_i}\|_2)\\
&\le \frac{\mu_+}{\mu_-^2K}\frac{CK}{n} +\frac{1}{\mu_-}\frac{C}{n}+\frac{\mu_+^2C}{\mu_-^3n}\lesssim \frac{1}{n}.
\end{align*}

Combine all we have the $t$-th entry of $\delta$ denoted by $\delta_t$
satisfies $\E|\delta_t|_2\lesssim \frac{1}{n}$ for all $t$.
$\hfill\square$

\subsection*{Proof of Lemma S.6.} 
Define the following events 
\[
\mathcal{E}_0 :=\{ \frac{1}{Kn}\sum_{j=1}^{K}\sum_{i=1}^{n}m_1(Y_{ij}) \le  2M\},
\]

\[
\mathcal{E}_1 := \{\|\frac{1}{K}\sum_{j=1}^{K}\{\nabla_{\beta\beta}L_j(\beta^*, \bar{\gamma}_j) -\bar H^{(j)}_{\beta\gamma} ({{{{\bar H}}_{\gamma\gamma}}^{(j)}})^{-1}\nabla_{\gamma\beta}L_j(\beta^*, \bar{\gamma}_j) + I^{(j)}_{\beta\beta}-I^{(j)}_{\beta\gamma}{I^{(j)}_{\gamma\gamma}}^{-1}I^{(j)}_{\gamma\beta}\} \|_2\le C_1\}
\]
\[
\mathcal{E}_2 := \{\|\frac{1}{K}\sum_{j=1}^{K}\{\nabla_{\beta}L_j(\beta^*, \bar{\gamma}_j) -\bar H^{(j)}_{\beta\gamma} ({{{{\bar H}}_{\gamma\gamma}}^{(j)}})^{-1}\nabla_{\gamma}L_j(\beta^*, \bar{\gamma}_j)\}\|_2 \le C_2\},
\]
for some constants $M$, $C_1$ and $C_2$ which satisfy $\E\{ m_k (Y_{ij})\}< M$ for all $j \in \{1, \dots, K\}$, and $k = 1, 2$, $C_1\le \rho\mu_-/2$ and $C_2<(1-\rho)\rho\mu_-^2/8M$. Applying Lemma 6 in \cite{zhang2012communication} we have under
event $\mathcal{E} = \{\cap_{i=0,1,2}\mathcal{E}_i\}$, 
\[
\|\check{\beta} -{\beta}^*\|_2 \le C\|\frac{1}{K}\sum_{j=1}^{K}\{\nabla_{\beta}L_j(\beta^*, \bar{\gamma}_j) -\bar H^{(j)}_{\beta\gamma} ({{{{\bar H}}_{\gamma\gamma}}^{(j)}})^{-1}\nabla_{\gamma}L_j(\beta^*, \bar{\gamma}_j)\}\|_2,
\]
which implies 
\[
\|\check{\beta} -{\beta}^*\|^8_2 \le C\|\frac{1}{K}\sum_{j=1}^{K}\{\nabla_{\beta}L_j(\beta^*, \bar{\gamma}_j) -\bar H^{(j)}_{\beta\gamma} ({{{{\bar H}}_{\gamma\gamma}}^{(j)}})^{-1}\nabla_{\gamma}L_j(\beta^*, \bar{\gamma}_j)\}\|_2^8,
\]
Now we control the term $\sum_{j=1}^{K}\{\nabla_{\beta}L_j(\beta^*, \bar{\gamma}_j) -\bar H^{(j)}_{\beta\gamma} ({{{{\bar H}}_{\gamma\gamma}}^{(j)}})^{-1}\nabla_{\gamma}L_j(\beta^*, \bar{\gamma}_j)\}/K$.
We have 
\begin{align}
&\frac{1}{K}\sum_{j=1}^{K}\{\nabla_{\beta}L_j(\beta^*, \bar{\gamma}_j) -\bar H^{(j)}_{\beta\gamma} ({{{{\bar H}}_{\gamma\gamma}}^{(j)}})^{-1}\nabla_{\gamma}L_j(\beta^*, \bar{\gamma}_j)\}\nonumber\\
=& \frac{1}{K}\sum_{j=1}^{K}\{\nabla_{\beta}L_j(\beta^*, {\gamma}_j^*) -I_{\beta\gamma}^{(j)}{I_{\gamma\gamma}^{(j)}}^{-1}\nabla_{\gamma}L_j(\beta^*, {\gamma}_j^*)\}\label{e11}\\&+\frac{1}{K}\sum_{j=1}^{K}\{\nabla_{\beta\gamma}L_j(\beta^*,\gamma_j')(\bar\gamma_j-\gamma_j^*)-I_{\beta\gamma}^{(j)}{I_{\gamma\gamma}^{(j)}}^{-1}\nabla_{\gamma\gamma}L_j(\beta^*, {\gamma}_j')(\bar\gamma_j-\gamma_j^*)\}\label{e12}\\
&- \frac{1}{K}\sum_{j=1}^{K}\{\bar H^{(j)}_{\beta\gamma} ({{{{\bar H}}_{\gamma\gamma}}^{(j)}})^{-1}- I^{(j)}_{\beta\gamma} (I_{\gamma\gamma}^{(j)})^{-1} \}\{ \nabla_\gamma L_j (\beta^*, \bar\gamma_j)\},\label{e13}
\end{align}
where $\gamma_j'$ satisfies $\|\gamma_j'-\gamma_j^*\|_2\le\|\bar\gamma_j-\gamma_j^*\|_2$.
For the last term in the right hand side of the above equation, we have  
\begin{align*}
&\|\bar H^{(j)}_{\beta\gamma} ({{{{\bar H}}_{\gamma\gamma}}^{(j)}})^{-1}- I^{(j)}_{\beta\gamma} (I_{\gamma\gamma}^{(j)})^{-1}\|_2\\\le &\|\bar H^{(j)}_{\beta\gamma}\|_2\|({{{{\bar H}}_{\gamma\gamma}}^{(j)}})^{-1}-(I_{\gamma\gamma}^{(j)})^{-1}\|_2+\|\bar H^{(j)}_{\beta\gamma}-I^{(j)}_{\beta\gamma}\|_2\|(I_{\gamma\gamma}^{(j)})^{-1}\|_2\\
\le &\|\bar H^{(j)}_{\beta\gamma}\|_2\|({{{{\bar H}}_{\gamma\gamma}}^{(j)}})^{-1}\|_2\|I_{\gamma\gamma}^{(j)}-{{{{\bar H}}_{\gamma\gamma}}^{(j)}}\|_2\|(I_{\gamma\gamma}^{(j)})^{-1}\|_2+\|\bar H^{(j)}_{\beta\gamma}-I^{(j)}_{\beta\gamma}\|_2\|(I_{\gamma\gamma}^{(j)})^{-1}\|_2.
\end{align*}
By Lemma S.9, we know that $\bar H\succeq (1-\rho)\mu_-$ with probability $1-Ce^{-n}$. And
\begin{align*}
&\E\|\bar H^{(j)}_{\beta\gamma} ({{{{\bar H}}_{\gamma\gamma}}^{(j)}})^{-1}- I^{(j)}_{\beta\gamma} (I_{\gamma\gamma}^{(j)})^{-1}\|^{16}_2\\
\le &C_3\E\|I_{\gamma\gamma}^{(j)}-{{{{\bar H}}_{\gamma\gamma}}^{(j)}}\|^{16}_2+C_4\|\bar H^{(j)}_{\beta\gamma}-I^{(j)}_{\beta\gamma}\|^{16}_2 \lesssim 1/n^8.
\end{align*}
In addition we have 
\begin{align*}
\E\|\nabla_\gamma L_j (\beta^*, \bar\gamma_j)\|_2^{16}\le C_5\E\|\nabla_\gamma L_j (\beta^*, \gamma_j^*)\|_2^{16} +C_6M\E\|\bar{\gamma}_j - \gamma_j^*\|_2^{16} \lesssim \frac{1}{n^8}.
\end{align*}
Thus, for the term in(\ref{e13}) we have 
\[
\E\|\frac{1}{K}\sum_{j=1}^{K}\{\bar H^{(j)}_{\beta\gamma} ({{{{\bar H}}_{\gamma\gamma}}^{(j)}})^{-1}- I^{(j)}_{\beta\gamma} (I_{\gamma\gamma}^{(j)})^{-1} \}\{ \nabla_\gamma L_j (\beta^*, \bar\gamma_j)\}\|_2^8\lesssim 1/n^8. 
\]
The term in (\ref{e12}) can be further decomposed to 
\begin{align*}
&\frac{1}{K}\sum_{j=1}^{K}\{\nabla_{\beta\gamma}L_j(\beta^*,\gamma_j')(\bar\gamma_j-\gamma_j^*)-I_{\beta\gamma}^{(j)}{I_{\gamma\gamma}^{(j)}}^{-1}\nabla_{\gamma\gamma}L_j(\beta^*, {\gamma}_j')(\bar\gamma_j-\gamma_j^*)\}\\
&=\frac{1}{K}\sum_{j=1}^{K}\Big\{\{\nabla_{\beta\gamma}L_j(\beta^*,\gamma_j')-I_{\beta\gamma}^{(j)}{I_{\gamma\gamma}^{(j)}}^{-1}\nabla_{\gamma\gamma}L_j(\beta^*, {\gamma}_j')\}(\bar\gamma_j-\gamma_j^*)\Big\}\\
&=\frac{1}{K}\sum_{j=1}^{K}\Big\{\{\nabla_{\beta\gamma}L_j(\beta^*,\gamma_j')+I_{\beta\gamma}^{(j)}-I_{\beta\gamma}^{(j)}-I_{\beta\gamma}^{(j)}{I_{\gamma\gamma}^{(j)}}^{-1}\nabla_{\gamma\gamma}L_j(\beta^*, {\gamma}_j')\}(\bar\gamma_j-\gamma_j^*)\Big\}. 
\end{align*}
From Lemma S.1, Assumption 5, and event $\mathcal{E}_0$ we have 
\begin{align}
&\E\|\nabla_{\beta\gamma}L_j(\beta^*,\gamma_j')+I_{\beta\gamma}^{(j)}-I_{\beta\gamma}^{(j)}-I_{\beta\gamma}^{(j)}{I_{\gamma\gamma}^{(j)}}^{-1}\nabla_{\gamma\gamma}L_j(\beta^*, {\gamma}_j')\|_2^{16}\nonumber\\
&\le \E\|\nabla_{\beta\gamma}L_j(\beta^*,\gamma_j')+I_{\beta\gamma}^{(j)}\|_2^{16}+\E\|I_{\beta\gamma}^{(j)}-I_{\beta\gamma}^{(j)}{I_{\gamma\gamma}^{(j)}}^{-1}\nabla_{\gamma\gamma}L_j(\beta^*, {\gamma}_j')\|_2^{16}
\lesssim\frac{1}{n^8}. \label{ee7}
\end{align}
Therefore we have 
\begin{align}
\label{e7}\E\Big\{\|\frac{1}{K}\sum_{j=1}^{K}\{\nabla_{\beta\gamma}L_j(\beta^*,\gamma_j')(\bar\gamma_j-\gamma_j^*)-I_{\beta\gamma}^{(j)}{I_{\gamma\gamma}^{(j)}}^{-1}\nabla_{\gamma\gamma}L_j(\beta^*, {\gamma}_j')(\bar\gamma_j-\gamma_j^*)\}\|_2^8\Big\}\lesssim \frac{1}{n^8}.
\end{align}
Also, we have 
\begin{align}\label{ee8}
\E\|\frac{1}{K}\sum_{j=1}^{K}\{\nabla_{\beta}L_j(\beta^*, {\gamma}_j^*) -I_{\beta\gamma}^{(j)}{I_{\gamma\gamma}^{(j)}}^{-1}\nabla_{\gamma}L_j(\beta^*, {\gamma}_j^*)\}\|_2^8\lesssim \frac{1}{(Kn)^4}.
\end{align}
Combine (\ref{e7}) and (\ref{ee8}) we obtain 
\[
\E\{\|\check{\beta}-\beta^*\|_2^8I(\mathcal{E})\}\lesssim \frac{1}{(Kn)^4}+\frac{1}{n^8}.
\]

Now we calculate the probability for $\mathcal{E}^c$. From the definition of $\mathcal{E}_0$ we have  $\p(\mathcal{E}_0^c)\lesssim \exp(-Kn)$. 

For $\mathcal{E}_1^c$, we have 
\begin{align*}
&\|\frac{1}{K}\sum_{j=1}^{K}\{\nabla_{\beta\beta}L_j(\beta^*, \bar{\gamma}_j) -\bar H^{(j)}_{\beta\gamma} ({{{{\bar H}}_{\gamma\gamma}}^{(j)}})^{-1}\nabla_{\gamma\beta}L_j(\beta^*, \bar{\gamma}_j) + I^{(j)}_{\beta\beta}-I^{(j)}_{\beta\gamma}{I^{(j)}_{\gamma\gamma}}^{-1}I^{(j)}_{\gamma\beta}\} \|_2 \\
\le& \|\frac{1}{K}\sum_{j=1}^{K}\{\nabla_{\beta\beta}L_j(\beta^*, {\gamma}_j^*)+ I^{(j)}_{\beta\beta}-I_{\beta\gamma}^{(j)}{I_{\gamma\gamma}^{(j)}}^{-1}\nabla_{\gamma\beta}L_j(\beta^*, {\gamma}_j)-I^{(j)}_{\beta\gamma}{I^{(j)}_{\gamma\gamma}}^{-1}I^{(j)}_{\gamma\beta}\}\|_2\\
&+\frac{1}{K}\sum_{j=1}^{K}\{\|\nabla_{\beta\beta}L_j(\beta^*, \bar{\gamma}_j)- \nabla_{\beta\beta}L_j(\beta^*, {\gamma}_j^*)\|_2+\|I_{\beta\gamma}^{(j)}{I_{\gamma\gamma}^{(j)}}^{-1}\|_2\|\nabla_{\gamma\beta}L_j(\beta^*, \bar{\gamma}_j) -\nabla_{\gamma\beta}L_j(\beta^*, {\gamma}_j^*) \|_2\}\\&
+\|\frac{1}{K}\sum_{j=1}^{K}\{\bar H^{(j)}_{\beta\gamma} ({{{{\bar H}}_{\gamma\gamma}}^{(j)}})^{-1}- I^{(j)}_{\beta\gamma} (I_{\gamma\gamma}^{(j)})^{-1} \}\{ \nabla_\gamma L_j (\beta^*, \bar\gamma_j)\}\|_2
\end{align*}
And further we have 
\begin{align*}
&\p(\|\frac{1}{K}\sum_{j=1}^{K}\{\nabla_{\beta\beta}L_j(\beta^*, {\gamma}_j^*)+ I^{(j)}_{\beta\beta}-I_{\beta\gamma}^{(j)}{I_{\gamma\gamma}^{(j)}}^{-1}\nabla_{\gamma\beta}L_j(\beta^*, {\gamma}_j)-I^{(j)}_{\beta\gamma}{I^{(j)}_{\gamma\gamma}}^{-1}I^{(j)}_{\gamma\beta}\}\|_2>C_1/3)\lesssim \exp(-Kn).
\end{align*}
Also,
\begin{align*}
&\p(\frac{1}{K}\sum_{j=1}^{K}\{\|\nabla_{\beta\beta}L_j(\beta^*, \bar{\gamma}_j)- \nabla_{\beta\beta}L_j(\beta^*, {\gamma}_j^*)\|_2+C_3\|\nabla_{\gamma\beta}L_j(\beta^*, \bar{\gamma}_j) -\nabla_{\gamma\beta}L_j(\beta^*, {\gamma}_j^*) \|_2\}>C_1/3)\\&\le 
\p(\frac{1}{K}\sum_{j=1}^{K}\|\bar{\gamma}_j- {{\gamma}_j^*}\|_2>C_4) =\p(\frac{1}{K}\sum_{j=1}^{K}\|\bar{\gamma}_j- {{\gamma}_j^*}\|^{16}_2>C_5) \lesssim \frac{1}{n^8}
\end{align*}
and 
\begin{align*}
&\p(\|\frac{1}{K}\sum_{j=1}^{K}\{\bar H^{(j)}_{\beta\gamma} ({{{{\bar H}}_{\gamma\gamma}}^{(j)}})^{-1}- I^{(j)}_{\beta\gamma} (I_{\gamma\gamma}^{(j)})^{-1} \}\{ \nabla_\gamma L_j (\beta^*, \bar\gamma_j)\}\|_2>C_1/3)\\&\le 
\E\|\frac{1}{K}\sum_{j=1}^{K}\{\bar H^{(j)}_{\beta\gamma} ({{{{\bar H}}_{\gamma\gamma}}^{(j)}})^{-1}- I^{(j)}_{\beta\gamma} (I_{\gamma\gamma}^{(j)})^{-1} \}\{ \nabla_\gamma L_j (\beta^*, \bar\gamma_j)\}\|^8_2/(C_1/3)^8\lesssim \frac{1}{n^8}.
\end{align*}
Thus
$
\p(\mathcal{E}_1^c)\lesssim{1}/{n^8}.
$
For $\mathcal{E}_2^c$, since we have under $\mathcal{E}_0$
\[
\E\{\frac{1}{K}\|\sum_{j=1}^{K}\{\nabla_{\beta}L_j(\beta^*, \bar{\gamma}_j) -\bar H^{(j)}_{\beta\gamma} ({{{{\bar H}}_{\gamma\gamma}}^{(j)}})^{-1}\nabla_{\gamma}L_j(\beta^*, \bar{\gamma}_j)\}\|_2^8I(\mathcal{E}_0)\}\lesssim \frac{1}{(Kn)^4}+\frac{1}{n^8}.
\]
Therefore we have under $\mathcal{E}_0$ 
\[
\p\{\frac{1}{K}\|\sum_{j=1}^{K}\{\nabla_{\beta}L_j(\beta^*, \bar{\gamma}_j) -\bar H^{(j)}_{\beta\gamma} ({{{{\bar H}}_{\gamma\gamma}}^{(j)}})^{-1}\nabla_{\gamma}L_j(\beta^*, \bar{\gamma}_j)\}\|_2^8>C_2^8\}\lesssim\frac{1}{(Kn)^4}+\frac{1}{n^8}, 
\]
which implies $\p\{\mathcal{E}_0\cap(\mathcal{E}_2^c)\} \lesssim{1}/{(Kn)^4}+{1}/{n^8}$.
Thus,
\[
\p\{\mathcal{E}^c\}\le\p\{\mathcal{E}_0^c\}+\p\{\mathcal{E}_1^c\}+\p\{\mathcal{E}_0\cap\mathcal{E}_2\}\lesssim{1}/{(Kn)^4}+{1}/{n^8}.
\]
Combine all, we have 
\[
\E\|\check{\beta}-\beta^*\|_2^2\lesssim{1}/{(Kn)^4}+{1}/{n^8}.
\]
By the definition of $\check \beta$, we have
\begin{align*}
0&=\sum_{j=1}^{K}\{\nabla_{\beta}L_j(\check\beta, \bar{\gamma}_j) -I_{\beta\gamma}^{(j)}{I_{\gamma\gamma}^{(j)}}^{-1}\nabla_{\gamma}L_j(\check\beta, \bar{\gamma}_j)\}\\
& = \sum_{j=1}^{K}\{\nabla_{\beta}L_j(\beta^*, {\gamma}_j^*) -I_{\beta\gamma}^{(j)}{I_{\gamma\gamma}^{(j)}}^{-1}\nabla_{\gamma}L_j(\beta^*, {\gamma}_j^*)\}\\&+\sum_{j=1}^{K}\{\nabla_{\beta\beta}L_j(\beta', {\gamma}_j')-I_{\beta\gamma}^{(j)}{I_{\gamma\gamma}^{(j)}}^{-1}\nabla_{\gamma\beta}L_j(\beta', {\gamma}_j')\}(\check{\beta}-\beta^*)\\
&+\sum_{j=1}^{K}\{\nabla_{\beta\gamma}L_j(\beta', {\gamma}_j')-I_{\beta\gamma}^{(j)}{I_{\gamma\gamma}^{(j)}}^{-1}\nabla_{\gamma\gamma}L_j(\beta', {\gamma}_j')\}(\bar{\gamma}_j-{\gamma}_j^*)\\&
- \frac{1}{K}\sum_{j=1}^{K}\{\bar H^{(j)}_{\beta\gamma} ({{{{\bar H}}_{\gamma\gamma}}^{(j)}})^{-1}- I^{(j)}_{\beta\gamma} (I_{\gamma\gamma}^{(j)})^{-1} \}\{ \nabla_\gamma L_j (\check\beta, \bar\gamma_j)\}
\end{align*}
where $\gamma_j'$ satisfies $\|\gamma_j'-\gamma_j^*\|_2\le\|\bar\gamma_j-\gamma_j^*\|_2$, and $\beta'$ satisfies $\|\beta'-\beta^*\|_2\le\|\check\beta-\beta^*\|_2$. 
Therefore we get 
\begin{align*}
\frac{1}{K}\sum_{j=1}^{K}\{I^{(j)}_{\beta|\gamma}\}(\check{\beta}-\beta^*) =& \frac{1}{K}\sum_{j=1}^{K}\{\nabla_{\beta}L_j(\beta^*, {\gamma}_j^*) -I_{\beta\gamma}^{(j)}{I_{\gamma\gamma}^{(j)}}^{-1}\nabla_{\gamma}L_j(\beta^*, {\gamma}_j^*)\} \\
&+\frac{1}{K}\sum_{j=1}^{K}\{\nabla_{\beta\gamma}L_j(\beta', {\gamma}_j')-I_{\beta\gamma}^{(j)}{I_{\gamma\gamma}^{(j)}}^{-1}\nabla_{\gamma\gamma}L_j(\beta', {\gamma}_j')\}(\bar{\gamma}_j-{\gamma}_j^*)\\
&+\frac{1}{K}\sum_{j=1}^{K}\{\nabla_{\beta\beta}L_j(\beta', {\gamma}_j')-I_{\beta\gamma}^{(j)}{I_{\gamma\gamma}^{(j)}}^{-1}\nabla_{\gamma\beta}L_j(\beta', {\gamma}_j')+I^{(j)}_{\beta|\gamma}\}(\check{\beta}-\beta^*)\\
&- \frac{1}{K}\sum_{j=1}^{K}\{\bar H^{(j)}_{\beta\gamma} ({{{{\bar H}}_{\gamma\gamma}}^{(j)}})^{-1}- I^{(j)}_{\beta\gamma} (I_{\gamma\gamma}^{(j)})^{-1} \}\{ \nabla_\gamma L_j (\check\beta, \bar\gamma_j)\}
\end{align*}
We denote 
\[
\delta_1 = \frac{1}{K}\sum_{j=1}^{K}\{\nabla_{\beta\gamma}L_j(\beta', {\gamma}_j')-I_{\beta\gamma}^{(j)}{I_{\gamma\gamma}^{(j)}}^{-1}\nabla_{\gamma\gamma}L_j(\beta', {\gamma}_j')\}(\bar{\gamma}_j-{\gamma}_j^*),
\] 
\[
\delta_2 = \frac{1}{K}\sum_{j=1}^{K}\{\nabla_{\beta\beta}L_j(\beta', {\gamma}_j')-I_{\beta\gamma}^{(j)}{I_{\gamma\gamma}^{(j)}}^{-1}\nabla_{\gamma\beta}L_j(\beta', {\gamma}_j')+I^{(j)}_{\beta|\gamma}\}(\check{\beta}-\beta^*),
\]
and 
\[
\delta_3 =- \frac{1}{K}\sum_{j=1}^{K}\{\bar H^{(j)}_{\beta\gamma} ({{{{\bar H}}_{\gamma\gamma}}^{(j)}})^{-1}- I^{(j)}_{\beta\gamma} (I_{\gamma\gamma}^{(j)})^{-1} \}\{ \nabla_\gamma L_j (\check\beta, \bar\gamma_j)\}
\]
and we only need to prove  $\E\|\delta_k\|_2^8\lesssim1/n^8$, for $k =1, 2, 3$. 
For $\delta_1$, we have 
\begin{align*}
\E\|\delta_1\|_2^8\le \frac{C}{K}\sum_{j=1}^{K}\left\{\E\|\nabla_{\beta\gamma}L_j(\beta', {\gamma}_j')-I_{\beta\gamma}^{(j)}{I_{\gamma\gamma}^{(j)}}^{-1}\nabla_{\gamma\gamma}L_j(\beta', {\gamma}_j')\|_2^{16}\E\|\bar{\gamma}_j-{\gamma}_j^*\|_2^{16}\right\}^{1/2},
\end{align*}
Following the same proof as (\ref{ee7}), we have
\begin{align*}
\E\{\|\nabla_{\beta\gamma}L_j(\beta', {\gamma}_j')-I_{\beta\gamma}^{(j)}{I_{\gamma\gamma}^{(j)}}^{-1}\nabla_{\gamma\gamma}L_j(\beta', {\gamma}_j')\|_2^{16}\}\lesssim \frac{1}{n^8}.
\end{align*}
and 
\begin{align*}
\E\{\|\nabla_{\beta\beta}L_j(\beta', {\gamma}_j')-I_{\beta\gamma}^{(j)}{I_{\gamma\gamma}^{(j)}}^{-1}\nabla_{\gamma\beta}L_j(\beta', {\gamma}_j')+I^{(j)}_{\beta|\gamma}\|_2^{16}\}\lesssim \frac{1}{n^8}.
\end{align*}
For $\delta_3$ we have 
\begin{align*}
\nabla_\gamma L_j (\check\beta, \bar\gamma_j) = \nabla_\gamma L_j (\beta^*, \gamma_j^*)+\nabla_{\gamma\beta} L_j (\beta', \gamma_j') (\check\beta-\beta^*)+\nabla_{\gamma\gamma} L_j (\beta', \gamma_j') (\bar\gamma_j-\gamma_j^*),
\end{align*}
and therefore we have 
\begin{align*}
\E\{\|\nabla_\gamma L_j (\check\beta, \bar\gamma_j)\|_2^{16}\}\lesssim\frac{1}{n^8}.
\end{align*}

Thus, $\E\|\delta_1\|^8_2 + \E\|\delta_2\|^8_2+\E\|\delta_3\|^8_2\lesssim 1/n^8$. And we have $\check \delta = [\sum_{j=1}^{K}\{I^{(j)}_{\beta|\gamma}\}/K]^{-1}\{\delta_1+\delta_2+\delta_3\}$ satisfies that $\E\|\check \delta\|_2^8\lesssim 1/n^8$.
$\hfill\square$

\subsection*{Proof of Lemma S.7} 
For simple notation, here in this proof we denote $\tilde{\beta}^{(1)}$  as $ \tilde{\beta}$. 
Similar as the previous proof, we define the following events:

\[
\mathcal{E}_{0j} :=\{ \frac{1}{n}\sum_{i=1}^{n}m_k(Y_{ij}) \le  2M, \text{ for } k = 1, 2\},
\]

\[
\mathcal{E}_1 := \{\|\nabla_\beta\tilde U(\check\beta)-\frac{1}{K}\sum_{j=1}^{K}\{\nabla_{\beta\beta}L_j(\check\beta, \bar{\gamma}_j) -\bar H^{(j)}_{\beta\gamma} ({{{{\bar H}}_{\gamma\gamma}}^{(j)}})^{-1}\nabla_{\gamma\beta}L_j(\check\beta, \bar{\gamma}_j)\}\|_2\le C_1\}
\]
\[
\mathcal{E}_2 := \{\|\tilde U(\check\beta)\|_2 \le C_2\},
\] 
for some constants $M$, $C_1$ and $C_2$ which satisfy $\E\{ m_k (Y_{ij})\}< M$ for all $j \in \{1, \dots, K\}$, and $k = 1, 2$, $C_1\le \rho\mu_-/2$ and $C_2<(1-\rho)\rho\mu_-^2/8M$. Let $\mathcal{E}_0 = \cap_{1\le j\le K}\mathcal{E}_{0j}$. Applying Lemma 6 in \cite{zhang2012communication} we have under
event $\mathcal{E} = \{\cap_{i=0,1,2}\mathcal{E}_i\}$, 
\[
\|\tilde{\beta} -\check{\beta}\|^4_2 \le C\|\tilde U(\check\beta)\|^4_2.
\]
Now we control the term $\E\{\|\tilde U(\check\beta)\|_2^4\}$. We have 
\begin{align*}
\tilde U(\check\beta) =  U_1(\check\beta) + \{\frac{1}{K}\sum_{j=1}^{K}\{\nabla_\beta L_j (\bar\beta, \bar\gamma_j)- \bar H^{(j)}_{\beta\gamma} ({{{{\bar H}}_{\gamma\gamma}}^{(j)}})^{-1}  \nabla_\gamma L_j (\bar\beta, \bar\gamma_j)\}-   U_1(\bar\beta)\}, 
\end{align*}
and since
\[
0 = \frac{1}{K}\sum_{j=1}^{K}\{\nabla_\beta L_j (\check\beta, \bar\gamma_j)- \bar H^{(j)}_{\beta\gamma} ({{{{\bar H}}_{\gamma\gamma}}^{(j)}})^{-1}  \nabla_\gamma L_j (\check\beta, \bar\gamma_j)\},
\]
we have 
\begin{align}
&\tilde U(\check\beta) =  U_1(\check\beta) - \frac{1}{K}\sum_{j=1}^{K}\{\nabla_\beta L_j (\check\beta, \bar\gamma_j)- \bar H^{(j)}_{\beta\gamma} ({{{{\bar H}}_{\gamma\gamma}}^{(j)}})^{-1}  \nabla_\gamma L_j (\check\beta, \bar\gamma_j)\}\nonumber\\& -\{  U_1(\bar\beta)-\frac{1}{K}\sum_{j=1}^{K}\{\nabla_\beta L_j (\bar\beta, \bar\gamma_j)- \bar H^{(j)}_{\beta\gamma} ({{{{\bar H}}_{\gamma\gamma}}^{(j)}})^{-1}  \nabla_\gamma L_j (\bar\beta, \bar\gamma_j)\}\}\nonumber\\
& = \{\nabla_\beta U_1(\beta')  - \frac{1}{K}\sum_{j=1}^{K}\{\nabla_\beta L_j (\beta', \bar\gamma_j)- \bar H^{(j)}_{\beta\gamma} ({{{{\bar H}}_{\gamma\gamma}}^{(j)}})^{-1}  \nabla_\gamma L_j (\beta', \bar\gamma_j)\}\}(\check{\beta}- \bar{\beta}),\label{eeee}
\end{align}
where $\beta'$ satisfies $\|\beta'-\check\beta\|_2\le \|\bar\beta-\check\beta\|_2$.
By Lemma S.2, we have 
\begin{align*}
\bar{\beta}-\beta^* = \frac{1}{K}\sum_{j=1}^{K}\bar\beta_j-\beta^* = \frac{1}{K}\sum_{j=1}^{K}\{(I^{(j)}_{\beta|\gamma})^{-1}\nabla_{\beta}L_j(\theta_j^*)-(I^{(j)}_{\beta|\gamma})^{-1}I^{(j)}_{\beta\gamma}(I^{(j)}_{\gamma\gamma})^{-1}\nabla_{\gamma}L_j(\theta_j^*)\} +\frac{1}{K}\sum_{j=1}^{K}\delta_{\beta,j}
\end{align*}
where $\delta_{\beta,j}$ is the subvector of $\delta_j$ defined in Lemma S.2. Thus, we have 
\[
\E \|\bar{\beta}-\beta^*\|_2^8 \lesssim \frac{1}{K^4n^4} +\frac{1}{n^8}.
\]
Combining with Lemma S.6, we have 
$\E\|\check{\beta}- \bar{\beta}\|^4_2\lesssim{1}/{K^4n^4} +{1}/{n^8}$.
Now we show that 
\begin{align}
\E\|\nabla_\beta U_1(\beta')  - \frac{1}{K}\sum_{j=1}^{K}\{\nabla_\beta L_j (\beta', \bar\gamma_j)- \bar H^{(j)}_{\beta\gamma} ({{{{\bar H}}_{\gamma\gamma}}^{(j)}})^{-1}  \nabla_\gamma L_j (\beta', \bar\gamma_j)\}\|_2^8\lesssim \frac{1}{n^4}. \label{ee4}
\end{align}
We have 
\begin{align}
&\nabla_\beta U_1(\beta')  - \frac{1}{K}\sum_{j=1}^{K}\{\nabla_\beta L_j (\beta', \bar\gamma_j)- \bar H^{(j)}_{\beta\gamma} ({{{{\bar H}}_{\gamma\gamma}}^{(j)}})^{-1}  \nabla_\gamma L_j (\beta', \bar\gamma_j)\}\label{ee5}\\ 
&= \frac{1}{Kn}\sum_{j=1}^{K} \sum_{i=1}^{n}\left\{\frac{f(y_{i1}; \bar\beta, \bar\gamma_j)}{f(y_{i1}; \bar\beta, \bar\gamma_1)}\nabla_{\beta\beta}\log f(y_{i1}; \beta', \bar\gamma_j) - \nabla_{\beta\beta}\log f(y_{ij}; \beta', \bar\gamma_j) \right\}\\&
-\frac{1}{Kn}\sum_{j=1}^{K} \sum_{i=1}^{n}\left\{\frac{f(y_{i1}; \bar\beta, \bar\gamma_j)}{f(y_{i1}; \bar\beta, \bar\gamma_1)}\tilde H_{\beta\gamma}^{(1,j)}\{ \tilde H_{\gamma\gamma}^{(1,j)}\}^{-1}\nabla_{\gamma\beta}\log f(y_{i1}; \beta', \bar\gamma_j) - \bar H^{(j)}_{\beta\gamma} ({{{{\bar H}}_{\gamma\gamma}}^{(j)}})^{-1}\nabla_{\gamma\beta}\log f(y_{ij}; \beta', \bar\gamma_j)\right\}
\end{align}
For the term in (19) we have 
\begin{align}
&\frac{1}{Kn}\sum_{j=1}^{K} \sum_{i=1}^{n}\left\{\frac{f(y_{i1}; \bar\beta, \bar\gamma_j)}{f(y_{i1}; \bar\beta, \bar\gamma_1)}\nabla_{\beta\beta}\log f(y_{i1}; \beta', \bar\gamma_j) - \nabla_{\beta\beta}\log f(y_{ij}; \beta', \bar\gamma_j) \right\}\\
= &\frac{1}{Kn}\sum_{j=1}^{K} \sum_{i=1}^{n}\left\{\frac{f(y_{i1}; \bar\beta, \bar\gamma_j)}{f(y_{i1}; \bar\beta, \bar\gamma_1)}\nabla_{\beta\beta}\log f(y_{i1}; \beta', \bar\gamma_j) - \frac{f(y_{i1}; \beta^*, \gamma_j^*)}{f(y_{i1}; \beta^*, \gamma_1^*)}\nabla_{\beta\beta}\log f(y_{i1}; \beta^*, \gamma_j^*) \right\}\\
&+\frac{1}{Kn}\sum_{j=1}^{K} \sum_{i=1}^{n}\left\{\nabla_{\beta\beta}\log f(y_{ij}; \beta^*, \gamma_j^*) - \nabla_{\beta\beta}\log f(y_{ij}; \beta', \bar\gamma_j) \right\}\\
& + \frac{1}{Kn}\sum_{j=1}^{K} \sum_{i=1}^{n}\left\{\frac{f(y_{i1}; \beta^*, \gamma_j^*)}{f(y_{i1}; \beta^*, \gamma_1^*)}\nabla_{\beta\beta}\log f(y_{i1}; \beta^*, \gamma_j^*)-\nabla_{\beta\beta}\log f(y_{ij}; \beta^*, \gamma_j^*)  \right\}.
\end{align}
By Assumption 5 and event $\mathcal{E}_0$ we have 
\[
\E\|\frac{1}{Kn}\sum_{j=1}^{K} \sum_{i=1}^{n}\left\{\frac{f(y_{i1}; \bar\beta, \bar\gamma_j)}{f(y_{i1}; \bar\beta, \bar\gamma_1)}\nabla_{\beta\beta}\log f(y_{i1}; \beta', \bar\gamma_j) - \nabla_{\beta\beta}\log f(y_{ij}; \beta', \bar\gamma_j) \right\}\|_2^8\lesssim\frac{1}{n^4}.
\]
In addition, we have 
\begin{align*}
&\frac{1}{Kn}\sum_{j=1}^{K} \sum_{i=1}^{n}\left\{\frac{f(y_{i1}; \bar\beta, \bar\gamma_j)}{f(y_{i1}; \bar\beta, \bar\gamma_1)}\tilde H_{\beta\gamma}^{(1,j)}\{ \tilde H_{\gamma\gamma}^{(1,j)}\}^{-1}\nabla_{\gamma\beta}\log f(y_{i1}; \beta', \bar\gamma_j) - \bar H^{(j)}_{\beta\gamma} ({{{{\bar H}}_{\gamma\gamma}}^{(j)}})^{-1}\nabla_{\gamma\beta}\log f(y_{ij}; \beta', \bar\gamma_j)\right\}\\
&= \frac{1}{K}\sum_{j=1}^{K}\Big\{\tilde H_{\beta\gamma}^{(1,j)}\{\tilde H_{\gamma\gamma}^{(1,j)}\}^{-1}\frac{1}{n} \sum_{i=1}^{n}\left\{\frac{f(y_{i1}; \bar\beta, \bar\gamma_j)}{f(y_{i1}; \bar\beta, \bar\gamma_1)}\nabla_{\gamma\beta}\log f(y_{i1}; \beta', \bar\gamma_j)\right\} -\bar H^{(j)}_{\beta\gamma} ({{{{\bar H}}_{\gamma\gamma}}^{(j)}})^{-1}\nabla_{\gamma\beta}L_j( \beta', \bar\gamma_j)\Big\}
\end{align*}
Denote $\tilde{A}_j = \tilde H_{\beta\gamma}^{(1,j)}\{\tilde H_{\gamma\gamma}^{(1,j)}\}^{-1}$, $\bar{A}_j = \bar H^{(j)}_{\beta\gamma} ({{{{\bar H}}_{\gamma\gamma}}^{(j)}})^{-1}$,  $$\tilde{B}_j = \frac{1}{n}\sum_{i=1}^{n}\left\{\frac{f(y_{i1}; \bar\beta, \bar\gamma_j)}{f(y_{i1}; \bar\beta, \bar\gamma_1)}\nabla_{\gamma\beta}\log f(y_{i1}; \beta', \bar\gamma_j)\right\}$$
and $\bar B_j = \nabla_{\gamma\beta}L_j( \beta', \bar\gamma_j)$, the above term can be written as 
\begin{align*}
\frac{1}{K}\sum_{j=1}^{K}\Big\{\tilde{A}_j\tilde{B}_j-\bar A_j\bar B_j\Big\} = \frac{1}{K}\sum_{j=1}^{K}\Big\{\tilde{A}_j(\tilde{B}_j-\bar B_j) +(\tilde{A}_j-\bar A_j)\bar B_j\Big\}. 
\end{align*}
Further by Lemma S.9 we have $\|\tilde A_j\|_2\le2\mu_+(\mu_-(1-\rho))^{-1}$, $\|\bar A_j\|_2\le2\mu_+(\mu_-(1-\rho))^{-1}$, $\|\tilde B_j\|_2\le2\mu_+$, $\|\bar B_j\|_2\le2\mu_+$ with probability at least $1-\exp(-Cn)$. Also, we have under $\mathcal{E}_0$
\begin{align*}
&\|\tilde{B}_j-\bar B_j\|^8_2 =\| \frac{1}{n}\sum_{i=1}^{n}\{\frac{f(y_{i1}; \bar\beta, \bar\gamma_j)}{f(y_{i1}; \bar\beta, \bar\gamma_1)}\nabla_{\gamma\beta}\log f(y_{i1}; \beta', \bar\gamma_j)-\nabla_{\gamma\beta}\log f(y_{i1}; \beta', \bar\gamma_j)\}\|\\
&= C\|\frac{1}{n}\sum_{i=1}^{n}\{\frac{f(y_{i1}; \bar\beta, \bar\gamma_j)}{f(y_{i1}; \bar\beta, \bar\gamma_1)}\nabla_{\gamma\beta}\log f(y_{i1}; \beta', \bar\gamma_j)-\frac{f(y_{i1}; \beta^*, \gamma_j^*)}{f(y_{i1}; \beta^*, \gamma_1^*)}\nabla_{\gamma\beta}\log f(y_{i1}; \beta^*, \gamma_j^*)\}\|_2^8\\
&+C\|\frac{1}{n}\sum_{i=1}^{n}\{\nabla_{\gamma\beta}\log f(y_{i1}; \beta', \bar\gamma_j)-\nabla_{\gamma\beta}\log f(y_{i1}; \beta^*, \gamma_j^*)\}\|_2^8\\
&+C\|\frac{1}{n}\sum_{i=1}^{n}\{\frac{f(y_{i1}; \beta^*, \gamma_j^*)}{f(y_{i1}; \beta^*, \gamma_1^*)}\nabla_{\gamma\beta}\log f(y_{i1}; \beta^*, \gamma_j^*)-\nabla_{\gamma\beta}\log f(y_{i1}; \beta^*, \gamma_j^*)\}\|_2^8\lesssim\frac{1}{n^4}.
\end{align*}
Thus, we have 
\beq\label{ee6}
\E\|\frac{1}{K}\sum_{j=1}^{K}\Big\{\tilde{A}_j\tilde{B}_j-\bar A_j\bar B_j\Big\}\|_2^8\lesssim\frac{1}{n^4},
\eeq
which proved (\ref{ee4}).
Combine all we have 
\begin{align*}
\E\{\|\tilde U(\check\beta)\|_2^8I(\mathcal{E})\}\le\frac{1}{K^2n^4}+\frac{1}{n^{6}}. 
\end{align*}

Next we calculate the probability of $\mathcal{E}^c$. We have 
\beq\label{ee10}
\p (\mathcal{E}_0^c) \lesssim K\exp(-n)
\eeq
and for $\mathcal{E}_1^c$, we have 
\begin{align*}
&\|\nabla_\beta\tilde U(\check\beta)-\frac{1}{K}\sum_{j=1}^{K}\{\nabla_{\beta\beta}L_j(\check\beta, \bar{\gamma}_j) -\bar H^{(j)}_{\beta\gamma} ({{{{\bar H}}_{\gamma\gamma}}^{(j)}})^{-1}\nabla_{\gamma\beta}L_j(\check\beta, \bar{\gamma}_j)\}\|_2\\
=&\|\nabla_\beta U_1(\check\beta)-\frac{1}{K}\sum_{j=1}^{K}\{\nabla_{\beta\beta}L_j(\check\beta, \bar{\gamma}_j) -\bar H^{(j)}_{\beta\gamma} ({{{{\bar H}}_{\gamma\gamma}}^{(j)}})^{-1}\nabla_{\gamma\beta}L_j(\check\beta, \bar{\gamma}_j)\}\|_2.
\end{align*}
Following the similar procedures from (\ref{ee5})-(\ref{ee6}), we have that under $\mathcal{E}_0$, we have 
\begin{align*}
&\E\|\nabla_\beta U_1(\check\beta)-\frac{1}{K}\sum_{j=1}^{K}\{\nabla_{\beta\beta}L_j(\check\beta, \bar{\gamma}_j) -\bar H^{(j)}_{\beta\gamma} ({{{{\bar H}}_{\gamma\gamma}}^{(j)}})^{-1}\nabla_{\gamma\beta}L_j(\check\beta, \bar{\gamma}_j)\}\|^{12}_2\lesssim1/n^6
\end{align*}
which implies
\beq
\p\{\|\nabla_\beta U_1(\check\beta)-\frac{1}{K}\sum_{j=1}^{K}\{\nabla_{\beta\beta}L_j(\check\beta, \bar{\gamma}_j) -\bar H^{(j)}_{\beta\gamma} ({{{{\bar H}}_{\gamma\gamma}}^{(j)}})^{-1}\nabla_{\gamma\beta}L_j(\check\beta, \bar{\gamma}_j)\}\|_2>C_1\}\lesssim1/n^6.
\eeq
In the meanwhile for $\mathcal{E}_2^c$, we have showed that under $\mathcal{E}_0$, we have 
\[
\E\|\tilde U(\check\beta)\|^4_2\lesssim \frac{1}{K^2n^4}+\frac{1}{n^{6}},
\]
and therefore 
\beq\label{ee9}
\p\{\tilde U(\check\beta)\|_2>C_2\}\lesssim\frac{1}{K^2n^4}+\frac{1}{n^{6}}
\eeq
Combine all, we have $$\p(\mathcal{E}^c)\lesssim \frac{1}{K^2n^4}+\frac{1}{n^{6}},$$ which completes the proof.

\subsection*{Proof of Lemma S.8} 
When updating the estimator of $\gamma_j$ within the $j$-th site, we have 
\[
\bar{\gamma}_j^{(2)} = \arg\max_{\beta, \Gamma}\beta_j L_j(\tilde\beta^{(1)}, \gamma_j). 
\]
Follow the same proof as Step 2 in the proof of Lemma S.4, we have 
\[
\E\|\bar{\gamma}_j^{(2)}-\gamma_j^*\|_2^4\le \frac{1}{n^2}.
\]
Also, we have 
\begin{align*}
0 = \nabla\gamma L_j(\tilde\beta^{(1)}, \bar\gamma_j^{(2)}) = \nabla_\gamma L_j(\beta^*, \gamma_j^*) +\nabla_{\gamma\beta} L_j(\beta', \gamma_j')(\tilde\beta^{(1)}-\beta^*)+\nabla_{\gamma\gamma} L_j(\beta', \gamma_j')(\bar{\gamma}_j^{(2)}-\gamma_j^*)
\end{align*}
and by reorganizing the above equation we have 
\begin{align*}
I_{\gamma\gamma}^{(j)}(\bar{\gamma}_j^{(2)}-\gamma_j^*) =& \nabla_\gamma L_j(\beta^*, \gamma_j^*) -I^{(j)}_{\gamma\beta} (\tilde\beta^{(1)}-\beta^*)+\{\nabla_{\gamma\gamma} L_j(\beta', \gamma_j')+I_{\gamma\gamma}^{(j)}\}(\bar{\gamma}_j^{(2)}-\gamma_j^*)\\&
+\{\nabla_{\gamma\beta} L_j(\beta', \gamma_j')+I^{(j)}_{\gamma\beta}\}(\tilde\beta^{(1)}-\beta^*).
\end{align*}
By Lemma S.7, we have 
\begin{align*}
\tilde\beta^{(1)}-\beta^*= \{\sum_{j=1}^{K}I^{(j)}_{\beta|\gamma}\}^{-1}\sum_{j=1}^{K}\{\nabla_\beta L_j (\beta^*, \gamma_j^*)-I^{(j)}_{\beta\gamma}{I^{(j)}_{\gamma\gamma}}^{-1}\nabla_{\gamma}L_j(\beta^*, \gamma_j^*)\}+\delta_\beta +\{\tilde\beta^{(1)}-\hat\beta\},
\end{align*}
where $\delta_\beta$ is the subvector corresponding to $\beta$ defined in Lemma S.5.
Thus, we have 
\begin{align*}
I_{\gamma\gamma}^{(j)}(\bar{\gamma}_j^{(2)}-\gamma_j^*) =& \nabla_\gamma L_j(\beta^*, \gamma_j^*) -I^{(j)}_{\gamma\beta}\{\sum_{j=1}^{K}I^{(j)}_{\beta|\gamma}\}^{-1}\sum_{j=1}^{K}\{\nabla_\beta L_j (\beta^*, \gamma_j^*)-I^{(j)}_{\beta\gamma}{I^{(j)}_{\gamma\gamma}}^{-1}\nabla_{\gamma}L_j(\beta^*, \gamma_j^*)\} \\&+\{\nabla_{\gamma\gamma} L_j(\beta', \gamma_j')+I_{\gamma\gamma}^{(j)}\}(\bar{\gamma}_j^{(2)}-\gamma_j^*)
+\{\nabla_{\gamma\beta} L_j(\beta', \gamma_j')+I^{(j)}_{\gamma\beta}\}(\tilde\beta^{(1)}-\beta^*)
\\&-I^{(j)}_{\gamma\beta}\{\delta_\beta +\tilde\beta^{(1)}-\hat\beta\}.
\end{align*}
Define $\bar \delta^{(2)} = (I_{\gamma\gamma}^{(j)})^{-1}[\{\nabla_{\gamma\gamma} L_j(\beta', \gamma_j')+I_{\gamma\gamma}^{(j)}\}(\bar{\gamma}_j^{(2)}-\gamma_j^*)
+\{\nabla_{\gamma\beta} L_j(\beta', \gamma_j')+I^{(j)}_{\gamma\beta}\}(\tilde\beta^{(1)}-\beta^*)
-I^{(j)}_{\gamma\beta}\{\delta_\beta +\tilde\beta^{(1)}-\hat\beta\}]$. We have 
\[
\E\|\{\nabla_{\gamma\gamma} L_j(\beta', \gamma_j')+I_{\gamma\gamma}^{(j)}\}(\bar{\gamma}_j^{(2)}-\gamma_j^*)\|_2^2\lesssim\frac{1}{n^2},
\]
\[
\E\|\{\nabla_{\gamma\beta} L_j(\beta', \gamma_j')+I^{(j)}_{\gamma\beta}\}(\tilde\beta^{(1)}-\beta^*)\|_2^2\lesssim\frac{1}{Kn^2}+\frac{1}{n^3},
\]
and 
\[
\E\|I^{(j)}_{\gamma\beta}\{\delta_\beta +\tilde\beta^{(1)}-\hat\beta\}\|^2_2 \lesssim \frac{1}{n^2}, 
\]
which implies $\E\|\bar \delta^{(2)}\|_2^2\lesssim \frac{1}{n^2}$. $\hfill\square$.

\subsection*{Proof of Lemma S.9}
\textit{Part I:}
For the $j$-th site, we define the following events

\[
\mathcal{E}_0 :=\{ \frac{1}{n}\sum_{i=1}^{n}m_1(Y_{ij}) \le  2M\},
\]

\[
\mathcal{E}_1 := \{\|\nabla^2 L_j(\theta_j^*) - \E\nabla^2L_j(\theta_j^*)\|_2 \le C_3\},
\]

and 

\[
\mathcal{E}_2 := \{\|\nabla L_j(\theta_j^*)\|_2 \le C_4\}.
\]
for some constants $C_3$ and $C_4$ which satisfy $C_3\le \rho\mu_-/2$ and $C_4<(1-\rho)\rho\mu_-^2/8M$.  By replacing $F_1(\theta)$, $F_0(\theta)$ by $L_j(\theta_j)$ and $F_j(\theta_j)$ to Lemma 6 in \cite{zhang2012communication}, we obtain that under event $\mathcal{E} = \cap_{i=0,1,2}\mathcal{E}_i$, we have
\[
\nabla^2 L_j(\theta_j) \succeq (1-\rho)\mu_-\text{I}_d
\]
for $\theta_j\in U(\delta_{\rho})$, where $\delta_{\rho}\le \mu_-\rho/4M$. Also, we have for any $\theta_j' \in U(\delta_{\rho})$, 
\begin{align*}
&\|-\nabla^2 L_j(\theta_j') - I^{(j)}\|_2\le\|-\nabla^2 L_j(\theta_j')+\nabla^2 L_j(\theta_j^*)\|_2 + \|\nabla^2 L_j(\theta_j^*)+I^{(j)}\|_2 \\&\le M\|\theta_j'-\theta_j^*\|_2+\rho\mu_-/2
\end{align*}
Since $\delta_{\rho}\le \mu_-\rho/4M$, we have 
\begin{align*}
\|-\nabla^2 L_j(\theta_j')\|_2 - \|I^{(j)}\|_2\le\rho \mu_-
\end{align*}
Thus we have $\|-\nabla^2 L_j(\theta_j')\|_2 \le2 \mu_+$. By Lemma S.2, we know that $\p\{\mathcal{E}^c\}\lesssim\exp(-n)$.

\textit{Part II:}
For the $j$-th site, we define 
\[
\tilde L_j = \frac{1}{n}\sum_{i=1}^{n}\log f(y_{i1};\beta,\gamma_j)\frac{f(y_{i1};\bar \beta, \bar \gamma_j)}{f(y_{i1}; \bar \beta, \bar \gamma_1)}
\]
and we define the following events

\[
\mathcal{E}'_0 :=\{ \frac{1}{n}\sum_{i=1}^{n}m_1(Y_{ij}) \le  2M, \text{ for } k = 1, 2\},
\]

\[
\mathcal{E}'_1 := \{\|\nabla^2 \tilde L_j - \E\nabla^2L_j(\theta_j^*)\|_2 \le C_3\},
\]

and 

\[
\mathcal{E}'_2 := \{\|\nabla \tilde L_j(\theta_j^*)\|_2 \le C_4\}.
\]
By replacing $F_1(\theta)$, $F_0(\theta)$ by $\tilde L_j(\theta_j)$ and $F_j(\theta_j)$ to Lemma 6 in \cite{zhang2012communication}, we obtain that under event $\mathcal{E}' = \cap_{i=0,1,2}\mathcal{E}'_i$, we have
\[
\nabla^2 \tilde L_j(\theta_j) \succeq (1-\rho)\mu_-\text{I}_d
\]
for $\theta_j\in U(\delta_{\rho})$, where $\delta_{\rho}\le \mu_-\rho/4M$. Also, for any $\theta_j' \in U(\delta_{\rho})$, similarly as Part I, we have $\|-\nabla^2 L_j(\theta_j')\|_2 \le2 \mu_+$. 
Now we calculate  $\p\{\mathcal{E}'^c\}$
We have $\p\{\mathcal{E}'^c_0\}\lesssim\exp(-n)$, and for , we have$\mathcal{E}'^c_1$,  
Under $\mathcal{E}'_0$, we have
\begin{align*}
&\|\nabla^2 \tilde L_j - I^{(j)}\|_2\le \|\frac{1}{n}\sum_{i=1}^{n}\{\nabla^2\log f(y_{i1};\beta,\gamma_j)\frac{f(y_{i1};\bar \beta, \bar \gamma_j)}{f(y_{i1}; \bar \beta, \bar \gamma_1)} -\nabla^2\log f(y_{i1};\beta,\gamma_j)\frac{f(y_{i1}; \beta^*,  \gamma_j^*)}{f(y_{i1}; \beta^*,  \gamma_1^*)}\}\|\\&
+\|\frac{1}{n}\sum_{i=1}^{n}\nabla^2\log f(y_{i1};\beta,\gamma_j)\frac{f(y_{i1}; \beta^*,  \gamma_j^*)}{f(y_{i1}; \beta^*,  \gamma_1^*)}- I^{(j)}\|_2\\&
\le 2M\{\|\bar{\beta}-\beta^*\|_2+\|\bar{\gamma}_j-\gamma_j^*\|_2\}+\|\frac{1}{n}\sum_{i=1}^{n}\nabla^2\log f(y_{i1};\beta,\gamma_j)\frac{f(y_{i1}; \beta^*,  \gamma_j^*)}{f(y_{i1}; \beta^*,  \gamma_1^*)}- I^{(j)}\|_2.
\end{align*}
We have 
\[
\p\{\|\frac{1}{n}\sum_{i=1}^{n}\nabla^2\log f(y_{i1};\beta,\gamma_j)\frac{f(y_{i1}; \beta^*,  \gamma_j^*)}{f(y_{i1}; \beta^*,  \gamma_1^*)}- I^{(j)}\|_2>C_1/3\}\lesssim\exp(-n).
\]
By Lemma S.2, we know that 
\begin{align*}
\bar{\theta}_j-\theta_j =  
{I^{(j)}}^{-1}\nabla L_j({\theta}_j^*)+\delta_j.
\end{align*}
Under $\mathcal{E}$,
\begin{align*}
\|\delta_j\|_2 \le \frac{MC_1^2}{\mu_-}\|\nabla L_j(\theta_j^*)\|_2^2 +\frac{C_1}{\mu_-}\|\nabla L_j(\theta_j^*)\|_2\|\nabla^2 L_j({\theta}_j^*)+{I^{(j)}}\|_2.
\end{align*}
Thus, we have for any $C>0$, 
\begin{align*}
&\p\{\|\delta_j\|_2 >C\}\le\p\{\frac{MC_1^2}{\mu_-}\|\nabla L_j(\theta_j^*)\|_2^2>C/2\}\\&+\p\{\frac{C_1}{\mu_-}\|\nabla L_j(\theta_j^*)\|_2\|\nabla^2 L_j({\theta}_j^*)+{I^{(j)}}\|_2>C/2\}\\
&\le\p\{\|\nabla L_j(\theta_j^*)\|_2>C_3\} +\p\{\frac{C_1}{\mu_-}\|\nabla L_j(\theta_j^*)\|_2\|>\surd{(C/2)}\}\\&+ \p\{\|\nabla^2 L_j({\theta}_j^*)+{I^{(j)}}\|_2>\surd{(C/2)}\}\lesssim \exp(-n).
\end{align*}

And we have, 
\begin{align*}
\bar{\beta}-\beta^* =  \frac{1}{K}\sum_{j=1}^{K}\{(I^{(j)}_{\beta|\gamma})^{-1}\nabla_{\beta}L_j(\theta_j^*)-(I^{(j)}_{\beta|\gamma})^{-1}I^{(j)}_{\beta\gamma}(I^{(j)}_{\gamma\gamma})^{-1}\nabla_{\gamma}L_j(\theta_j^*)\} +\frac{1}{K}\sum_{j=1}^{K}\delta_{\beta,j}
\end{align*}
and 
\begin{align*}
\|\bar{\beta}-\beta^*\|_2 \le& \|\frac{1}{K}\sum_{j=1}^{K}\{(I^{(j)}_{\beta|\gamma})^{-1}\nabla_{\beta}L_j(\theta_j^*)-(I^{(j)}_{\beta|\gamma})^{-1}I^{(j)}_{\beta\gamma}(I^{(j)}_{\gamma\gamma})^{-1}\nabla_{\gamma}L_j(\theta_j^*)\}\|_2 \\&+\|\frac{1}{K}\sum_{j=1}^{K}\delta_{\beta,j}\|_2.
\end{align*}
Thus, we have 
\begin{align*}
&\p\{2M\|\bar{\beta}-\beta^*\|>C_1/3\}\\&\le \p\{\|\frac{1}{K}\sum_{j=1}^{K}\{(I^{(j)}_{\beta|\gamma})^{-1}\nabla_{\beta}L_j(\theta_j^*)-(I^{(j)}_{\beta|\gamma})^{-1}I^{(j)}_{\beta\gamma}(I^{(j)}_{\gamma\gamma})^{-1}\nabla_{\gamma}L_j(\theta_j^*)\}\|_2\|>C_1/(12M)\}\\& +\p\{\|\frac{1}{K}\sum_{j=1}^{K}\delta_{j}\|_2>C_1/(12M)\}
\lesssim \exp(-n),
\end{align*}
and  similarly, we have 
\begin{align*}
&\p\{2M\|\bar{\gamma}_j-\gamma_j^*\|>C_1/3\}
\lesssim \exp(-n).
\end{align*}
In summary
\[
\p(\mathcal{E}'^c)\le \p(\mathcal{E}_0'^c)+\p(\mathcal{E}_0\cap\mathcal{E}_1^c)+\p(\mathcal{E}\cap\mathcal{E}_2^c)+\p(\mathcal{E}^c)\lesssim\exp(-n)
.\]
$\hfill\square$

\subsection*{Proof of Lemma S.10}
By Lemma S.2, we have 
\begin{align*}
\bar{\beta}-\beta^* = \frac{1}{K}\sum_{j=1}^{K}\bar\beta_j = \frac{1}{K}\sum_{j=1}^{K}\{(I^{(j)}_{\beta|\gamma})^{-1}\nabla_{\beta}L_j(\theta_j^*)-(I^{(j)}_{\beta|\gamma})^{-1}I^{(j)}_{\beta\gamma}(I^{(j)}_{\gamma\gamma})^{-1}\nabla_{\gamma}L_j(\theta_j^*)\} +\frac{1}{K}\sum_{j=1}^{K}\delta_{\beta,j},
\end{align*}
where $\delta_{\beta,j}$ is the subvector of $\delta_j$ defined in Lemma S.2. 

By Theorem 3 in \cite{zhang2018non}, we have 
\begin{align*}
&\E\| \frac{1}{K}\sum_{j=1}^{K}\{(I^{(j)}_{\beta|\gamma})^{-1}\nabla_{\beta}L_j(\theta_j^*)-(I^{(j)}_{\beta|\gamma})^{-1}I^{(j)}_{\beta\gamma}(I^{(j)}_{\gamma\gamma})^{-1}\nabla_{\gamma}L_j(\theta_j^*)\}\|_2 \\& = \int_0^\infty\p\{\| \frac{1}{K}\sum_{j=1}^{K}\{(I^{(j)}_{\beta|\gamma})^{-1}\nabla_{\beta}L_j(\theta_j^*)-(I^{(j)}_{\beta|\gamma})^{-1}I^{(j)}_{\beta\gamma}(I^{(j)}_{\gamma\gamma})^{-1}\nabla_{\gamma}L_j(\theta_j^*)\}\|_2>t\}dt\\
&\ge\int_0^\infty  C\exp(-Knt)dt \gtrsim \frac{1}{Kn}. 
\end{align*}
Also, by Lemma S.2, we have 
\[
\E \|\frac{1}{K}\sum_{j=1}^{K}\delta_{\beta,j}\|_2 \lesssim \frac{1}{n}. 
\]
Thus, when $K/n \rightarrow 0$, we have 
\begin{align*}
&\E\|\bar{\beta}-\beta^*\|_2\ge\E\| \frac{1}{K}\sum_{j=1}^{K}\{(I^{(j)}_{\beta|\gamma})^{-1}\nabla_{\beta}L_j(\theta_j^*)-(I^{(j)}_{\beta|\gamma})^{-1}I^{(j)}_{\beta\gamma}(I^{(j)}_{\gamma\gamma})^{-1}\nabla_{\gamma}L_j(\theta_j^*)\}\|_2\\
&-\E \|\frac{1}{K}\sum_{j=1}^{K}\delta_{\beta,j}\|_2 \gtrsim \frac{1}{Kn}.
\end{align*}
$\hfill\square$

\subsection*{Proof of Lemma S.11}
We have 
\begin{align*}
\E_{\theta_j^*} g(Y_j) = \int_{y} g(y)f(y; \theta_j^*)dy = \int_{y} g(y)\frac{f(y; \theta_j^*)}{f(y; \theta_1^*)}f(y; \theta_1^*)dy = \E_{\theta_1^*} g(Y_1).
\end{align*}
$\hfill\square$

\subsection*{Proof of Lemma S.12} 
In Lemma S.7 already showed that 
\[
\E\{\|\tilde{\beta}^{(1)}- \check{\beta}\|_2^4\}\lesssim\frac{1}{K^2n^4}+\frac{1}{n^6}.
\] 
Now we only need to show that 
\[
\E\{\|\tilde{\beta}^{(O)}- \check{\beta}\|_2^4\}\lesssim\frac{1}{K^2n^4}+\frac{1}{n^6},
\] 
which will imply the desired result.

According to the definition of $\tilde{\beta}^{(O)}$, it is the solution of the estimating equation
\[
\{\frac{1}{K}\sum_{j=1}^{K}\{\nabla_\beta L_j (\bar\beta, \bar\gamma_j)- \bar H^{(j)}_{\beta\gamma} ({{{{\bar H}}_{\gamma\gamma}}^{(j)}})^{-1}  \nabla_\gamma L_j (\bar\beta, \bar\gamma_j)\}+\nabla_\beta U_1(\bar \beta) (\beta-\bar{\beta})=0
\]
and $\check\beta$  is the solution of the estimating equation
\[
0 = \frac{1}{K}\sum_{j=1}^{K}\{\nabla_\beta L_j (\beta, \bar\gamma_j)- \bar H^{(j)}_{\beta\gamma} ({{{{\bar H}}_{\gamma\gamma}}^{(j)}})^{-1}  \nabla_\gamma L_j (\beta, \bar\gamma_j)\}.
\]
We now define
\[
\mathcal{E}'_{0j} :=\{ \frac{1}{n}\sum_{i=1}^{n}m_k(Y_{ij}) \le  2M, \text{ for } k = 1, 2\},
\]

\[
\mathcal{E}'_1 := \{\|\nabla_\beta U_1(\bar\beta)-\frac{1}{K}\sum_{j=1}^{K}\{\nabla_{\beta\beta}L_j(\check\beta, \bar{\gamma}_j) -\bar H^{(j)}_{\beta\gamma} ({{{{\bar H}}_{\gamma\gamma}}^{(j)}})^{-1}\nabla_{\gamma\beta}L_j(\check\beta, \bar{\gamma}_j)\}\|_2\le C_1\}
\]
\[
\mathcal{E}'_2 := \{\|\frac{1}{K}\sum_{j=1}^{K}\{\nabla_\beta L_j (\bar\beta, \bar\gamma_j)- \bar H^{(j)}_{\beta\gamma} ({{{{\bar H}}_{\gamma\gamma}}^{(j)}})^{-1}  \nabla_\gamma L_j (\bar\beta, \bar\gamma_j)\}+\nabla_\beta U_1(\bar \beta) (\check\beta-\bar{\beta})\|_2 \le C_2\},
\] 
for some constants $M$, $C_1$ and $C_2$ which satisfy $\E\{ m_k (Y_{ij})\}< M$ for all $j \in \{1, \dots, K\}$, and $k = 1, 2$, $C_1\le \rho\mu_-/2$ and $C_2<(1-\rho)\rho\mu_-^2/8M$. Let $\mathcal{E}_0 = \cap_{1\le j\le K}\mathcal{E}_{0j}$. Applying Lemma 6 in \cite{zhang2012communication} we have under
event $\mathcal{E'} = \{\cap_{i=0,1,2}\mathcal{E}'_i\}$, 
\[
\|\tilde{\beta}^{O} -\check{\beta}\|^4_2 \le C\|\frac{1}{K}\sum_{j=1}^{K}\{\nabla_\beta L_j (\bar\beta, \bar\gamma_j)- \bar H^{(j)}_{\beta\gamma} ({{{{\bar H}}_{\gamma\gamma}}^{(j)}})^{-1}  \nabla_\gamma L_j (\bar\beta, \bar\gamma_j)\}+\nabla_\beta U_1(\bar \beta) (\check\beta-\bar{\beta})\|^4_2.
\]
Since
\[
0 = \frac{1}{K}\sum_{j=1}^{K}\{\nabla_\beta L_j (\check\beta, \bar\gamma_j)- \bar H^{(j)}_{\beta\gamma} ({{{{\bar H}}_{\gamma\gamma}}^{(j)}})^{-1}  \nabla_\gamma L_j (\check\beta, \bar\gamma_j)\},
\]
we have 
\begin{align*}
&\frac{1}{K}\sum_{j=1}^{K}\{\nabla_\beta L_j (\bar\beta, \bar\gamma_j)- \bar H^{(j)}_{\beta\gamma} ({{{{\bar H}}_{\gamma\gamma}}^{(j)}})^{-1}  \nabla_\gamma L_j (\bar\beta, \bar\gamma_j)\}+\nabla_\beta U_1(\bar \beta) (\check\beta-\bar{\beta}) \\
& = \{\nabla_\beta U_1(\bar\beta)  - \frac{1}{K}\sum_{j=1}^{K}\{\nabla_\beta L_j (\beta', \bar\gamma_j)- \bar H^{(j)}_{\beta\gamma} ({{{{\bar H}}_{\gamma\gamma}}^{(j)}})^{-1}  \nabla_\gamma L_j (\beta', \bar\gamma_j)\}\}(\check{\beta}- \bar{\beta}),
\end{align*}
where $\beta'$ satisfies $\|\beta'-\check\beta\|_2\le \|\bar\beta-\check\beta\|_2$.

We note that the above equation is the same as equation (\ref{eeee}) in the proof of Lemma S.7, only with $\nabla_\beta U_1(\beta')$ changed to $\nabla_\beta U_1(\bar\beta)$, which satisfies $\|\beta'-\check\beta\|_2\le \|\bar\beta-\check\beta\|_2$. Follow the same procedure, we are able to obtain the same conclusion 
under
event $\mathcal{E'}$
\begin{align*}
\E\{\|\frac{1}{K}\sum_{j=1}^{K}\{\nabla_\beta L_j (\bar\beta, \bar\gamma_j)- \bar H^{(j)}_{\beta\gamma} ({{{{\bar H}}_{\gamma\gamma}}^{(j)}})^{-1}  \nabla_\gamma L_j (\bar\beta, \bar\gamma_j)\}+\nabla_\beta U_1(\bar \beta) (\check\beta-\bar{\beta})\|_2^8I(\mathcal{E'})\}\le\frac{1}{K^2n^4}+\frac{1}{n^{6}}. 
\end{align*}

We also observe that $\mathcal{E}'_{0j} $ is the same as $\mathcal{E}_{0j} $ defined in the proof of Lemma S.7. In the definition of $\mathcal{E}'_{1} $, it is the same as $\mathcal{E}_{1} $ with $\nabla_\beta U_1(\bar\beta)$ replaced by $\nabla_\beta U_1(\check\beta)$. 
By Lemma S.2, we have 
\begin{align*}
\bar{\beta}-\beta^* = \frac{1}{K}\sum_{j=1}^{K}\bar\beta_j-\beta^* = \frac{1}{K}\sum_{j=1}^{K}\{(I^{(j)}_{\beta|\gamma})^{-1}\nabla_{\beta}L_j(\theta_j^*)-(I^{(j)}_{\beta|\gamma})^{-1}I^{(j)}_{\beta\gamma}(I^{(j)}_{\gamma\gamma})^{-1}\nabla_{\gamma}L_j(\theta_j^*)\} +\frac{1}{K}\sum_{j=1}^{K}\delta_{\beta,j}
\end{align*}
where $\delta_{\beta,j}$ is the subvector of $\delta_j$ defined in Lemma S.2. Thus, we have 
\[
\E \|\bar{\beta}-\beta^*\|_2^8 \lesssim \frac{1}{K^4n^4} +\frac{1}{n^8}.
\]
Follow the same derivation of Lemma S.7, we can replace $\check\beta$ by $\bar\beta$ and use the above property of $\bar\beta$ whenever we need to use the property \[
\E \|\check{\beta}-\beta^*\|_2^8 \lesssim \frac{1}{K^4n^4} +\frac{1}{n^8}.
\]
We can show that  $$\p(\mathcal{E'}^c)\lesssim \frac{1}{K^2n^4}+\frac{1}{n^{6}},$$ which completes the proof.

\end{document}